\documentclass[
 reprint,
 showpacs, preprintnumbers,
 amsmath, amssymb,
 aps,
 pra,
 floatfix,
]{revtex4-1}

\usepackage{pdfpages}
\makeatletter
\AtBeginDocument{\let\LS@rot\@undefined}
\makeatother

\usepackage{graphicx}
\usepackage{tikz}
\usepackage{xcolor}
\usetikzlibrary{arrows}

\makeatletter
\newcommand{\colorcaption}[2][]{%
  \begingroup%
  \renewcommand{\@caption@fignum@sep}{ (Color online). }%
  \caption[#1]{#2}%
  \endgroup%
}
\makeatother

\usepackage{dcolumn}
\usepackage{bm}
\usepackage{bbm}
\usepackage{upgreek}
\usepackage{hyperref}
\usepackage{blindtext}

\usepackage{changes}

\usepackage{dsfont}

\definecolor{darkgreen}{RGB}{0 100 0}

\begin{document}

	\preprint{}

	\title{Connection between zero-energy Yu-Shiba-Rusinov states and $ \pmb{0} $-$ \pmb{\pi} $~transitions in~magnetic Josephson~junctions}
	
	\author{Andreas Costa}%
	\email[Corresponding author: ]{andreas.costa@physik.uni-regensburg.de}
 	\affiliation{Institute for Theoretical Physics, University of Regensburg, 93040 Regensburg, Germany}
 	
	\author{Jaroslav Fabian}%
 	\affiliation{Institute for Theoretical Physics, University of Regensburg, 93040 Regensburg, Germany}
 	
 	\author{Denis Kochan}%
 	\affiliation{Institute for Theoretical Physics, University of Regensburg, 93040 Regensburg, Germany}

	\date{\today}

\begin{abstract}
	We study theoretically the bound state spectrum and $ 0 $-$ \pi $ transitions in ballistic quasi-one-dimensional superconductor/ferromagnetic insulator/superconductor Josephson junctions. In addition to the Andreev bound states, stemming from the phase coherence, the magnetic barrier gives rise to qualitatively different Yu-Shiba-Rusinov~(YSR) bound states with genuine spectral features and spin characteristics. We show that zero-energy YSR states are much more robust against the presence of scalar tunneling than their Andreev counterparts and also fingerprint a quantum phase transition from the junctions' $ 0 $ into the $ \pi $ phase, connected to a measurable reversal of the Josephson current; this evidence persists also in the presence of Rashba spin-orbit coupling. 

\end{abstract}

\maketitle

\section{Introduction  \label{SecI}}

Since its discovery, superconductivity evolved into a most influential area in fundamental science and technology. While the exchange interaction in metals favors parallel spins~\cite{Stoner1939}, the $ s $-wave pairing in superconductors promotes the formation of Cooper pairs with antiparallel spin alignments~\cite{Bardeen1957a, Bardeen1957b}. The competition of those two antagonistic interactions within one system leads to interesting physical phenomena~\cite{Eschrig2011, Linder2015, Gingrich2016}. Prominent examples are S/F/S Josephson junctions~\cite{Bulaevskii1977,*Bulaevskii1977alt,Buzdin1982,*Buzdin1982alt,Andreev1991,Demler1997,Golubov2004,Buzdin2005,Bergeret2005,Annunziata2011,Costa2017}, in which a leakage of Cooper pairs from the superconducting (S) electrodes  introduces a nontrivial pairing in the proximitized ferromagnet (F). In response to the spin-selective exchange splitting in the F, the induced order parameter oscillates with a characteristic spatial length~\cite{Andreev1991,Demler1997}; depending on the thickness of the F, the phase difference between the two S electrodes can accumulate an intrinsic $ \pi $~shift. That is responsible for the reversal of the Josephson current direction in such a $ \pi $~state junction regime as compared to its (usual) $ 0 $~state counterpart.

Another realization of $ \pi $~Josephson junctions relies on the coupling of S electrodes via interacting quantum dots~(QDs). Several theoretical works~\cite{Glazman1989,*Glazman1989alt,Clerk2000,Vecino2003,Choi2004,Siano2004,*Siano2004alt,Bauer2007,Karrasch2008,Meng2009,Luitz2010,Rodero2011,Luitz2012,Kirsanskas2015} showed that the junction regimes can be controlled by the strength of the lead-QD coupling and the QD charging energy. Experimental observations of $ 0 $-$ \pi $~transitions in S/F/S~\cite{Ryazanov2001,Kontos2002,Robinson2006,Feofanov2010} and S/QD/S~\cite{Buitelaar2002,Jorgensen2007,Eichler2009,Deacon2010,Pillet2010,Franke2011,Lee2012,Maurand2012,*Maurand2012alt,Chang2013,Kim2013,Pillet2013,Kumar2014,Lee2014,Delagrange2015,Gingrich2016,Assouline2017,Li2017,VanWoerkom2017} Josephson junctions boosted hopes for their engineering and designed technological applications, counting qubits~\cite{Yamashita2005}, quantum computing~\cite{Ioffe1999,Mooij1999,Devoret2013}, and spintronics~\cite{Eschrig2011,Fabian2004,Fabian2007}.

An unambiguous spectroscopic fingerprint of Josephson junctions is the formation of subgap Andreev bound states (ABSs)~\cite{Andreev1964,*Andreev1964alt,Andreev1966,*Andreev1966alt}, which have been studied in single~\cite{Lee2014,Delagrange2015,Li2017,VanWoerkom2017} and double~\cite{Su2017} QD-coupled Josephson junctions. In the latter case, the ABSs hybridized to novel Andreev molecular states, which can eventually launch a platform for realizing 
Majorana physics~\cite{Fu2008,Oreg2010,Duckheim2011,Mourik2012,Rokhinson2012,Nadj-Perge2014}. However, understanding the spectral features of Josephson junctions in magnetic systems becomes more intricate since the magnetism breaks Cooper pairs and allows a creation of additional subgap bound states, commonly known as Yu-Shiba-Rusinov (YSR) states~\cite{Yu1965,Shiba1968,Shiba1969,Rusinov1968, *Rusinov1968alt}. YSR states have been intensively studied in various systems, e.g., in S~substrates hosting magnetic adatoms~\cite{Satori1992,Simonin1995,Salkola1997,Balatsky2006,Tsai2009,Ruby2016,Kaladzhyan2017,Koerber2018} or nanowires connecting normal/superconductor junctions~\cite{Zitko2015,Jellinggaard2016}.

In this paper, we investigate the subgap bound states in ballistic S/FI/S Josephson junctions with ultrathin barriers containing ferromagnetic insulators (FIs).  The subgap states possess unique spectral~\cite{Vecino2003,Kawabata2012} and \emph{spin} characteristics that are tunable by tunneling strengths or the S phase difference. Moreover, magnetic tunneling causes an interesting interplay between ABSs and YSR states, modifying, for example, the quasiparticle density of states (DOS)~\cite{Kirsanskas2015,Lee2012}. We pay special attention to zero-energy YSR states which can signal topological superconductivity~\cite{Sau2013,Pientka2015,Hatter2015} and  ground state phase transitions~\cite{Sakurai1970}. We clearly distinguish the $ 0 $ and $ \pi $ phases of the ballistic S/FI/S Josephson junctions~\cite{Andersen2006,Kawabata2010,Kawabata2012} and unravel the $ 0 $-$ \pi $~transition mechanism on the microscopic level: the reversal of the tunneling of Cooper pairs in the YSR channel near zero energy stems from the ground state phase transition. Therefore, besides the two conventional mechanisms explaining $ 0 $-$ \pi $~transitions---(1)~proximity-induced effects in S/F/S and (2)~the interplay between the Fermi statistics and strong
correlations in S/QD/S junctions---we concentrate on the third one: pair tunneling via the spectrally distinct YSR~states at the interface of the magnetic FI barrier.
To demonstrate the universality and the robustness of our mechanism, we also investigate $ 0 $-$ \pi $~transitions in  S/FI/S Josephson junctions in the presence of Rashba spin-orbit coupling (SOC)~\cite{Bychkov1984,Fabian2007} in the S leads, eventually showing the same qualitative behavior. Our findings offer a comprehensive understanding of $ 0 $-$ \pi $~transitions in Josephson junctions.

The paper is organized in the following way. In Sec.~\ref{SecII}, we introduce our theoretical model and study the bound state spectra for some important limiting cases. Section~\ref{SecIII} briefly comments on the states' impact on the quasiparticle DOS. The main part of the paper is the analysis of the connection between the bound state spectrum and the Josephson current reversing $ 0 $-$ \pi $~transitions, which can be found in Secs.~\ref{SecIV}--\ref{SecVI}.

\section{Theoretical model and bound~state~spectrum  \label{SecII}}
\begin{figure}
	\includegraphics[width=0.45\textwidth]{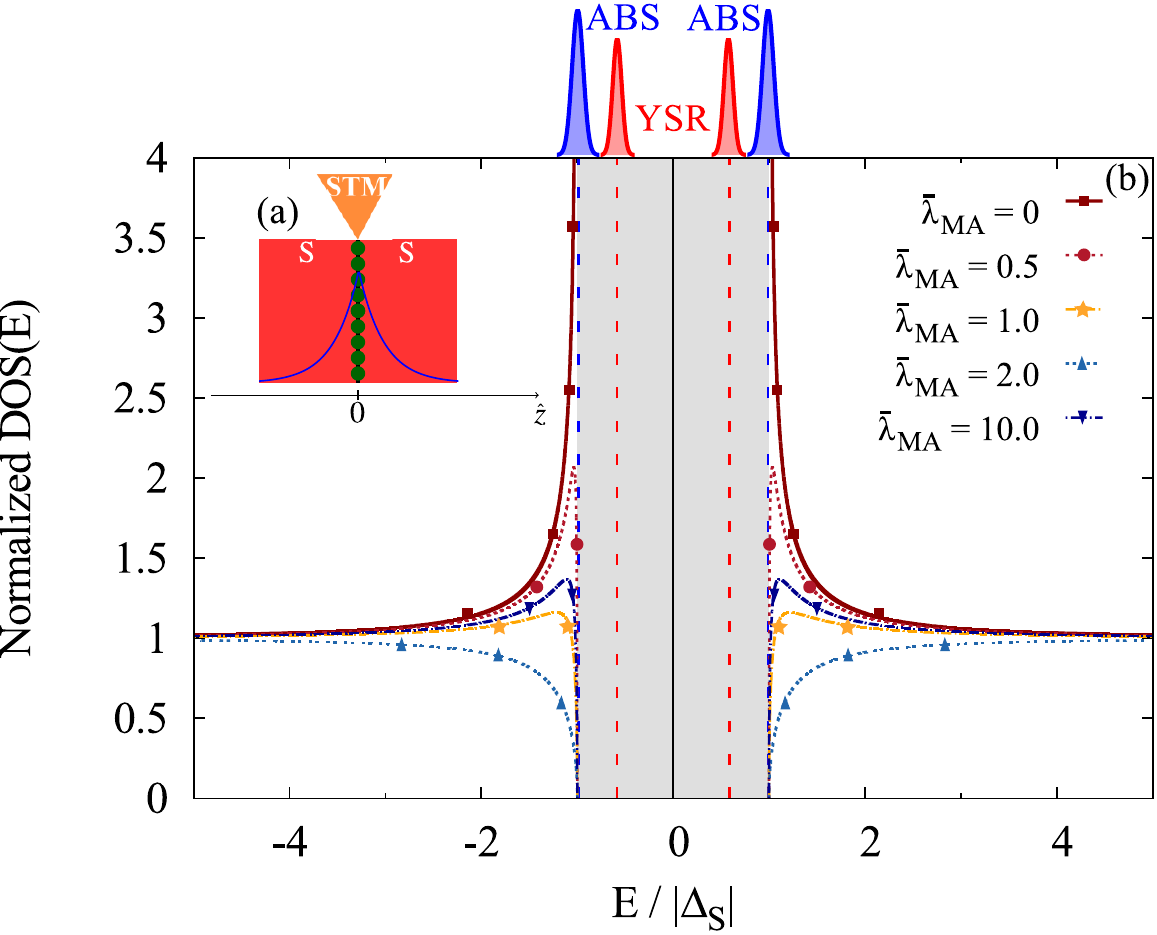}
	\colorcaption{(a)~Josephson junction geometry; two superconductors (red) are separated by a thin ferromagnetic insulator (green dots). (b)~Calculated quasiparticle DOS for $ \overline{\lambda}_\mathrm{SC}=2 $, zero phase difference, and various $ \overline{\lambda}_\mathrm{MA} $'s with pronounced coherence peak modulations. ABSs and YSR states inside the gap are schematically illustrated.
	\label{FigScheme}}
\end{figure}

We consider a vertical ballistic S/FI/S Josephson junction, consisting of two semi-infinite S regions that are separated by a thin deltalike tunnel barrier, simulating scalar and magnetic tunneling; see Fig.~\ref{FigScheme}(a). 
To analyze its spectral properties, we model the junction in terms of the stationary Bogoljubov--de~Gennes Hamiltonian~\cite{DeGennes1989},
\begin{equation}
\hat{H}_{\rm BdG}=\left[ \begin{matrix} \hat{H}_\mathrm{e} & \hat{\Delta}_\mathrm{S}(z) \\ \hat{\Delta}_\mathrm{S}^\dagger(z) & \hat{H}_\mathrm{h} \end{matrix} \right] \,.  
\label{tEqBdG}
\end{equation}
Here,
$ \hat{H}_\mathrm{e} = ( -\hbar^2/2m \boldsymbol{\nabla}^2 - \mu ) \hat{\sigma}_0 + ( \lambda_\mathrm{SC} \hat{\sigma}_0+ \lambda_\mathrm{MA} \hat{\sigma}_z ) \delta(z) + \hat{H}_\mathrm{R} $
represents the single-electron Hamiltonian and $ \hat{H}_\mathrm{h} = -\hat{\sigma}_y {\hat{H}_\mathrm{e}}^* \hat{\sigma}_y $ its hole counterpart. Scalar and magnetic tunneling at the interface are modeled by deltalike potentials~\cite{DeJong1995, Zutic1999, Zutic2000} with effective coupling strengths $ \lambda_\mathrm{SC}$ and $ \lambda_\mathrm{MA} $, respectively, while $ \hat{H}_\mathrm{R} = -\lambda_\mathrm{R} k_z \hat{\sigma}_y \Theta(z) $ accounts for Rashba SOC in the right lead, realized by a semiconducting electrode with proximity-induced superconductivity~\cite{Yang2012}. Spin matrices $ \hat{\sigma}_0 $ and $ \hat{\sigma}_i $ stand for the two-by-two identity and the $ i $th Pauli matrix. For the sake of simplicity, we assume a symmetric S/FI/S junction with equal quasiparticle masses~$ m $, the same chemical potential~$ \mu $, and the transverse~$ \hat{z} $-dependent S order parameter 
$\hat{\Delta}_\mathrm{S}(z) = | \Delta_\mathrm{S} | \hat{\sigma}_0 \left[ \Theta(-z) + \mathrm{e}^{\mathrm{i} \phi_\mathrm{S}} \Theta(z) \right]$, where $\phi_\mathrm{S} $ is the phase difference.

The general procedure for finding the eigencharacteristics of the Hamiltonian $\hat{H}_{\rm BdG}$, given by Eq.~(\ref{tEqBdG}), is outlined in 
the Supplemental Material (SM)~\footnote{See attached Supplemental Material, including Refs.~\cite{DeGennes1989,McMillan1968,Beenakker1991,Golubov2004,Oreg2010,Duckheim2011,Mourik2012,Rokhinson2012,Nadj-Perge2014,Fu2008,Yu1965,Shiba1968,Rusinov1968,*Rusinov1968alt,Costa2017,Carbotte1990,Andreev1991,Demler1997,Fulde1964,Larkin1964,*Larkin1964alt,Ryazanov2001,Brinkman2000,Kawabata2010}, for more details.}. 
In what follows, we focus on quasi-one-dimensional~(quasi-1D) junctions for which the thickness in the transverse $ \hat{x} $ and $ \hat{y} $~directions is 
much shorter than in the longitudinal $ \hat{z} $~direction (higher-dimensional junctions bring no new features; see SM~\cite{Note1}). The particle-hole symmetric subgap eigenspectrum $ E_{n}^\pm $ ($ n=1,2 $) in the absence of SOC reads
\begin{widetext}
	\footnotesize
	\begin{align}
\frac{E_{n}^\pm}{\left| \Delta_\mathrm{S} \right|} &= \pm\sqrt{  
\frac{ \left( \overline{\lambda}_\mathrm{SC}^2 - \overline{\lambda}_\mathrm{MA}^2 \right)^2+
8\,\left[2\cos^2\frac{\phi_\mathrm{S}}{2} + \frac{\overline{\lambda}_\mathrm{SC}^2}{2} \left( \cos^2 \frac{\phi_\mathrm{S}}{2} + 1 \right) + \frac{\overline{\lambda}_\mathrm{MA}^2}{2} \sin^2 \frac{\phi_\mathrm{S}}{2} + 
(-1)^n |\overline{\lambda}_\mathrm{MA}|\sqrt{\sin^2 \phi_\mathrm{S} + \overline{\lambda}_\mathrm{SC}^2\sin^2\frac{\phi_\mathrm{S}}{2}+ \overline{\lambda}_\mathrm{MA}^2\cos^2\frac{\phi_\mathrm{S}}{2}}\right]}
{ \left( \overline{\lambda}_\mathrm{SC}^2 - \overline{\lambda}_\mathrm{MA}^2+ 4 \right)^2 +16\overline{\lambda}_\mathrm{MA}^2 }}\,,  \label{tEqSpectrum}
	\end{align}

\end{widetext}
where 
$ \overline{\lambda}_\mathrm{SC} = 2m\lambda_\mathrm{SC} / (\hbar^2 q_\mathrm{F}) $ and 
$ \overline{\lambda}_\mathrm{MA} = 2m\lambda_\mathrm{MA} / (\hbar^2 q_\mathrm{F}) $ represent effective tunneling strengths with respect to the Fermi level (chemical potential~$ \mu $); $ q_\mathrm{F}=\sqrt{2m\mu}/\hbar $ stands for the corresponding Fermi momentum.
\begin{figure}
	\includegraphics[width=0.4\textwidth]{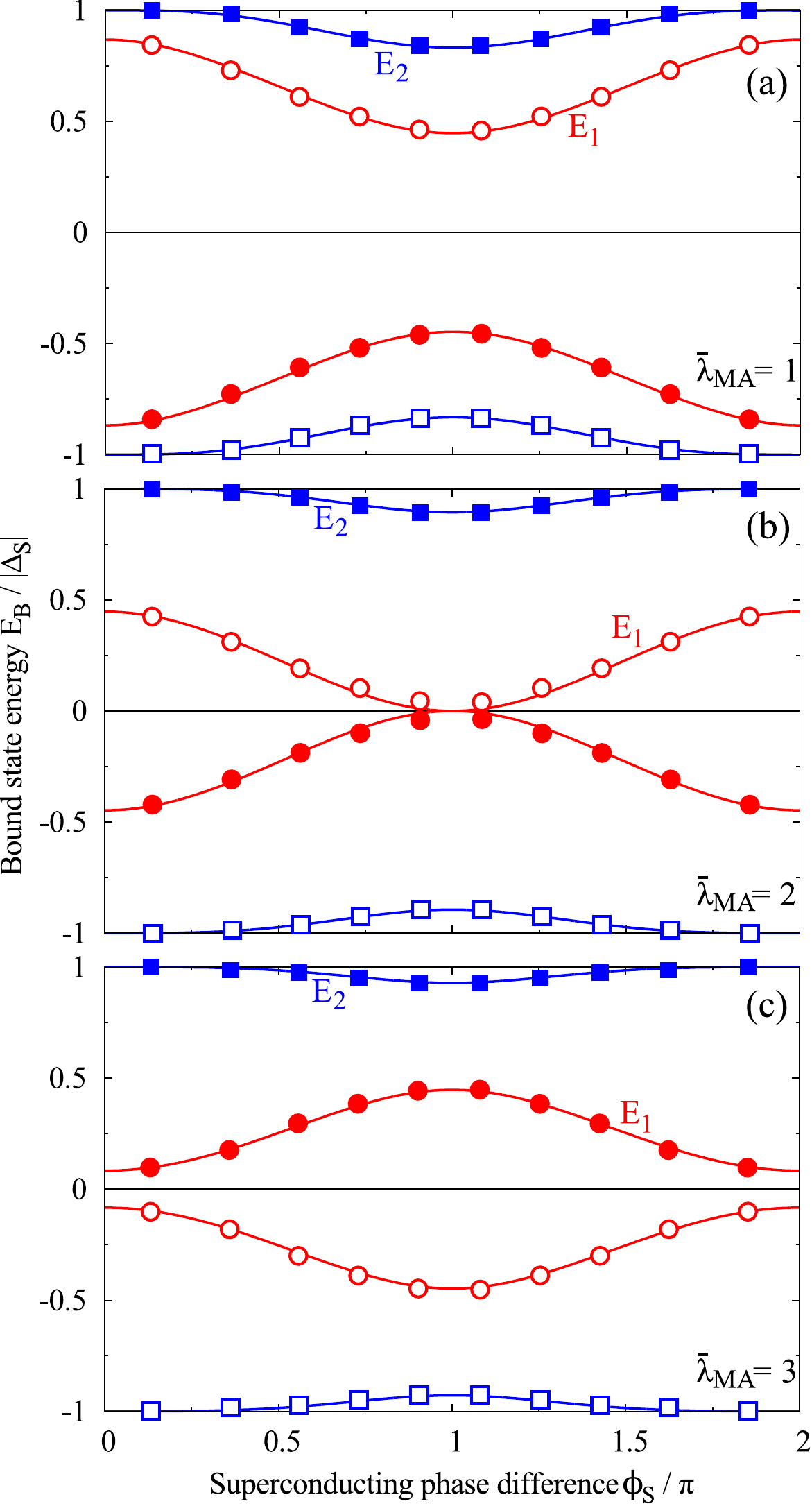}
	\colorcaption{Spin-resolved bound state energies $ E_1 $ and $ E_2 $ of the YSR (red) and the Andreev (blue) states as functions of $ \phi_\mathrm{S} $ for $ \overline{\lambda}_\mathrm{SC}=2 $ and various $ \overline{\lambda}_\mathrm{MA} $'s given in the plots; Rashba SOC is absent. Filled (empty) circles indicate spin up (down) YSR and filled (empty) squares spin up (down) Andreev states. \label{Fig2t}}
\end{figure}

Let us briefly examine the main spectral characteristics of such Josephson junctions. Taking $ \overline{\lambda}_\mathrm{MA}\rightarrow 0 $, we recover the Andreev limit~\cite{Beenakker1991,Golubov2004},
\begin{equation}
E_1^\pm = E_2^\pm = \pm \left| \Delta_\mathrm{S} \right| \sqrt{\frac{\overline{\lambda}_\mathrm{SC}^2 + 4 \cos^2(\phi_\mathrm{S}/2)}{\overline{\lambda}_\mathrm{SC}^2 + 4}} .  
\label{tEqABS}
\end{equation}
Letting $ \overline{\lambda}_\mathrm{SC}\rightarrow 0 $ and $ \phi_\mathrm{S} \rightarrow 0 $, the spectrum complementary yields the celebrated YSR states~\cite{Yu1965, Shiba1968, Rusinov1968,*Rusinov1968alt}
\begin{equation}
	E_1^\pm = \pm \left| \Delta_\mathrm{S} \right|\, \frac{\overline{\lambda}_\mathrm{MA}^2 - 4 }{\overline{\lambda}_\mathrm{MA}^2 + 4}
	\label{tEqYSRS}
\end{equation}
and two remaining states at $ E_2^\pm = \pm | \Delta_\mathrm{S} | $, which coincide with the ABS at $ \phi_\mathrm{S}=0 $; see Eq.~\eqref{tEqABS}.
For that reason, we will refer to $ E_1^\pm $ as the \emph{YSR} and to $ E_2^\pm $ as the \emph{Andreev} branch.

Figure~\ref{Fig2t} shows a generic spin-resolved spectrum as a function of $ \phi_\mathrm{S} $ for various $ \overline{\lambda}_\mathrm{MA} $. Generally, the Andreev branch is always closer to the gap edges than the YSR branch, serving as a spectroscopic fingerprint for distinguishing those states. While the Andreev states, given by Eq.~\eqref{tEqABS}, cross zero energy only for a transparent interface and phase differences $ \phi_\mathrm{S}=\pi \, (\mathrm{mod} \, 2\pi) $~\cite{Beenakker1991}, Eq.~\eqref{tEqSpectrum} suggests that additional magnetic tunneling supports zero-energy YSR states in a wide range of parameters. Analyzing Eq.~(\ref{tEqSpectrum}), one sees that for $ 0 \leq \overline{\lambda}_\mathrm{MA}^2 - \overline{\lambda}_\mathrm{SC}^2 \leq 4 $, there always exists a $ \phi_\mathrm{S} $ for which the YSR branch crosses zero energy; see Fig.~\ref{Fig2t}. The corresponding $ \phi_\mathrm{S} $ comes as a solution of
\begin{equation}
	\overline{\lambda}_\mathrm{MA} = \pm \sqrt{\overline{\lambda}_\mathrm{SC}^2+4 \cos^2(\phi_\mathrm{S}/2)}\,.  
\label{tEqCondition}
\end{equation}

\section{Modulation of quasiparticle DOS  \label{SecIII}}
The subgap states strongly impact the quasiparticle DOS. Figure~\ref{FigScheme}(b) shows the DOS for $ \overline{\lambda}_\mathrm{SC}=2 $ and different $ \overline{\lambda}_\mathrm{MA} $'s at zero phase difference (for methodology, see SM~\cite{Note1}). Without magnetic tunneling, the spectrum only consists of ABS and the quasiparticle spectrum shows the standard BCS-like DOS. Gradually growing $ \overline{\lambda}_\mathrm{MA} $, the spectrum also contains YSR states that move to the gap center. As a consequence, a part of spectral weight is taken into the gap and the quasiparticle coherence peaks modify. A similar peak structure was identified in quantum-dot experiments~\cite{Kumar2014, Li2017}. Raising $ \overline{\lambda}_\mathrm{MA} $, the quasiparticle spectral peaks become unprecedentedly suppressed and disappear when the YSR states cross zero energy. A further increase of $ \overline{\lambda}_\mathrm{MA} $ shifts the bound states back towards the gap edges and, simultaneously, the quasiparticle DOS rises again.

\section{Qualitative analysis---ground state phase transition  \label{SecIV}}
We see from Fig.~\ref{Fig2t} that the ABSs do not change spins with evolving $ \phi_\mathrm{S} $, whereas the YSR states effectively `flip' spins when crossing zero energy. The states' spin ordering reflects important ground state properties, which we briefly discuss. Without the deltalike terms in $ \hat{H}_\mathrm{e} $ and $ \hat{H}_\mathrm{h} $, absent SOC, and for $ \phi_\mathrm{S}=0 $, the BdG Hamiltonian $ \hat{H}_\mathrm{BdG} $, given by Eq.~(\ref{tEqBdG}), can be easily diagonalized. Denoting by $\alpha^\dagger$ ($\alpha$) the creation (annihilation) operators for the BdG eigenmodes with positive energies and by $\beta^\dagger$ ($\beta$) the corresponding operators for negative energies, the BCS ground state $|\Omega\rangle$ is a Slater product over the occupied and unoccupied eigenmodes,
\begin{equation}
	|\Omega\rangle =\prod\limits_{\mathbf{k}} \alpha^{\phantom{\dagger}}_{\mathbf{k},\uparrow}\alpha^{\phantom{\dagger}}_{-\mathbf{k},\downarrow} \beta^\dagger_{\mathbf{k},\uparrow} \beta^\dagger_{-\mathbf{k},\downarrow}|0\rangle\, ,
\end{equation}
where $ |0\rangle $ represents the Fermi vacuum.

\begin{figure}
	\includegraphics[width=0.45\textwidth]{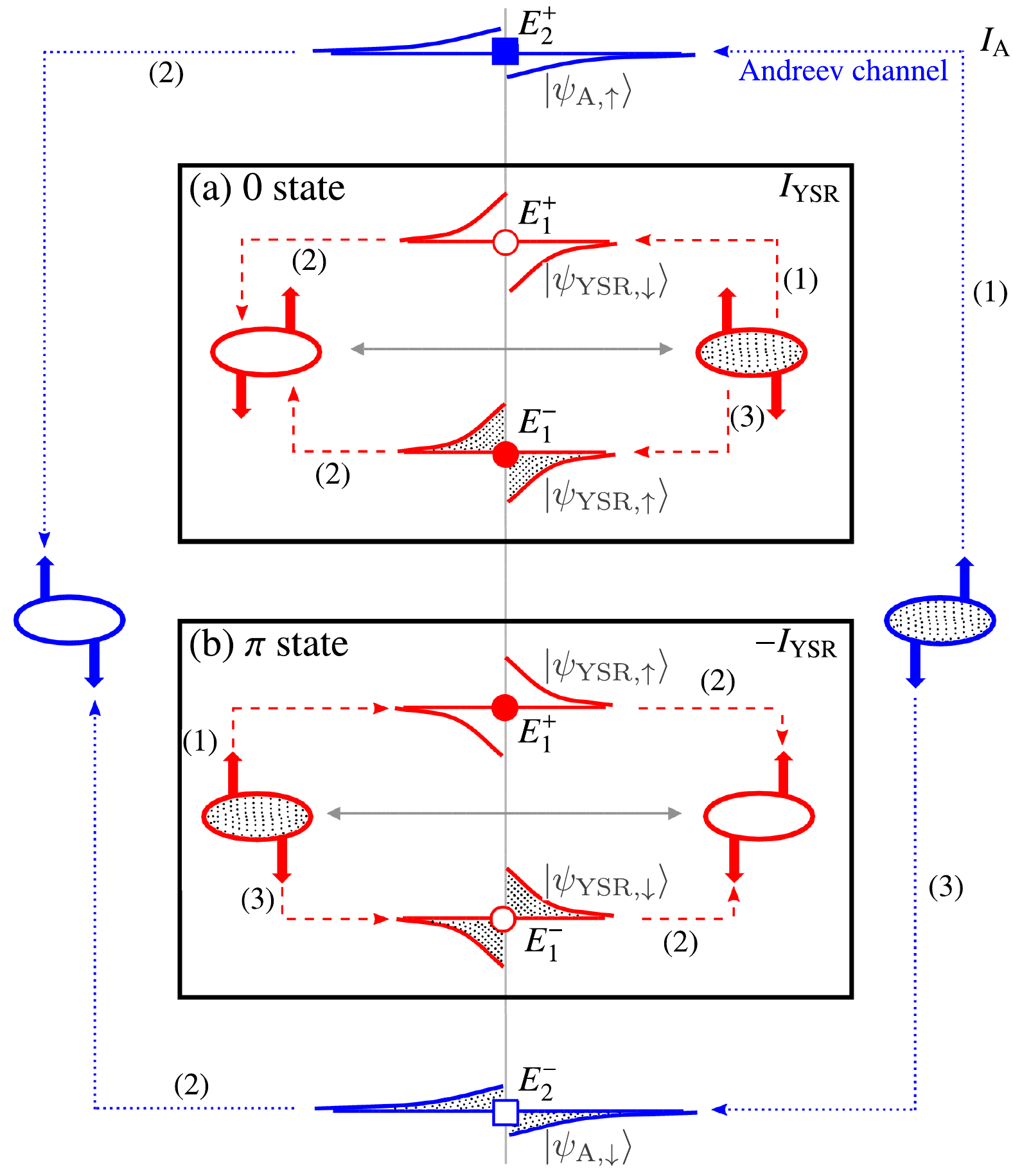}
	\colorcaption{Schematic picture of $ 0 $ and $ \pi $ ground states, and the associated transfer of Cooper pairs. Full (empty) squares/circles display spin~up (down) Andreev/YSR states; the states at energies $ E_2^\pm $ and $ E_1^\pm $ are labeled as $ |\psi_{\mathrm{A},\sigma}\rangle $ and $ |\psi_{\mathrm{YSR},\sigma}\rangle $. The sketched evanescent tails indicate the spatial electron-hole density difference for a given $ |\psi\rangle $. Occupied states and Cooper pairs are shaded. The dotted (blue) lines describe the Cooper pair transfer via the Andreev channel (current $ I_\mathrm{A} $) and the dashed (red) lines via the YSR channel (current $ I_\mathrm{YSR} $). The individual steps~(1)--(3) are described in the text. The $ 0 $ state junction is displayed in~(a) and the $ \pi $ state junction in~(b). The change of the YSR spectral properties at the $ 0 $-$ \pi $ transition reverses $ I_\mathrm{YSR} $, while $ I_\mathrm{A} $ remains unchanged.  \label{tFigXXX}}
\end{figure}

What happens to $ |\Omega\rangle $ when adding a deltalike exchange interaction, parameterized by $ \lambda_{\mathrm{MA}} $? For simplicity, let us start with zero $\lambda_{\mathrm{SC}}$ and $\phi_{\rm S}$.
Assuming a small positive $ \lambda_{\mathrm{MA}}$, spin up states have higher energy than their spin down counterparts. This means that initially degenerate quasiparticle eigenmodes spin split and form spatially (quasi)localized impurity subgap states. Anticipating the results of Eq.~\eqref{tEqSpectrum} for $ \lambda_{\mathrm{SC}}=0$ and $\phi_{\rm S}=0 $, we denote the localized states with positive energies $ 0<E_{1}^+ < E_2^+ \simeq |\Delta_\mathrm{S}| $ as $ |\psi_{\rm{YSR},\downarrow}\rangle $ and $ |\psi_{\rm{A},\uparrow}\rangle $, and states with negative energies $ -|\Delta_\mathrm{S}| \simeq E_2^- < E_1^- < 0$ as $ |\psi_{\rm{A},\downarrow}\rangle $ and $ |\psi_{\rm{YSR},\uparrow}\rangle $; see Fig.~\ref{tFigXXX}.
Generally, the states $ |\psi_{\rm{YSR},\uparrow}\rangle $ and $ |\psi_{\rm{YSR},\downarrow}\rangle$ are shifted more towards the center of the gap than $ |\psi_{\rm{A},\downarrow}\rangle $ and $ |\psi_{\rm{A},\uparrow}\rangle$ for $ \lambda_{\mathrm{MA}}>0$, and hence become spatially more localized around the impurity. 
With a further increase of $ \lambda_\mathrm{MA} $, the energy of $ |\psi_{\rm{YSR},\downarrow}\rangle $ continuously lowers, while that of $ |\psi_{\rm{YSR},\uparrow}\rangle $ rises. Before reaching the critical value $ \lambda_{\mathrm{MA}}^\mathrm{crit.} \sim \xi_{\rm{BCS}} |\Delta_\mathrm{S}| $ ($ \xi_{\rm{BCS}} $ is the BCS coherence length), $ |\psi_{\rm{YSR},\downarrow}\rangle $ remains unoccupied and $ |\psi_{\rm{YSR},\uparrow}\rangle $ occupied; see Fig.~\ref{tFigXXX}(a). Denoting the corresponding creation (annihilation) operators for $ |\psi_{\rm{A},\sigma}\rangle $ and $ |\psi_{\rm{YSR},\sigma}\rangle $ by $ {\mathrm{A}}^\dagger_\sigma $ (${\mathrm{A}}_\sigma $) and $ {\mathrm{Y}}^\dagger_\sigma $ ($ {\mathrm{Y}}_\sigma $),
respectively, we expect the ground state below $ \lambda_{\mathrm{MA}}^\mathrm{crit.}$ to be in the form
\begin{equation}
	|\Omega_{<}\rangle \sim {\mathrm{A}}^{\phantom{\dagger}}_\uparrow {\mathrm{Y}}^{\phantom{\dagger}}_\downarrow{\mathrm{Y}}^\dagger_\uparrow {\mathrm{A}}^\dagger_\downarrow \prod \limits_\mathbf{n} \tilde{\alpha}^{\phantom{\dagger}}_{\mathbf{n}, \uparrow} \tilde{\alpha}^{\phantom{\dagger}}_{-\mathbf{n}, \downarrow} \tilde{\beta}^\dagger_{\mathbf{n}, \uparrow} \tilde{\beta}^\dagger_{-\mathbf{n}, \downarrow}|0\rangle \, .  
\label{tEqLowerState}
\end{equation}
The tilde operators have the same meaning as before---quasiparticle eigenmodes (now perturbed) with energies above and below the gap~\footnote{Since momentum is not a good quantum number, we rather use a general index $\mathbf{n}$ that labels perturbed eigenmodes above and below the S gap.}.   
Increasing $ \lambda_\mathrm{MA} $ over the critical $ \lambda_{\mathrm{MA}}^\mathrm{crit.} $, the energy of $ |\psi_{\rm{YSR},\downarrow}\rangle $ becomes smaller than that of $ |\psi_{\rm{YSR},\uparrow}\rangle $ and, therefore, the relative occupations of both states interchange; see 
Fig.~\ref{tFigXXX}(b). The ground state wave function above $ \lambda_{\mathrm{MA}}^\mathrm{crit.}$ is now expected to be
\begin{equation}
	|\Omega_{>}\rangle \sim {\mathrm{A}}^{\phantom{\dagger}}_\uparrow {\mathrm{Y}}^{\phantom{\dagger}}_\uparrow {\mathrm{Y}}^\dagger_\downarrow {\mathrm{A}}^\dagger_\downarrow \prod\limits_\mathbf{n} \tilde{\alpha}^{\phantom{\dagger}}_{\mathbf{n}, \uparrow} \tilde{\alpha}^{\phantom{\dagger}}_{-\mathbf{n}, \downarrow} \tilde{\beta}^\dagger_{\mathbf{n}, \uparrow} \tilde{\beta}^\dagger_{-\mathbf{n}, \downarrow}|0\rangle \, .  \label{EqUpperState}
\end{equation}
The former ground state $ |\Omega_{<}\rangle $ has a modified spin content inside the gap when compared to $ |\Omega_{>}\rangle $, so both are in distinct quantum states, which we correspondingly call the $0$ and $\pi$ phases. Turning on $ \lambda_{\mathrm{SC}} $ and $ \phi_\mathrm{S} $, the bound state energies evolve in a complex way---see Eq.~(\ref{tEqSpectrum}) for $\lambda_{\mathrm{SC}}\neq0$ and $\phi_{\rm S}\neq0$---nevertheless, crossings at zero energy again indicate changes in the spin ordering of the ground states. Thus, reversals of the Josephson current at $ 0 $-$ \pi $~transitions in ballistic S/FI/S Josephson junctions are interpreted in terms of a \emph{change of the ground state spin order}.

\section{0-$ \pi $ transitions in Josephson~current  \label{SecV}}
\begin{figure}
	\includegraphics[width=0.45\textwidth]{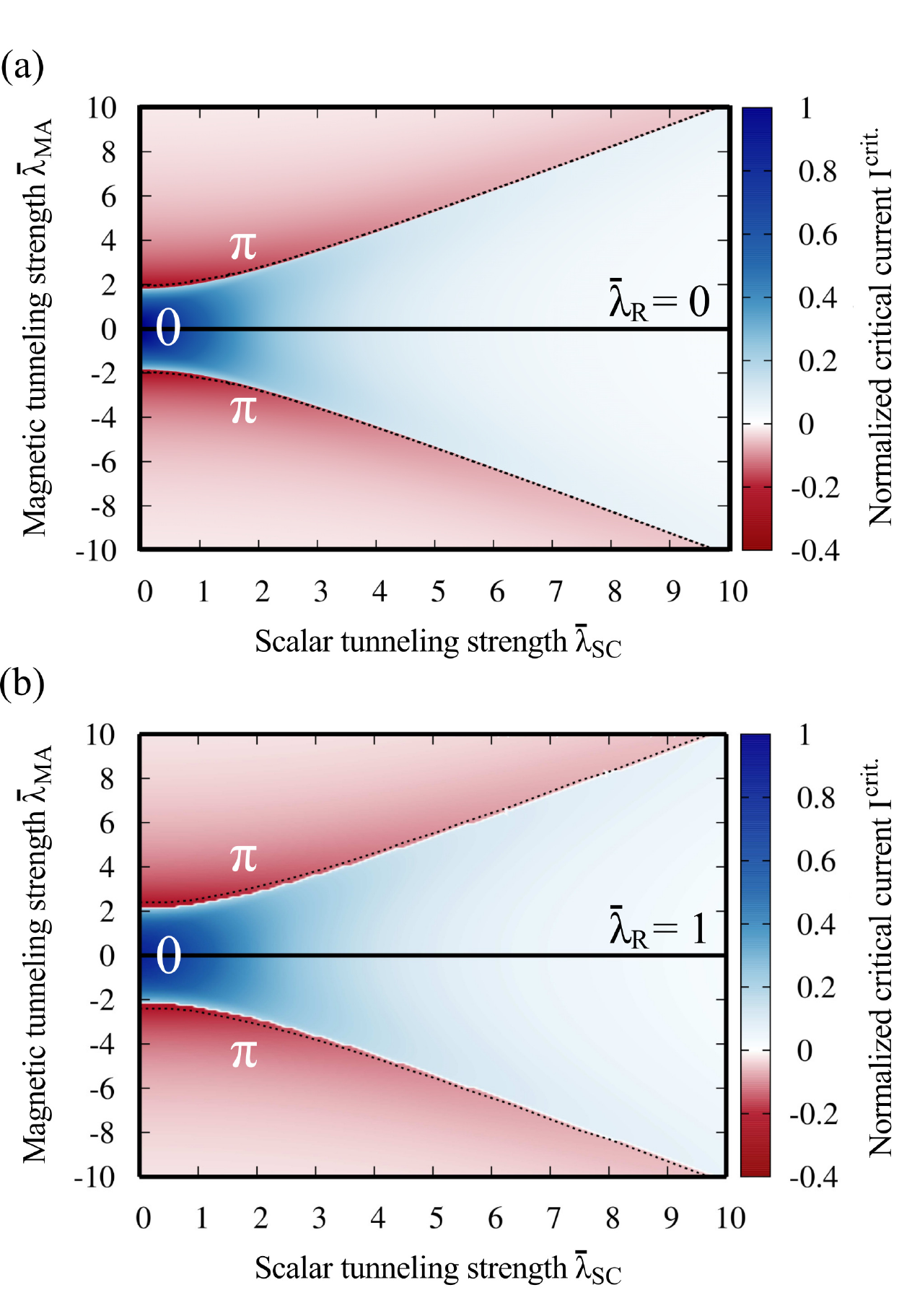}
	\colorcaption{(a)~Contour plot of the normalized critical current $ e I^\mathrm{crit.} R_\mathrm{S} / (\pi |\Delta_\mathrm{S}|) $ ($ R_\mathrm{S} $ is Sharvin's resistance) as a function of $ \overline{\lambda}_\mathrm{SC} $ and $ \overline{\lambda}_\mathrm{MA} $ in the absence of Rashba SOC at zero temperature. Blue and red regions represent $0$ and $\pi$ regimes with positive and negative $I^\mathrm{crit.}$. The transition lines separating the two phases are displayed by white borderlines; the parameters at which zero-energy YSR states emerge are shown by dashed lines. (b)~Same calculations in the additional presence of moderate Rashba SOC, $ \overline{\lambda}_\mathrm{R}=m\lambda_\mathrm{R}/(\hbar^2q_\mathrm{F})=1 $.
	\label{Fig3t}}
\end{figure}

Knowing the phase dependence of the bound state spectrum, we can obtain the Josephson current~\cite{Kulik1969, *Kulik1969alt}
$ I_\mathrm{J} (\phi_\mathrm{S}) = -\frac{e}{\hbar}\sum_{n=1}^2 \left[ \left( \frac{\partial E_n^+}{\partial \phi_\mathrm{S}} \right) \tanh \left( \frac{E_n^+}{2 k_\mathrm{B} T} \right) \right] $; $ e $ stands for the (positive) elementary charge and $ k_\mathrm{B} $ is Boltzmann's constant. From the experimental point of view, it is common to measure the critical Josephson current
$I^\mathrm{crit.} = \max_{\phi_\mathrm{S}} \left\{ \left| I_\mathrm{J} (\phi_\mathrm{S}) \right| \right\}$ and the corresponding critical S phase  $ \phi_\mathrm{S}^\mathrm{crit.} $. By tuning $\phi_\mathrm{S} $ to its critical value and performing scanning tunneling spectroscopy/scanning tunneling microscopy (STS/STM) in the vicinity of the interface, one could explore the spectral properties of the subgap states. 
Figure~\ref{Fig3t}(a) displays the dependence of $ I^\mathrm{crit.} $ on the tunneling strengths. For each $ \overline{\lambda}_\mathrm{SC} $, one finds a $ \overline{\lambda}_\mathrm{MA} $ at which the Josephson current's direction reverses, indicating transitions from $ 0 $ to $ \pi $~regimes. The maximal current in the $ 0 $ state is twice as large as in the $ \pi $ state; moreover, as expected, the $ 0 $-$ \pi $ transition lines coincide with the contours signifying the formation of zero-energy YSR states. In Fig.~\ref{Fig3t}(b), we show similar, fully numerical calculations in the presence of moderate Rashba SOC in the right electrode. Modulating the Rashba SOC by electrical gating~\cite{Nitta1997, Koga2002} can efficiently tune $ 0 $-$ \pi $~transitions. Nevertheless, there is still a clear coincidence between the $ 0 $-$ \pi $ transition lines and the zero-energy YSR states, although SOC inevitably introduces an intrinsic shift to the current-phase~relation. This causes slight deviations at weak tunnelings.

\section{Zero-energy YSR states \& reversal of Josephson current  \label{SecVI}}
To connect the zero-energy YSR states with the reversal of the Josephson current, one needs the bound state wave functions. The full calculation is rather technical; see SM~\cite{Note1}. Here, we qualitatively illustrate the main mechanism. Figure~\ref{tFigXXX} schematically shows the Andreev (blue dotted lines) and YSR (red dashed lines) channels, transporting Cooper pairs across the barrier. To understand the transport direction of one Cooper pair, we split each occupied and unoccupied bound state into its electronlike and holelike component and read out the corresponding electron-hole density difference. This is qualitatively illustrated by the sketched evanescent tails in Fig.~\ref{tFigXXX} for the corresponding $ |\psi_{\mathrm{A},\sigma}\rangle $ and $ |\psi_{\mathrm{YSR},\sigma}\rangle $ states; upward (downward) tails indicate electron (hole) dominance. For the ABS, there is always an electronlike excess in the left and a holelike excess in the right S. This means that the electrons forming a Cooper pair are transferred via the Andreev channel from the right into the left superconductor by the individual steps (1)--(3). In~(1), a spin~up electron tunnels to an empty state at positive energy $ E_2^+ $. In~(2), this electron pairs with a spin~down electron residing in the occupied state at energy $ E_2^- $, creating a spin-singlet Cooper pair in the left superconductor; finally, the remaining electron from the initial Cooper pair fills the vacant state at $ E_2^- $. The situation for the YSR channel is different. In the $ 0 $ state, shown by Fig.~\ref{tFigXXX}(a), the YSR states also show electronlike excess on the left side and holelike excess on the right side of the barrier. Hence, the previous mechanism still holds and Cooper pairs are also transferred from right to left. In total, both current contributions add together, $ I_\mathrm{A}+I_\mathrm{YSR} $. Contrarily, in the $ \pi $ phase, the spin-resolved electron and hole content of the YSR states spatially change and, therefore, the YSR states now drive Cooper pairs from left to right, i.e., against the direction of the Andreev channel. The total current is then $ I_\mathrm{A}-I_\mathrm{YSR} $. Since the transition probabilities of the activation step~(1) are proportional to $ \mathrm{e}^{-E_n^+/|\Delta_\mathrm{S}|} $ and $ E_1^+<E_2^+ $, the YSR contribution  is generally dominant and therefore, a reversal of $ I_\mathrm{YSR} $ also reverses the total Josephson current. This qualitative explanation is fully consistent with our presented calculations. \\[-5 pt]

\section{Summary  \label{SecVII}} 
We analyzed spectral and transport characteristics associated with Andreev and YSR states in magnetic Josephson junctions in terms of experimentally tunable parameters and showed that certain combinations of these parameters lead to zero-energy YSR states. Such states serve as a clear fingerprint of quantum phase transitions in the junctions' ground state. Particularly, we demonstrated that this phase transition coincides with the Josephson current reversing $ 0 $-$ \pi $ transitions. This coincidence between zero-energy YSR states and $ 0 $-$ \pi $ transitions persists also in the presence of Rashba SOC in one electrode.

\begin{acknowledgments}
	This work was supported by DFG~SFB Grant~No.~689 and DFG~SFB Grant~No.~1277 (B07). This project has received funding from the European~Union's Horizon~2020 research and innovation~programme under Grant~Agreement No.~696656 and the International Doctorate Program Topological Insulators of the Elite Network of Bavaria.
\end{acknowledgments}

\appendix
\bibliography{paper}

\begin{thebibliography}{111}%
\makeatletter
\providecommand \@ifxundefined [1]{%
 \@ifx{#1\undefined}
}%
\providecommand \@ifnum [1]{%
 \ifnum #1\expandafter \@firstoftwo
 \else \expandafter \@secondoftwo
 \fi
}%
\providecommand \@ifx [1]{%
 \ifx #1\expandafter \@firstoftwo
 \else \expandafter \@secondoftwo
 \fi
}%
\providecommand \natexlab [1]{#1}%
\providecommand \enquote  [1]{``#1''}%
\providecommand \bibnamefont  [1]{#1}%
\providecommand \bibfnamefont [1]{#1}%
\providecommand \citenamefont [1]{#1}%
\providecommand \href@noop [0]{\@secondoftwo}%
\providecommand \href [0]{\begingroup \@sanitize@url \@href}%
\providecommand \@href[1]{\@@startlink{#1}\@@href}%
\providecommand \@@href[1]{\endgroup#1\@@endlink}%
\providecommand \@sanitize@url [0]{\catcode `\\12\catcode `\$12\catcode
  `\&12\catcode `\#12\catcode `\^12\catcode `\_12\catcode `\%12\relax}%
\providecommand \@@startlink[1]{}%
\providecommand \@@endlink[0]{}%
\providecommand \url  [0]{\begingroup\@sanitize@url \@url }%
\providecommand \@url [1]{\endgroup\@href {#1}{\urlprefix }}%
\providecommand \urlprefix  [0]{URL }%
\providecommand \Eprint [0]{\href }%
\providecommand \doibase [0]{http://dx.doi.org/}%
\providecommand \selectlanguage [0]{\@gobble}%
\providecommand \bibinfo  [0]{\@secondoftwo}%
\providecommand \bibfield  [0]{\@secondoftwo}%
\providecommand \translation [1]{[#1]}%
\providecommand \BibitemOpen [0]{}%
\providecommand \bibitemStop [0]{}%
\providecommand \bibitemNoStop [0]{.\EOS\space}%
\providecommand \EOS [0]{\spacefactor3000\relax}%
\providecommand \BibitemShut  [1]{\csname bibitem#1\endcsname}%
\let\auto@bib@innerbib\@empty
\bibitem [{\citenamefont {Stoner}(1939)}]{Stoner1939}%
  \BibitemOpen
  \bibfield  {author} {\bibinfo {author} {\bibfnamefont {E.~C.}\ \bibnamefont
  {Stoner}},\ }\href {\doibase 10.1098/rspa.1939.0003} {\bibfield  {journal}
  {\bibinfo  {journal} {Proc. R. Soc. London A: Math. Phys. Engineer. Sci.}\
  }\textbf {\bibinfo {volume} {169}},\ \bibinfo {pages} {339} (\bibinfo {year}
  {1939})}\BibitemShut {NoStop}%
\bibitem [{\citenamefont {Bardeen}\ \emph
  {et~al.}(1957{\natexlab{a}})\citenamefont {Bardeen}, \citenamefont {Cooper},\
  and\ \citenamefont {Schrieffer}}]{Bardeen1957a}%
  \BibitemOpen
  \bibfield  {author} {\bibinfo {author} {\bibfnamefont {J.}~\bibnamefont
  {Bardeen}}, \bibinfo {author} {\bibfnamefont {L.~N.}\ \bibnamefont {Cooper}},
  \ and\ \bibinfo {author} {\bibfnamefont {J.~R.}\ \bibnamefont {Schrieffer}},\
  }\href {\doibase 10.1103/PhysRev.106.162} {\bibfield  {journal} {\bibinfo
  {journal} {Phys. Rev.}\ }\textbf {\bibinfo {volume} {106}},\ \bibinfo {pages}
  {162} (\bibinfo {year} {1957}{\natexlab{a}})}\BibitemShut {NoStop}%
\bibitem [{\citenamefont {Bardeen}\ \emph
  {et~al.}(1957{\natexlab{b}})\citenamefont {Bardeen}, \citenamefont {Cooper},\
  and\ \citenamefont {Schrieffer}}]{Bardeen1957b}%
  \BibitemOpen
  \bibfield  {author} {\bibinfo {author} {\bibfnamefont {J.}~\bibnamefont
  {Bardeen}}, \bibinfo {author} {\bibfnamefont {L.~N.}\ \bibnamefont {Cooper}},
  \ and\ \bibinfo {author} {\bibfnamefont {J.~R.}\ \bibnamefont {Schrieffer}},\
  }\href {\doibase 10.1103/PhysRev.108.1175} {\bibfield  {journal} {\bibinfo
  {journal} {Phys. Rev.}\ }\textbf {\bibinfo {volume} {108}},\ \bibinfo {pages}
  {1175} (\bibinfo {year} {1957}{\natexlab{b}})}\BibitemShut {NoStop}%
\bibitem [{\citenamefont {Eschrig}(2011)}]{Eschrig2011}%
  \BibitemOpen
  \bibfield  {author} {\bibinfo {author} {\bibfnamefont {M.}~\bibnamefont
  {Eschrig}},\ }\href
  {http://scitation.aip.org/content/aip/magazine/physicstoday/article/64/1/10.1063/1.3541944}
  {\bibfield  {journal} {\bibinfo  {journal} {Phys. Today}\ }\textbf {\bibinfo
  {volume} {64}},\ \bibinfo {pages} {43} (\bibinfo {year} {2011})}\BibitemShut
  {NoStop}%
\bibitem [{\citenamefont {Linder}\ and\ \citenamefont
  {Robinson}(2015)}]{Linder2015}%
  \BibitemOpen
  \bibfield  {author} {\bibinfo {author} {\bibfnamefont {J.}~\bibnamefont
  {Linder}}\ and\ \bibinfo {author} {\bibfnamefont {J.~W.~A.}\ \bibnamefont
  {Robinson}},\ }\href
  {http://www.nature.com/nphys/journal/v11/n4/abs/nphys3242.html} {\bibfield
  {journal} {\bibinfo  {journal} {Nature Phys.}\ }\textbf {\bibinfo {volume}
  {11}},\ \bibinfo {pages} {307} (\bibinfo {year} {2015})}\BibitemShut
  {NoStop}%
\bibitem [{\citenamefont {Gingrich}\ \emph {et~al.}(2016)\citenamefont
  {Gingrich}, \citenamefont {Niedzielski}, \citenamefont {Glick}, \citenamefont
  {Wang}, \citenamefont {Miller}, \citenamefont {Loloee}, \citenamefont {{Pratt
  Jr}},\ and\ \citenamefont {Birge}}]{Gingrich2016}%
  \BibitemOpen
  \bibfield  {author} {\bibinfo {author} {\bibfnamefont {E.~C.}\ \bibnamefont
  {Gingrich}}, \bibinfo {author} {\bibfnamefont {B.~M.}\ \bibnamefont
  {Niedzielski}}, \bibinfo {author} {\bibfnamefont {J.~A.}\ \bibnamefont
  {Glick}}, \bibinfo {author} {\bibfnamefont {Y.}~\bibnamefont {Wang}},
  \bibinfo {author} {\bibfnamefont {D.~L.}\ \bibnamefont {Miller}}, \bibinfo
  {author} {\bibfnamefont {R.}~\bibnamefont {Loloee}}, \bibinfo {author}
  {\bibfnamefont {W.~P.}\ \bibnamefont {{Pratt Jr}}}, \ and\ \bibinfo {author}
  {\bibfnamefont {N.~O.}\ \bibnamefont {Birge}},\ }\href
  {http://dx.doi.org/10.1038/nphys3681} {\bibfield  {journal} {\bibinfo
  {journal} {Nat. Phys.}\ }\textbf {\bibinfo {volume} {12}},\ \bibinfo {pages}
  {564} (\bibinfo {year} {2016})}\BibitemShut {NoStop}%
\bibitem [{\citenamefont {Bulaevskii}\ \emph
  {et~al.}(1977{\natexlab{a}})\citenamefont {Bulaevskii}, \citenamefont
  {Kuzii},\ and\ \citenamefont {Sobyanin}}]{Bulaevskii1977}%
  \BibitemOpen
  \bibfield  {author} {\bibinfo {author} {\bibfnamefont {L.~N.}\ \bibnamefont
  {Bulaevskii}}, \bibinfo {author} {\bibfnamefont {V.~V.}\ \bibnamefont
  {Kuzii}}, \ and\ \bibinfo {author} {\bibfnamefont {A.~A.}\ \bibnamefont
  {Sobyanin}},\ }\href@noop {} {\bibfield  {journal} {\bibinfo  {journal}
  {Pis'ma Zh. Eksp. Teor. Fiz.}\ }\textbf {\bibinfo {volume} {25}},\ \bibinfo
  {pages} {314} (\bibinfo {year} {1977}{\natexlab{a}})}\BibitemShut {NoStop}%
\bibitem [{\citenamefont {Bulaevskii}\ \emph
  {et~al.}(1977{\natexlab{b}})\citenamefont {Bulaevskii}, \citenamefont
  {Kuzii},\ and\ \citenamefont {Sobyanin}}]{Bulaevskii1977alt}%
  \BibitemOpen
  \bibfield  {author} {\bibinfo {author} {\bibfnamefont {L.~N.}\ \bibnamefont
  {Bulaevskii}}, \bibinfo {author} {\bibfnamefont {V.~V.}\ \bibnamefont
  {Kuzii}}, \ and\ \bibinfo {author} {\bibfnamefont {A.~A.}\ \bibnamefont
  {Sobyanin}},\ }\href
  {http://www.jetpletters.ac.ru/ps/1410/article_21163.shtml} {\bibfield
  {journal} {\bibinfo  {journal} {JETP Lett.}\ }\textbf {\bibinfo {volume}
  {25}},\ \bibinfo {pages} {290} (\bibinfo {year}
  {1977}{\natexlab{b}})}\BibitemShut {NoStop}%
\bibitem [{\citenamefont {Buzdin}\ \emph
  {et~al.}(1982{\natexlab{a}})\citenamefont {Buzdin}, \citenamefont
  {Bulaevskii},\ and\ \citenamefont {Panyukov}}]{Buzdin1982}%
  \BibitemOpen
  \bibfield  {author} {\bibinfo {author} {\bibfnamefont {A.~I.}\ \bibnamefont
  {Buzdin}}, \bibinfo {author} {\bibfnamefont {L.~N.}\ \bibnamefont
  {Bulaevskii}}, \ and\ \bibinfo {author} {\bibfnamefont {S.~V.}\ \bibnamefont
  {Panyukov}},\ }\href@noop {} {\bibfield  {journal} {\bibinfo  {journal}
  {Pis'ma Zh. Eksp. Teor. Fiz.}\ }\textbf {\bibinfo {volume} {35}},\ \bibinfo
  {pages} {147} (\bibinfo {year} {1982}{\natexlab{a}})}\BibitemShut {NoStop}%
\bibitem [{\citenamefont {Buzdin}\ \emph
  {et~al.}(1982{\natexlab{b}})\citenamefont {Buzdin}, \citenamefont
  {Bulaevskii},\ and\ \citenamefont {Panyukov}}]{Buzdin1982alt}%
  \BibitemOpen
  \bibfield  {author} {\bibinfo {author} {\bibfnamefont {A.~I.}\ \bibnamefont
  {Buzdin}}, \bibinfo {author} {\bibfnamefont {L.~N.}\ \bibnamefont
  {Bulaevskii}}, \ and\ \bibinfo {author} {\bibfnamefont {S.~V.}\ \bibnamefont
  {Panyukov}},\ }\href
  {http://www.jetpletters.ac.ru/ps/1314/article_19853.shtml} {\bibfield
  {journal} {\bibinfo  {journal} {JETP Lett.}\ }\textbf {\bibinfo {volume}
  {35}},\ \bibinfo {pages} {178} (\bibinfo {year}
  {1982}{\natexlab{b}})}\BibitemShut {NoStop}%
\bibitem [{\citenamefont {Andreev}\ \emph {et~al.}(1991)\citenamefont
  {Andreev}, \citenamefont {Buzdin},\ and\ \citenamefont
  {Osgood}}]{Andreev1991}%
  \BibitemOpen
  \bibfield  {author} {\bibinfo {author} {\bibfnamefont {A.~V.}\ \bibnamefont
  {Andreev}}, \bibinfo {author} {\bibfnamefont {A.~I.}\ \bibnamefont {Buzdin}},
  \ and\ \bibinfo {author} {\bibfnamefont {R.~M.}\ \bibnamefont {Osgood}},\
  }\href {http://link.aps.org/doi/10.1103/PhysRevB.43.10124} {\bibfield
  {journal} {\bibinfo  {journal} {Phys. Rev. B}\ }\textbf {\bibinfo {volume}
  {43}},\ \bibinfo {pages} {10124} (\bibinfo {year} {1991})}\BibitemShut
  {NoStop}%
\bibitem [{\citenamefont {Demler}\ \emph {et~al.}(1997)\citenamefont {Demler},
  \citenamefont {Arnold},\ and\ \citenamefont {Beasley}}]{Demler1997}%
  \BibitemOpen
  \bibfield  {author} {\bibinfo {author} {\bibfnamefont {E.~A.}\ \bibnamefont
  {Demler}}, \bibinfo {author} {\bibfnamefont {G.~B.}\ \bibnamefont {Arnold}},
  \ and\ \bibinfo {author} {\bibfnamefont {M.~R.}\ \bibnamefont {Beasley}},\
  }\href {http://link.aps.org/doi/10.1103/PhysRevB.55.15174} {\bibfield
  {journal} {\bibinfo  {journal} {Phys. Rev. B}\ }\textbf {\bibinfo {volume}
  {55}},\ \bibinfo {pages} {15174} (\bibinfo {year} {1997})}\BibitemShut
  {NoStop}%
\bibitem [{\citenamefont {Golubov}\ \emph {et~al.}(2004)\citenamefont
  {Golubov}, \citenamefont {Kupriyanov},\ and\ \citenamefont
  {Il'ichev}}]{Golubov2004}%
  \BibitemOpen
  \bibfield  {author} {\bibinfo {author} {\bibfnamefont {A.~A.}\ \bibnamefont
  {Golubov}}, \bibinfo {author} {\bibfnamefont {M.~Y.}\ \bibnamefont
  {Kupriyanov}}, \ and\ \bibinfo {author} {\bibfnamefont {E.}~\bibnamefont
  {Il'ichev}},\ }\href {http://link.aps.org/doi/10.1103/RevModPhys.76.411}
  {\bibfield  {journal} {\bibinfo  {journal} {Rev. Mod. Phys.}\ }\textbf
  {\bibinfo {volume} {76}},\ \bibinfo {pages} {411} (\bibinfo {year}
  {2004})}\BibitemShut {NoStop}%
\bibitem [{\citenamefont {Buzdin}(2005)}]{Buzdin2005}%
  \BibitemOpen
  \bibfield  {author} {\bibinfo {author} {\bibfnamefont {A.~I.}\ \bibnamefont
  {Buzdin}},\ }\href {http://link.aps.org/doi/10.1103/RevModPhys.77.935}
  {\bibfield  {journal} {\bibinfo  {journal} {Rev. Mod. Phys.}\ }\textbf
  {\bibinfo {volume} {77}},\ \bibinfo {pages} {935} (\bibinfo {year}
  {2005})}\BibitemShut {NoStop}%
\bibitem [{\citenamefont {Bergeret}\ \emph {et~al.}(2005)\citenamefont
  {Bergeret}, \citenamefont {Volkov},\ and\ \citenamefont
  {Efetov}}]{Bergeret2005}%
  \BibitemOpen
  \bibfield  {author} {\bibinfo {author} {\bibfnamefont {F.~S.}\ \bibnamefont
  {Bergeret}}, \bibinfo {author} {\bibfnamefont {A.~F.}\ \bibnamefont
  {Volkov}}, \ and\ \bibinfo {author} {\bibfnamefont {K.~B.}\ \bibnamefont
  {Efetov}},\ }\href {http://link.aps.org/doi/10.1103/RevModPhys.77.1321}
  {\bibfield  {journal} {\bibinfo  {journal} {Rev. Mod. Phys.}\ }\textbf
  {\bibinfo {volume} {77}},\ \bibinfo {pages} {1321} (\bibinfo {year}
  {2005})}\BibitemShut {NoStop}%
\bibitem [{\citenamefont {Annunziata}\ \emph {et~al.}(2011)\citenamefont
  {Annunziata}, \citenamefont {Enoksen}, \citenamefont {Linder}, \citenamefont
  {Cuoco}, \citenamefont {Noce},\ and\ \citenamefont
  {Sudb\o{}}}]{Annunziata2011}%
  \BibitemOpen
  \bibfield  {author} {\bibinfo {author} {\bibfnamefont {G.}~\bibnamefont
  {Annunziata}}, \bibinfo {author} {\bibfnamefont {H.}~\bibnamefont {Enoksen}},
  \bibinfo {author} {\bibfnamefont {J.}~\bibnamefont {Linder}}, \bibinfo
  {author} {\bibfnamefont {M.}~\bibnamefont {Cuoco}}, \bibinfo {author}
  {\bibfnamefont {C.}~\bibnamefont {Noce}}, \ and\ \bibinfo {author}
  {\bibfnamefont {A.}~\bibnamefont {Sudb\o{}}},\ }\href {\doibase
  10.1103/PhysRevB.83.144520} {\bibfield  {journal} {\bibinfo  {journal} {Phys.
  Rev. B}\ }\textbf {\bibinfo {volume} {83}},\ \bibinfo {pages} {144520}
  (\bibinfo {year} {2011})}\BibitemShut {NoStop}%
\bibitem [{\citenamefont {Costa}\ \emph {et~al.}(2017)\citenamefont {Costa},
  \citenamefont {H\"ogl},\ and\ \citenamefont {Fabian}}]{Costa2017}%
  \BibitemOpen
  \bibfield  {author} {\bibinfo {author} {\bibfnamefont {A.}~\bibnamefont
  {Costa}}, \bibinfo {author} {\bibfnamefont {P.}~\bibnamefont {H\"ogl}}, \
  and\ \bibinfo {author} {\bibfnamefont {J.}~\bibnamefont {Fabian}},\ }\href
  {\doibase 10.1103/PhysRevB.95.024514} {\bibfield  {journal} {\bibinfo
  {journal} {Phys. Rev. B}\ }\textbf {\bibinfo {volume} {95}},\ \bibinfo
  {pages} {024514} (\bibinfo {year} {2017})}\BibitemShut {NoStop}%
\bibitem [{\citenamefont {Glazman}\ and\ \citenamefont
  {Matveev}(1989{\natexlab{a}})}]{Glazman1989}%
  \BibitemOpen
  \bibfield  {author} {\bibinfo {author} {\bibfnamefont {L.~I.}\ \bibnamefont
  {Glazman}}\ and\ \bibinfo {author} {\bibfnamefont {K.~A.}\ \bibnamefont
  {Matveev}},\ }\href@noop {} {\bibfield  {journal} {\bibinfo  {journal}
  {Pis'ma Zh. Eksp. Teor. Fiz.}\ }\textbf {\bibinfo {volume} {49}},\ \bibinfo
  {pages} {570} (\bibinfo {year} {1989}{\natexlab{a}})}\BibitemShut {NoStop}%
\bibitem [{\citenamefont {Glazman}\ and\ \citenamefont
  {Matveev}(1989{\natexlab{b}})}]{Glazman1989alt}%
  \BibitemOpen
  \bibfield  {author} {\bibinfo {author} {\bibfnamefont {L.~I.}\ \bibnamefont
  {Glazman}}\ and\ \bibinfo {author} {\bibfnamefont {K.~A.}\ \bibnamefont
  {Matveev}},\ }\href
  {http://www.jetpletters.ac.ru/ps/1121/article_16988.shtml} {\bibfield
  {journal} {\bibinfo  {journal} {JETP Lett.}\ }\textbf {\bibinfo {volume}
  {49}},\ \bibinfo {pages} {659} (\bibinfo {year}
  {1989}{\natexlab{b}})}\BibitemShut {NoStop}%
\bibitem [{\citenamefont {Clerk}\ and\ \citenamefont
  {Ambegaokar}(2000)}]{Clerk2000}%
  \BibitemOpen
  \bibfield  {author} {\bibinfo {author} {\bibfnamefont {A.~A.}\ \bibnamefont
  {Clerk}}\ and\ \bibinfo {author} {\bibfnamefont {V.}~\bibnamefont
  {Ambegaokar}},\ }\href {\doibase 10.1103/PhysRevB.61.9109} {\bibfield
  {journal} {\bibinfo  {journal} {Phys. Rev. B}\ }\textbf {\bibinfo {volume}
  {61}},\ \bibinfo {pages} {9109} (\bibinfo {year} {2000})}\BibitemShut
  {NoStop}%
\bibitem [{\citenamefont {Vecino}\ \emph {et~al.}(2003)\citenamefont {Vecino},
  \citenamefont {Mart\'{\i}n-Rodero},\ and\ \citenamefont
  {Levy~Yeyati}}]{Vecino2003}%
  \BibitemOpen
  \bibfield  {author} {\bibinfo {author} {\bibfnamefont {E.}~\bibnamefont
  {Vecino}}, \bibinfo {author} {\bibfnamefont {A.}~\bibnamefont
  {Mart\'{\i}n-Rodero}}, \ and\ \bibinfo {author} {\bibfnamefont
  {A.}~\bibnamefont {Levy~Yeyati}},\ }\href {\doibase
  10.1103/PhysRevB.68.035105} {\bibfield  {journal} {\bibinfo  {journal} {Phys.
  Rev. B}\ }\textbf {\bibinfo {volume} {68}},\ \bibinfo {pages} {035105}
  (\bibinfo {year} {2003})}\BibitemShut {NoStop}%
\bibitem [{\citenamefont {Choi}\ \emph {et~al.}(2004)\citenamefont {Choi},
  \citenamefont {Lee}, \citenamefont {Kang},\ and\ \citenamefont
  {Belzig}}]{Choi2004}%
  \BibitemOpen
  \bibfield  {author} {\bibinfo {author} {\bibfnamefont {M.-S.}\ \bibnamefont
  {Choi}}, \bibinfo {author} {\bibfnamefont {M.}~\bibnamefont {Lee}}, \bibinfo
  {author} {\bibfnamefont {K.}~\bibnamefont {Kang}}, \ and\ \bibinfo {author}
  {\bibfnamefont {W.}~\bibnamefont {Belzig}},\ }\href {\doibase
  10.1103/PhysRevB.70.020502} {\bibfield  {journal} {\bibinfo  {journal} {Phys.
  Rev. B}\ }\textbf {\bibinfo {volume} {70}},\ \bibinfo {pages} {020502}
  (\bibinfo {year} {2004})}\BibitemShut {NoStop}%
\bibitem [{\citenamefont {Siano}\ and\ \citenamefont
  {Egger}(2004)}]{Siano2004}%
  \BibitemOpen
  \bibfield  {author} {\bibinfo {author} {\bibfnamefont {F.}~\bibnamefont
  {Siano}}\ and\ \bibinfo {author} {\bibfnamefont {R.}~\bibnamefont {Egger}},\
  }\href {\doibase 10.1103/PhysRevLett.93.047002} {\bibfield  {journal}
  {\bibinfo  {journal} {Phys. Rev. Lett.}\ }\textbf {\bibinfo {volume} {93}},\
  \bibinfo {pages} {047002} (\bibinfo {year} {2004})}\BibitemShut {NoStop}%
\bibitem [{\citenamefont {Siano}\ and\ \citenamefont
  {Egger}(2005)}]{Siano2004alt}%
  \BibitemOpen
  \bibfield  {author} {\bibinfo {author} {\bibfnamefont {F.}~\bibnamefont
  {Siano}}\ and\ \bibinfo {author} {\bibfnamefont {R.}~\bibnamefont {Egger}},\
  }\href {\doibase 10.1103/PhysRevLett.94.039902} {\ \textbf {\bibinfo {volume}
  {94}},\ \bibinfo {pages} {039902(E)} (\bibinfo {year} {2005})}\BibitemShut
  {NoStop}%
\bibitem [{\citenamefont {Bauer}\ \emph {et~al.}(2007)\citenamefont {Bauer},
  \citenamefont {Oguri},\ and\ \citenamefont {Hewson}}]{Bauer2007}%
  \BibitemOpen
  \bibfield  {author} {\bibinfo {author} {\bibfnamefont {J.}~\bibnamefont
  {Bauer}}, \bibinfo {author} {\bibfnamefont {A.}~\bibnamefont {Oguri}}, \ and\
  \bibinfo {author} {\bibfnamefont {A.~C.}\ \bibnamefont {Hewson}},\ }\href
  {http://stacks.iop.org/0953-8984/19/i=48/a=486211} {\bibfield  {journal}
  {\bibinfo  {journal} {J. Phys. Condens. Matter}\ }\textbf {\bibinfo {volume}
  {19}},\ \bibinfo {pages} {486211} (\bibinfo {year} {2007})}\BibitemShut
  {NoStop}%
\bibitem [{\citenamefont {Karrasch}\ \emph {et~al.}(2008)\citenamefont
  {Karrasch}, \citenamefont {Oguri},\ and\ \citenamefont
  {Meden}}]{Karrasch2008}%
  \BibitemOpen
  \bibfield  {author} {\bibinfo {author} {\bibfnamefont {C.}~\bibnamefont
  {Karrasch}}, \bibinfo {author} {\bibfnamefont {A.}~\bibnamefont {Oguri}}, \
  and\ \bibinfo {author} {\bibfnamefont {V.}~\bibnamefont {Meden}},\ }\href
  {\doibase 10.1103/PhysRevB.77.024517} {\bibfield  {journal} {\bibinfo
  {journal} {Phys. Rev. B}\ }\textbf {\bibinfo {volume} {77}},\ \bibinfo
  {pages} {024517} (\bibinfo {year} {2008})}\BibitemShut {NoStop}%
\bibitem [{\citenamefont {Meng}\ \emph {et~al.}(2009)\citenamefont {Meng},
  \citenamefont {Florens},\ and\ \citenamefont {Simon}}]{Meng2009}%
  \BibitemOpen
  \bibfield  {author} {\bibinfo {author} {\bibfnamefont {T.}~\bibnamefont
  {Meng}}, \bibinfo {author} {\bibfnamefont {S.}~\bibnamefont {Florens}}, \
  and\ \bibinfo {author} {\bibfnamefont {P.}~\bibnamefont {Simon}},\ }\href
  {\doibase 10.1103/PhysRevB.79.224521} {\bibfield  {journal} {\bibinfo
  {journal} {Phys. Rev. B}\ }\textbf {\bibinfo {volume} {79}},\ \bibinfo
  {pages} {224521} (\bibinfo {year} {2009})}\BibitemShut {NoStop}%
\bibitem [{\citenamefont {Luitz}\ and\ \citenamefont
  {Assaad}(2010)}]{Luitz2010}%
  \BibitemOpen
  \bibfield  {author} {\bibinfo {author} {\bibfnamefont {D.~J.}\ \bibnamefont
  {Luitz}}\ and\ \bibinfo {author} {\bibfnamefont {F.~F.}\ \bibnamefont
  {Assaad}},\ }\href {\doibase 10.1103/PhysRevB.81.024509} {\bibfield
  {journal} {\bibinfo  {journal} {Phys. Rev. B}\ }\textbf {\bibinfo {volume}
  {81}},\ \bibinfo {pages} {024509} (\bibinfo {year} {2010})}\BibitemShut
  {NoStop}%
\bibitem [{\citenamefont {Mart{\'{i}}n-Rodero}\ and\ \citenamefont
  {Yeyati}(2011)}]{Rodero2011}%
  \BibitemOpen
  \bibfield  {author} {\bibinfo {author} {\bibfnamefont {A.}~\bibnamefont
  {Mart{\'{i}}n-Rodero}}\ and\ \bibinfo {author} {\bibfnamefont {A.~L.}\
  \bibnamefont {Yeyati}},\ }\href {\doibase 10.1080/00018732.2011.624266}
  {\bibfield  {journal} {\bibinfo  {journal} {Adv. Phys.}\ }\textbf {\bibinfo
  {volume} {60}},\ \bibinfo {pages} {899} (\bibinfo {year} {2011})}\BibitemShut
  {NoStop}%
\bibitem [{\citenamefont {Luitz}\ \emph {et~al.}(2012)\citenamefont {Luitz},
  \citenamefont {Assaad}, \citenamefont {Novotn\'y}, \citenamefont {Karrasch},\
  and\ \citenamefont {Meden}}]{Luitz2012}%
  \BibitemOpen
  \bibfield  {author} {\bibinfo {author} {\bibfnamefont {D.~J.}\ \bibnamefont
  {Luitz}}, \bibinfo {author} {\bibfnamefont {F.~F.}\ \bibnamefont {Assaad}},
  \bibinfo {author} {\bibfnamefont {T.}~\bibnamefont {Novotn\'y}}, \bibinfo
  {author} {\bibfnamefont {C.}~\bibnamefont {Karrasch}}, \ and\ \bibinfo
  {author} {\bibfnamefont {V.}~\bibnamefont {Meden}},\ }\href {\doibase
  10.1103/PhysRevLett.108.227001} {\bibfield  {journal} {\bibinfo  {journal}
  {Phys. Rev. Lett.}\ }\textbf {\bibinfo {volume} {108}},\ \bibinfo {pages}
  {227001} (\bibinfo {year} {2012})}\BibitemShut {NoStop}%
\bibitem [{\citenamefont {Kir\ifmmode~\check{s}\else \v{s}\fi{}anskas}\ \emph
  {et~al.}(2015)\citenamefont {Kir\ifmmode~\check{s}\else \v{s}\fi{}anskas},
  \citenamefont {Goldstein}, \citenamefont {Flensberg}, \citenamefont
  {Glazman},\ and\ \citenamefont {Paaske}}]{Kirsanskas2015}%
  \BibitemOpen
  \bibfield  {author} {\bibinfo {author} {\bibfnamefont {G.}~\bibnamefont
  {Kir\ifmmode~\check{s}\else \v{s}\fi{}anskas}}, \bibinfo {author}
  {\bibfnamefont {M.}~\bibnamefont {Goldstein}}, \bibinfo {author}
  {\bibfnamefont {K.}~\bibnamefont {Flensberg}}, \bibinfo {author}
  {\bibfnamefont {L.~I.}\ \bibnamefont {Glazman}}, \ and\ \bibinfo {author}
  {\bibfnamefont {J.}~\bibnamefont {Paaske}},\ }\href {\doibase
  10.1103/PhysRevB.92.235422} {\bibfield  {journal} {\bibinfo  {journal} {Phys.
  Rev. B}\ }\textbf {\bibinfo {volume} {92}},\ \bibinfo {pages} {235422}
  (\bibinfo {year} {2015})}\BibitemShut {NoStop}%
\bibitem [{\citenamefont {Ryazanov}\ \emph {et~al.}(2001)\citenamefont
  {Ryazanov}, \citenamefont {Oboznov}, \citenamefont {Rusanov}, \citenamefont
  {Veretennikov}, \citenamefont {Golubov},\ and\ \citenamefont
  {Aarts}}]{Ryazanov2001}%
  \BibitemOpen
  \bibfield  {author} {\bibinfo {author} {\bibfnamefont {V.~V.}\ \bibnamefont
  {Ryazanov}}, \bibinfo {author} {\bibfnamefont {V.~A.}\ \bibnamefont
  {Oboznov}}, \bibinfo {author} {\bibfnamefont {A.~Y.}\ \bibnamefont
  {Rusanov}}, \bibinfo {author} {\bibfnamefont {A.~V.}\ \bibnamefont
  {Veretennikov}}, \bibinfo {author} {\bibfnamefont {A.~A.}\ \bibnamefont
  {Golubov}}, \ and\ \bibinfo {author} {\bibfnamefont {J.}~\bibnamefont
  {Aarts}},\ }\href {http://link.aps.org/doi/10.1103/PhysRevLett.86.2427}
  {\bibfield  {journal} {\bibinfo  {journal} {Phys. Rev. Lett.}\ }\textbf
  {\bibinfo {volume} {86}},\ \bibinfo {pages} {2427} (\bibinfo {year}
  {2001})}\BibitemShut {NoStop}%
\bibitem [{\citenamefont {Kontos}\ \emph {et~al.}(2002)\citenamefont {Kontos},
  \citenamefont {Aprili}, \citenamefont {Lesueur}, \citenamefont {Gen{\^{e}}t},
  \citenamefont {Stephanidis},\ and\ \citenamefont {Boursier}}]{Kontos2002}%
  \BibitemOpen
  \bibfield  {author} {\bibinfo {author} {\bibfnamefont {T.}~\bibnamefont
  {Kontos}}, \bibinfo {author} {\bibfnamefont {M.}~\bibnamefont {Aprili}},
  \bibinfo {author} {\bibfnamefont {J.}~\bibnamefont {Lesueur}}, \bibinfo
  {author} {\bibfnamefont {F.}~\bibnamefont {Gen{\^{e}}t}}, \bibinfo {author}
  {\bibfnamefont {B.}~\bibnamefont {Stephanidis}}, \ and\ \bibinfo {author}
  {\bibfnamefont {R.}~\bibnamefont {Boursier}},\ }\href
  {http://link.aps.org/doi/10.1103/PhysRevLett.89.137007} {\bibfield  {journal}
  {\bibinfo  {journal} {Phys. Rev. Lett.}\ }\textbf {\bibinfo {volume} {89}},\
  \bibinfo {pages} {137007} (\bibinfo {year} {2002})}\BibitemShut {NoStop}%
\bibitem [{\citenamefont {Robinson}\ \emph {et~al.}(2006)\citenamefont
  {Robinson}, \citenamefont {Piano}, \citenamefont {Burnell}, \citenamefont
  {Bell},\ and\ \citenamefont {Blamire}}]{Robinson2006}%
  \BibitemOpen
  \bibfield  {author} {\bibinfo {author} {\bibfnamefont {J.~W.~A.}\
  \bibnamefont {Robinson}}, \bibinfo {author} {\bibfnamefont {S.}~\bibnamefont
  {Piano}}, \bibinfo {author} {\bibfnamefont {G.}~\bibnamefont {Burnell}},
  \bibinfo {author} {\bibfnamefont {C.}~\bibnamefont {Bell}}, \ and\ \bibinfo
  {author} {\bibfnamefont {M.~G.}\ \bibnamefont {Blamire}},\ }\href
  {http://link.aps.org/doi/10.1103/PhysRevLett.97.177003} {\bibfield  {journal}
  {\bibinfo  {journal} {Phys. Rev. Lett.}\ }\textbf {\bibinfo {volume} {97}},\
  \bibinfo {pages} {177003} (\bibinfo {year} {2006})}\BibitemShut {NoStop}%
\bibitem [{\citenamefont {Feofanov}\ \emph {et~al.}(2010)\citenamefont
  {Feofanov}, \citenamefont {Oboznov}, \citenamefont {Bol'ginov}, \citenamefont
  {Lisenfeld}, \citenamefont {Poletto}, \citenamefont {Ryazanov}, \citenamefont
  {Rossolenko}, \citenamefont {Khabipov}, \citenamefont {Balashov},
  \citenamefont {Zorin}, \citenamefont {Dmitriev}, \citenamefont {Koshelets},\
  and\ \citenamefont {Ustinov}}]{Feofanov2010}%
  \BibitemOpen
  \bibfield  {author} {\bibinfo {author} {\bibfnamefont {A.~K.}\ \bibnamefont
  {Feofanov}}, \bibinfo {author} {\bibfnamefont {V.~A.}\ \bibnamefont
  {Oboznov}}, \bibinfo {author} {\bibfnamefont {V.~V.}\ \bibnamefont
  {Bol'ginov}}, \bibinfo {author} {\bibfnamefont {J.}~\bibnamefont
  {Lisenfeld}}, \bibinfo {author} {\bibfnamefont {S.}~\bibnamefont {Poletto}},
  \bibinfo {author} {\bibfnamefont {V.~V.}\ \bibnamefont {Ryazanov}}, \bibinfo
  {author} {\bibfnamefont {A.~N.}\ \bibnamefont {Rossolenko}}, \bibinfo
  {author} {\bibfnamefont {M.}~\bibnamefont {Khabipov}}, \bibinfo {author}
  {\bibfnamefont {D.}~\bibnamefont {Balashov}}, \bibinfo {author}
  {\bibfnamefont {A.~B.}\ \bibnamefont {Zorin}}, \bibinfo {author}
  {\bibfnamefont {P.~N.}\ \bibnamefont {Dmitriev}}, \bibinfo {author}
  {\bibfnamefont {V.~P.}\ \bibnamefont {Koshelets}}, \ and\ \bibinfo {author}
  {\bibfnamefont {A.~V.}\ \bibnamefont {Ustinov}},\ }\href
  {http://www.nature.com/doifinder/10.1038/nphys1700} {\bibfield  {journal}
  {\bibinfo  {journal} {Nat. Phys.}\ }\textbf {\bibinfo {volume} {6}},\
  \bibinfo {pages} {593} (\bibinfo {year} {2010})}\BibitemShut {NoStop}%
\bibitem [{\citenamefont {Buitelaar}\ \emph {et~al.}(2002)\citenamefont
  {Buitelaar}, \citenamefont {Nussbaumer},\ and\ \citenamefont
  {Sch\"onenberger}}]{Buitelaar2002}%
  \BibitemOpen
  \bibfield  {author} {\bibinfo {author} {\bibfnamefont {M.~R.}\ \bibnamefont
  {Buitelaar}}, \bibinfo {author} {\bibfnamefont {T.}~\bibnamefont
  {Nussbaumer}}, \ and\ \bibinfo {author} {\bibfnamefont {C.}~\bibnamefont
  {Sch\"onenberger}},\ }\href {\doibase 10.1103/PhysRevLett.89.256801}
  {\bibfield  {journal} {\bibinfo  {journal} {Phys. Rev. Lett.}\ }\textbf
  {\bibinfo {volume} {89}},\ \bibinfo {pages} {256801} (\bibinfo {year}
  {2002})}\BibitemShut {NoStop}%
\bibitem [{\citenamefont {J{\o}rgensen}\ \emph {et~al.}(2007)\citenamefont
  {J{\o}rgensen}, \citenamefont {Novotný}, \citenamefont {Grove-Rasmussen},
  \citenamefont {Flensberg},\ and\ \citenamefont {Lindelof}}]{Jorgensen2007}%
  \BibitemOpen
  \bibfield  {author} {\bibinfo {author} {\bibfnamefont {H.~I.}\ \bibnamefont
  {J{\o}rgensen}}, \bibinfo {author} {\bibfnamefont {T.}~\bibnamefont
  {Novotný}}, \bibinfo {author} {\bibfnamefont {K.}~\bibnamefont
  {Grove-Rasmussen}}, \bibinfo {author} {\bibfnamefont {K.}~\bibnamefont
  {Flensberg}}, \ and\ \bibinfo {author} {\bibfnamefont {P.~E.}\ \bibnamefont
  {Lindelof}},\ }\href {\doibase 10.1021/nl071152w} {\bibfield  {journal}
  {\bibinfo  {journal} {Nano Lett.}\ }\textbf {\bibinfo {volume} {7}},\
  \bibinfo {pages} {2441} (\bibinfo {year} {2007})}\BibitemShut {NoStop}%
\bibitem [{\citenamefont {Eichler}\ \emph {et~al.}(2009)\citenamefont
  {Eichler}, \citenamefont {Deblock}, \citenamefont {Weiss}, \citenamefont
  {Karrasch}, \citenamefont {Meden}, \citenamefont {Sch\"onenberger},\ and\
  \citenamefont {Bouchiat}}]{Eichler2009}%
  \BibitemOpen
  \bibfield  {author} {\bibinfo {author} {\bibfnamefont {A.}~\bibnamefont
  {Eichler}}, \bibinfo {author} {\bibfnamefont {R.}~\bibnamefont {Deblock}},
  \bibinfo {author} {\bibfnamefont {M.}~\bibnamefont {Weiss}}, \bibinfo
  {author} {\bibfnamefont {C.}~\bibnamefont {Karrasch}}, \bibinfo {author}
  {\bibfnamefont {V.}~\bibnamefont {Meden}}, \bibinfo {author} {\bibfnamefont
  {C.}~\bibnamefont {Sch\"onenberger}}, \ and\ \bibinfo {author} {\bibfnamefont
  {H.}~\bibnamefont {Bouchiat}},\ }\href {\doibase 10.1103/PhysRevB.79.161407}
  {\bibfield  {journal} {\bibinfo  {journal} {Phys. Rev. B}\ }\textbf {\bibinfo
  {volume} {79}},\ \bibinfo {pages} {161407} (\bibinfo {year}
  {2009})}\BibitemShut {NoStop}%
\bibitem [{\citenamefont {Deacon}\ \emph {et~al.}(2010)\citenamefont {Deacon},
  \citenamefont {Tanaka}, \citenamefont {Oiwa}, \citenamefont {Sakano},
  \citenamefont {Yoshida}, \citenamefont {Shibata}, \citenamefont {Hirakawa},\
  and\ \citenamefont {Tarucha}}]{Deacon2010}%
  \BibitemOpen
  \bibfield  {author} {\bibinfo {author} {\bibfnamefont {R.~S.}\ \bibnamefont
  {Deacon}}, \bibinfo {author} {\bibfnamefont {Y.}~\bibnamefont {Tanaka}},
  \bibinfo {author} {\bibfnamefont {A.}~\bibnamefont {Oiwa}}, \bibinfo {author}
  {\bibfnamefont {R.}~\bibnamefont {Sakano}}, \bibinfo {author} {\bibfnamefont
  {K.}~\bibnamefont {Yoshida}}, \bibinfo {author} {\bibfnamefont
  {K.}~\bibnamefont {Shibata}}, \bibinfo {author} {\bibfnamefont
  {K.}~\bibnamefont {Hirakawa}}, \ and\ \bibinfo {author} {\bibfnamefont
  {S.}~\bibnamefont {Tarucha}},\ }\href {\doibase
  10.1103/PhysRevLett.104.076805} {\bibfield  {journal} {\bibinfo  {journal}
  {Phys. Rev. Lett.}\ }\textbf {\bibinfo {volume} {104}},\ \bibinfo {pages}
  {076805} (\bibinfo {year} {2010})}\BibitemShut {NoStop}%
\bibitem [{\citenamefont {Pillet}\ \emph {et~al.}(2010)\citenamefont {Pillet},
  \citenamefont {Quay}, \citenamefont {Morfin}, \citenamefont {Bena},
  \citenamefont {Yeyati},\ and\ \citenamefont {Joyez}}]{Pillet2010}%
  \BibitemOpen
  \bibfield  {author} {\bibinfo {author} {\bibfnamefont {J.-D.}\ \bibnamefont
  {Pillet}}, \bibinfo {author} {\bibfnamefont {C.~H.~L.}\ \bibnamefont {Quay}},
  \bibinfo {author} {\bibfnamefont {P.}~\bibnamefont {Morfin}}, \bibinfo
  {author} {\bibfnamefont {C.}~\bibnamefont {Bena}}, \bibinfo {author}
  {\bibfnamefont {A.~L.}\ \bibnamefont {Yeyati}}, \ and\ \bibinfo {author}
  {\bibfnamefont {P.}~\bibnamefont {Joyez}},\ }\href {\doibase
  10.1038/nphys1811} {\bibfield  {journal} {\bibinfo  {journal} {Nat. Phys.}\
  }\textbf {\bibinfo {volume} {6}},\ \bibinfo {pages} {965} (\bibinfo {year}
  {2010})}\BibitemShut {NoStop}%
\bibitem [{\citenamefont {Franke}\ \emph {et~al.}(2011)\citenamefont {Franke},
  \citenamefont {Schulze},\ and\ \citenamefont {Pascual}}]{Franke2011}%
  \BibitemOpen
  \bibfield  {author} {\bibinfo {author} {\bibfnamefont {K.~J.}\ \bibnamefont
  {Franke}}, \bibinfo {author} {\bibfnamefont {G.}~\bibnamefont {Schulze}}, \
  and\ \bibinfo {author} {\bibfnamefont {J.~I.}\ \bibnamefont {Pascual}},\
  }\href {\doibase 10.1126/science.1202204} {\bibfield  {journal} {\bibinfo
  {journal} {Science}\ }\textbf {\bibinfo {volume} {332}},\ \bibinfo {pages}
  {940} (\bibinfo {year} {2011})}\BibitemShut {NoStop}%
\bibitem [{\citenamefont {Lee}\ \emph {et~al.}(2012)\citenamefont {Lee},
  \citenamefont {Jiang}, \citenamefont {Aguado}, \citenamefont {Katsaros},
  \citenamefont {Lieber},\ and\ \citenamefont {De~Franceschi}}]{Lee2012}%
  \BibitemOpen
  \bibfield  {author} {\bibinfo {author} {\bibfnamefont {E.~J.~H.}\
  \bibnamefont {Lee}}, \bibinfo {author} {\bibfnamefont {X.}~\bibnamefont
  {Jiang}}, \bibinfo {author} {\bibfnamefont {R.}~\bibnamefont {Aguado}},
  \bibinfo {author} {\bibfnamefont {G.}~\bibnamefont {Katsaros}}, \bibinfo
  {author} {\bibfnamefont {C.~M.}\ \bibnamefont {Lieber}}, \ and\ \bibinfo
  {author} {\bibfnamefont {S.}~\bibnamefont {De~Franceschi}},\ }\href {\doibase
  10.1103/PhysRevLett.109.186802} {\bibfield  {journal} {\bibinfo  {journal}
  {Phys. Rev. Lett.}\ }\textbf {\bibinfo {volume} {109}},\ \bibinfo {pages}
  {186802} (\bibinfo {year} {2012})}\BibitemShut {NoStop}%
\bibitem [{\citenamefont {Maurand}\ \emph
  {et~al.}(2012{\natexlab{a}})\citenamefont {Maurand}, \citenamefont {Meng},
  \citenamefont {Bonet}, \citenamefont {Florens}, \citenamefont {Marty},\ and\
  \citenamefont {Wernsdorfer}}]{Maurand2012}%
  \BibitemOpen
  \bibfield  {author} {\bibinfo {author} {\bibfnamefont {R.}~\bibnamefont
  {Maurand}}, \bibinfo {author} {\bibfnamefont {T.}~\bibnamefont {Meng}},
  \bibinfo {author} {\bibfnamefont {E.}~\bibnamefont {Bonet}}, \bibinfo
  {author} {\bibfnamefont {S.}~\bibnamefont {Florens}}, \bibinfo {author}
  {\bibfnamefont {L.}~\bibnamefont {Marty}}, \ and\ \bibinfo {author}
  {\bibfnamefont {W.}~\bibnamefont {Wernsdorfer}},\ }\href {\doibase
  10.1103/PhysRevX.2.011009} {\bibfield  {journal} {\bibinfo  {journal} {Phys.
  Rev. X}\ }\textbf {\bibinfo {volume} {2}},\ \bibinfo {pages} {011009}
  (\bibinfo {year} {2012}{\natexlab{a}})}\BibitemShut {NoStop}%
\bibitem [{\citenamefont {Maurand}\ \emph
  {et~al.}(2012{\natexlab{b}})\citenamefont {Maurand}, \citenamefont {Meng},
  \citenamefont {Bonet}, \citenamefont {Florens}, \citenamefont {Marty},\ and\
  \citenamefont {Wernsdorfer}}]{Maurand2012alt}%
  \BibitemOpen
  \bibfield  {author} {\bibinfo {author} {\bibfnamefont {R.}~\bibnamefont
  {Maurand}}, \bibinfo {author} {\bibfnamefont {T.}~\bibnamefont {Meng}},
  \bibinfo {author} {\bibfnamefont {E.}~\bibnamefont {Bonet}}, \bibinfo
  {author} {\bibfnamefont {S.}~\bibnamefont {Florens}}, \bibinfo {author}
  {\bibfnamefont {L.}~\bibnamefont {Marty}}, \ and\ \bibinfo {author}
  {\bibfnamefont {W.}~\bibnamefont {Wernsdorfer}},\ }\href {\doibase
  10.1103/PhysRevX.2.019901} {\ \textbf {\bibinfo {volume} {2}},\ \bibinfo
  {pages} {19901(E)} (\bibinfo {year} {2012}{\natexlab{b}})}\BibitemShut
  {NoStop}%
\bibitem [{\citenamefont {Chang}\ \emph {et~al.}(2013)\citenamefont {Chang},
  \citenamefont {Manucharyan}, \citenamefont {Jespersen}, \citenamefont
  {Nyg\aa{}rd},\ and\ \citenamefont {Marcus}}]{Chang2013}%
  \BibitemOpen
  \bibfield  {author} {\bibinfo {author} {\bibfnamefont {W.}~\bibnamefont
  {Chang}}, \bibinfo {author} {\bibfnamefont {V.~E.}\ \bibnamefont
  {Manucharyan}}, \bibinfo {author} {\bibfnamefont {T.~S.}\ \bibnamefont
  {Jespersen}}, \bibinfo {author} {\bibfnamefont {J.}~\bibnamefont
  {Nyg\aa{}rd}}, \ and\ \bibinfo {author} {\bibfnamefont {C.~M.}\ \bibnamefont
  {Marcus}},\ }\href {\doibase 10.1103/PhysRevLett.110.217005} {\bibfield
  {journal} {\bibinfo  {journal} {Phys. Rev. Lett.}\ }\textbf {\bibinfo
  {volume} {110}},\ \bibinfo {pages} {217005} (\bibinfo {year}
  {2013})}\BibitemShut {NoStop}%
\bibitem [{\citenamefont {Kim}\ \emph {et~al.}(2013)\citenamefont {Kim},
  \citenamefont {Ahn}, \citenamefont {Kim}, \citenamefont {Choi}, \citenamefont
  {Bae}, \citenamefont {Kang}, \citenamefont {Lim}, \citenamefont {L\'opez},\
  and\ \citenamefont {Kim}}]{Kim2013}%
  \BibitemOpen
  \bibfield  {author} {\bibinfo {author} {\bibfnamefont {B.-K.}\ \bibnamefont
  {Kim}}, \bibinfo {author} {\bibfnamefont {Y.-H.}\ \bibnamefont {Ahn}},
  \bibinfo {author} {\bibfnamefont {J.-J.}\ \bibnamefont {Kim}}, \bibinfo
  {author} {\bibfnamefont {M.-S.}\ \bibnamefont {Choi}}, \bibinfo {author}
  {\bibfnamefont {M.-H.}\ \bibnamefont {Bae}}, \bibinfo {author} {\bibfnamefont
  {K.}~\bibnamefont {Kang}}, \bibinfo {author} {\bibfnamefont {J.~S.}\
  \bibnamefont {Lim}}, \bibinfo {author} {\bibfnamefont {R.}~\bibnamefont
  {L\'opez}}, \ and\ \bibinfo {author} {\bibfnamefont {N.}~\bibnamefont
  {Kim}},\ }\href {\doibase 10.1103/PhysRevLett.110.076803} {\bibfield
  {journal} {\bibinfo  {journal} {Phys. Rev. Lett.}\ }\textbf {\bibinfo
  {volume} {110}},\ \bibinfo {pages} {076803} (\bibinfo {year}
  {2013})}\BibitemShut {NoStop}%
\bibitem [{\citenamefont {Pillet}\ \emph {et~al.}(2013)\citenamefont {Pillet},
  \citenamefont {Joyez}, \citenamefont {\ifmmode~\check{Z}\else
  \v{Z}\fi{}itko},\ and\ \citenamefont {Goffman}}]{Pillet2013}%
  \BibitemOpen
  \bibfield  {author} {\bibinfo {author} {\bibfnamefont {J.-D.}\ \bibnamefont
  {Pillet}}, \bibinfo {author} {\bibfnamefont {P.}~\bibnamefont {Joyez}},
  \bibinfo {author} {\bibfnamefont {R.}~\bibnamefont {\ifmmode~\check{Z}\else
  \v{Z}\fi{}itko}}, \ and\ \bibinfo {author} {\bibfnamefont {M.~F.}\
  \bibnamefont {Goffman}},\ }\href {\doibase 10.1103/PhysRevB.88.045101}
  {\bibfield  {journal} {\bibinfo  {journal} {Phys. Rev. B}\ }\textbf {\bibinfo
  {volume} {88}},\ \bibinfo {pages} {045101} (\bibinfo {year}
  {2013})}\BibitemShut {NoStop}%
\bibitem [{\citenamefont {Kumar}\ \emph {et~al.}(2014)\citenamefont {Kumar},
  \citenamefont {Gaim}, \citenamefont {Steininger}, \citenamefont
  {Levy~Yeyati}, \citenamefont {Mart\'{\i}n-Rodero}, \citenamefont {H\"uttel},\
  and\ \citenamefont {Strunk}}]{Kumar2014}%
  \BibitemOpen
  \bibfield  {author} {\bibinfo {author} {\bibfnamefont {A.}~\bibnamefont
  {Kumar}}, \bibinfo {author} {\bibfnamefont {M.}~\bibnamefont {Gaim}},
  \bibinfo {author} {\bibfnamefont {D.}~\bibnamefont {Steininger}}, \bibinfo
  {author} {\bibfnamefont {A.}~\bibnamefont {Levy~Yeyati}}, \bibinfo {author}
  {\bibfnamefont {A.}~\bibnamefont {Mart\'{\i}n-Rodero}}, \bibinfo {author}
  {\bibfnamefont {A.~K.}\ \bibnamefont {H\"uttel}}, \ and\ \bibinfo {author}
  {\bibfnamefont {C.}~\bibnamefont {Strunk}},\ }\href {\doibase
  10.1103/PhysRevB.89.075428} {\bibfield  {journal} {\bibinfo  {journal} {Phys.
  Rev. B}\ }\textbf {\bibinfo {volume} {89}},\ \bibinfo {pages} {075428}
  (\bibinfo {year} {2014})}\BibitemShut {NoStop}%
\bibitem [{\citenamefont {Lee}\ \emph {et~al.}(2014)\citenamefont {Lee},
  \citenamefont {Jiang}, \citenamefont {Houzet}, \citenamefont {Aguado},
  \citenamefont {Lieber},\ and\ \citenamefont {De~Franceschi}}]{Lee2014}%
  \BibitemOpen
  \bibfield  {author} {\bibinfo {author} {\bibfnamefont {E.~J.~H.}\
  \bibnamefont {Lee}}, \bibinfo {author} {\bibfnamefont {X.}~\bibnamefont
  {Jiang}}, \bibinfo {author} {\bibfnamefont {M.}~\bibnamefont {Houzet}},
  \bibinfo {author} {\bibfnamefont {R.}~\bibnamefont {Aguado}}, \bibinfo
  {author} {\bibfnamefont {C.~M.}\ \bibnamefont {Lieber}}, \ and\ \bibinfo
  {author} {\bibfnamefont {S.}~\bibnamefont {De~Franceschi}},\ }\href {\doibase
  10.1038/nnano.2013.267} {\bibfield  {journal} {\bibinfo  {journal} {Nat.
  Nanotechnol.}\ }\textbf {\bibinfo {volume} {9}},\ \bibinfo {pages} {79}
  (\bibinfo {year} {2014})}\BibitemShut {NoStop}%
\bibitem [{\citenamefont {Delagrange}\ \emph {et~al.}(2015)\citenamefont
  {Delagrange}, \citenamefont {Luitz}, \citenamefont {Weil}, \citenamefont
  {Kasumov}, \citenamefont {Meden}, \citenamefont {Bouchiat},\ and\
  \citenamefont {Deblock}}]{Delagrange2015}%
  \BibitemOpen
  \bibfield  {author} {\bibinfo {author} {\bibfnamefont {R.}~\bibnamefont
  {Delagrange}}, \bibinfo {author} {\bibfnamefont {D.~J.}\ \bibnamefont
  {Luitz}}, \bibinfo {author} {\bibfnamefont {R.}~\bibnamefont {Weil}},
  \bibinfo {author} {\bibfnamefont {A.}~\bibnamefont {Kasumov}}, \bibinfo
  {author} {\bibfnamefont {V.}~\bibnamefont {Meden}}, \bibinfo {author}
  {\bibfnamefont {H.}~\bibnamefont {Bouchiat}}, \ and\ \bibinfo {author}
  {\bibfnamefont {R.}~\bibnamefont {Deblock}},\ }\href {\doibase
  10.1103/PhysRevB.91.241401} {\bibfield  {journal} {\bibinfo  {journal} {Phys.
  Rev. B}\ }\textbf {\bibinfo {volume} {91}},\ \bibinfo {pages} {241401}
  (\bibinfo {year} {2015})}\BibitemShut {NoStop}%
\bibitem [{\citenamefont {Assouline}\ \emph {et~al.}(2017)\citenamefont
  {Assouline}, \citenamefont {Feuillet-Palma}, \citenamefont {Zimmers},
  \citenamefont {Aubin}, \citenamefont {Aprili},\ and\ \citenamefont
  {Harmand}}]{Assouline2017}%
  \BibitemOpen
  \bibfield  {author} {\bibinfo {author} {\bibfnamefont {A.}~\bibnamefont
  {Assouline}}, \bibinfo {author} {\bibfnamefont {C.}~\bibnamefont
  {Feuillet-Palma}}, \bibinfo {author} {\bibfnamefont {A.}~\bibnamefont
  {Zimmers}}, \bibinfo {author} {\bibfnamefont {H.}~\bibnamefont {Aubin}},
  \bibinfo {author} {\bibfnamefont {M.}~\bibnamefont {Aprili}}, \ and\ \bibinfo
  {author} {\bibfnamefont {J.-C.}\ \bibnamefont {Harmand}},\ }\href {\doibase
  10.1103/PhysRevLett.119.097701} {\bibfield  {journal} {\bibinfo  {journal}
  {Phys. Rev. Lett.}\ }\textbf {\bibinfo {volume} {119}},\ \bibinfo {pages}
  {097701} (\bibinfo {year} {2017})}\BibitemShut {NoStop}%
\bibitem [{\citenamefont {Li}\ \emph {et~al.}(2017)\citenamefont {Li},
  \citenamefont {Kang}, \citenamefont {Caroff},\ and\ \citenamefont
  {Xu}}]{Li2017}%
  \BibitemOpen
  \bibfield  {author} {\bibinfo {author} {\bibfnamefont {S.}~\bibnamefont
  {Li}}, \bibinfo {author} {\bibfnamefont {N.}~\bibnamefont {Kang}}, \bibinfo
  {author} {\bibfnamefont {P.}~\bibnamefont {Caroff}}, \ and\ \bibinfo {author}
  {\bibfnamefont {H.~Q.}\ \bibnamefont {Xu}},\ }\href {\doibase
  10.1103/PhysRevB.95.014515} {\bibfield  {journal} {\bibinfo  {journal} {Phys.
  Rev. B}\ }\textbf {\bibinfo {volume} {95}},\ \bibinfo {pages} {014515}
  (\bibinfo {year} {2017})}\BibitemShut {NoStop}%
\bibitem [{\citenamefont {van Woerkom}\ \emph {et~al.}(2017)\citenamefont {van
  Woerkom}, \citenamefont {Proutski}, \citenamefont {van Heck}, \citenamefont
  {Bouman}, \citenamefont {Vayrynen}, \citenamefont {Glazman}, \citenamefont
  {Krogstrup}, \citenamefont {Nygard}, \citenamefont {Kouwenhoven},\ and\
  \citenamefont {Geresdi}}]{VanWoerkom2017}%
  \BibitemOpen
  \bibfield  {author} {\bibinfo {author} {\bibfnamefont {D.~J.}\ \bibnamefont
  {van Woerkom}}, \bibinfo {author} {\bibfnamefont {A.}~\bibnamefont
  {Proutski}}, \bibinfo {author} {\bibfnamefont {B.}~\bibnamefont {van Heck}},
  \bibinfo {author} {\bibfnamefont {D.}~\bibnamefont {Bouman}}, \bibinfo
  {author} {\bibfnamefont {J.~I.}\ \bibnamefont {Vayrynen}}, \bibinfo {author}
  {\bibfnamefont {L.~I.}\ \bibnamefont {Glazman}}, \bibinfo {author}
  {\bibfnamefont {P.}~\bibnamefont {Krogstrup}}, \bibinfo {author}
  {\bibfnamefont {J.}~\bibnamefont {Nygard}}, \bibinfo {author} {\bibfnamefont
  {L.~P.}\ \bibnamefont {Kouwenhoven}}, \ and\ \bibinfo {author} {\bibfnamefont
  {A.}~\bibnamefont {Geresdi}},\ }\href {\doibase 10.1038/nphys4150} {\bibfield
   {journal} {\bibinfo  {journal} {Nat. Phys.}\ }\textbf {\bibinfo {volume}
  {13}},\ \bibinfo {pages} {876} (\bibinfo {year} {2017})}\BibitemShut
  {NoStop}%
\bibitem [{\citenamefont {Yamashita}\ \emph {et~al.}(2005)\citenamefont
  {Yamashita}, \citenamefont {Tanikawa}, \citenamefont {Takahashi},\ and\
  \citenamefont {Maekawa}}]{Yamashita2005}%
  \BibitemOpen
  \bibfield  {author} {\bibinfo {author} {\bibfnamefont {T.}~\bibnamefont
  {Yamashita}}, \bibinfo {author} {\bibfnamefont {K.}~\bibnamefont {Tanikawa}},
  \bibinfo {author} {\bibfnamefont {S.}~\bibnamefont {Takahashi}}, \ and\
  \bibinfo {author} {\bibfnamefont {S.}~\bibnamefont {Maekawa}},\ }\href
  {http://link.aps.org/doi/10.1103/PhysRevLett.95.097001} {\bibfield  {journal}
  {\bibinfo  {journal} {Phys. Rev. Lett.}\ }\textbf {\bibinfo {volume} {95}},\
  \bibinfo {pages} {097001} (\bibinfo {year} {2005})}\BibitemShut {NoStop}%
\bibitem [{\citenamefont {Ioffe}\ \emph {et~al.}(1999)\citenamefont {Ioffe},
  \citenamefont {Geshkenbein}, \citenamefont {Feigel'man}, \citenamefont
  {Fauch{\`e}re},\ and\ \citenamefont {Blatter}}]{Ioffe1999}%
  \BibitemOpen
  \bibfield  {author} {\bibinfo {author} {\bibfnamefont {L.~B.}\ \bibnamefont
  {Ioffe}}, \bibinfo {author} {\bibfnamefont {V.~B.}\ \bibnamefont
  {Geshkenbein}}, \bibinfo {author} {\bibfnamefont {M.~V.}\ \bibnamefont
  {Feigel'man}}, \bibinfo {author} {\bibfnamefont {A.~L.}\ \bibnamefont
  {Fauch{\`e}re}}, \ and\ \bibinfo {author} {\bibfnamefont {G.}~\bibnamefont
  {Blatter}},\ }\href
  {http://www.nature.com/nature/journal/v398/n6729/abs/398679a0.html}
  {\bibfield  {journal} {\bibinfo  {journal} {Nature (London)}\ }\textbf
  {\bibinfo {volume} {398}},\ \bibinfo {pages} {679} (\bibinfo {year}
  {1999})}\BibitemShut {NoStop}%
\bibitem [{\citenamefont {Mooij}\ \emph {et~al.}(1999)\citenamefont {Mooij},
  \citenamefont {Orlando}, \citenamefont {Levitov}, \citenamefont {Tian},
  \citenamefont {van~der Wal},\ and\ \citenamefont {Lloyd}}]{Mooij1999}%
  \BibitemOpen
  \bibfield  {author} {\bibinfo {author} {\bibfnamefont {J.~E.}\ \bibnamefont
  {Mooij}}, \bibinfo {author} {\bibfnamefont {T.~P.}\ \bibnamefont {Orlando}},
  \bibinfo {author} {\bibfnamefont {L.}~\bibnamefont {Levitov}}, \bibinfo
  {author} {\bibfnamefont {L.}~\bibnamefont {Tian}}, \bibinfo {author}
  {\bibfnamefont {C.~H.}\ \bibnamefont {van~der Wal}}, \ and\ \bibinfo {author}
  {\bibfnamefont {S.}~\bibnamefont {Lloyd}},\ }\href
  {http://science.sciencemag.org/content/285/5430/1036} {\bibfield  {journal}
  {\bibinfo  {journal} {Science}\ }\textbf {\bibinfo {volume} {285}},\ \bibinfo
  {pages} {1036} (\bibinfo {year} {1999})}\BibitemShut {NoStop}%
\bibitem [{\citenamefont {Devoret}\ and\ \citenamefont
  {Schoelkopf}(2013)}]{Devoret2013}%
  \BibitemOpen
  \bibfield  {author} {\bibinfo {author} {\bibfnamefont {M.~H.}\ \bibnamefont
  {Devoret}}\ and\ \bibinfo {author} {\bibfnamefont {R.~J.}\ \bibnamefont
  {Schoelkopf}},\ }\href {http://science.sciencemag.org/content/339/6124/1169}
  {\bibfield  {journal} {\bibinfo  {journal} {Science}\ }\textbf {\bibinfo
  {volume} {339}},\ \bibinfo {pages} {1169} (\bibinfo {year}
  {2013})}\BibitemShut {NoStop}%
\bibitem [{\citenamefont {{\v{Z}}uti{\'{c}}}\ \emph {et~al.}(2004)\citenamefont
  {{\v{Z}}uti{\'{c}}}, \citenamefont {Fabian},\ and\ \citenamefont {{Das
  Sarma}}}]{Fabian2004}%
  \BibitemOpen
  \bibfield  {author} {\bibinfo {author} {\bibfnamefont {I.}~\bibnamefont
  {{\v{Z}}uti{\'{c}}}}, \bibinfo {author} {\bibfnamefont {J.}~\bibnamefont
  {Fabian}}, \ and\ \bibinfo {author} {\bibfnamefont {S.}~\bibnamefont {{Das
  Sarma}}},\ }\href {http://link.aps.org/doi/10.1103/RevModPhys.76.323}
  {\bibfield  {journal} {\bibinfo  {journal} {Rev. Mod. Phys.}\ }\textbf
  {\bibinfo {volume} {76}},\ \bibinfo {pages} {323} (\bibinfo {year}
  {2004})}\BibitemShut {NoStop}%
\bibitem [{\citenamefont {Fabian}\ \emph {et~al.}(2007)\citenamefont {Fabian},
  \citenamefont {Matos-Abiague}, \citenamefont {Ertler}, \citenamefont
  {Stano},\ and\ \citenamefont {{\v{Z}}uti{\'{c}}}}]{Fabian2007}%
  \BibitemOpen
  \bibfield  {author} {\bibinfo {author} {\bibfnamefont {J.}~\bibnamefont
  {Fabian}}, \bibinfo {author} {\bibfnamefont {A.}~\bibnamefont
  {Matos-Abiague}}, \bibinfo {author} {\bibfnamefont {C.}~\bibnamefont
  {Ertler}}, \bibinfo {author} {\bibfnamefont {P.}~\bibnamefont {Stano}}, \
  and\ \bibinfo {author} {\bibfnamefont {I.}~\bibnamefont
  {{\v{Z}}uti{\'{c}}}},\ }\href
  {http://www.physics.sk/aps/pub.php?y=2007&pub=aps-07-04} {\bibfield
  {journal} {\bibinfo  {journal} {Acta Phys. Slovaca}\ }\textbf {\bibinfo
  {volume} {57}},\ \bibinfo {pages} {565} (\bibinfo {year} {2007})}\BibitemShut
  {NoStop}%
\bibitem [{\citenamefont {Andreev}(1964{\natexlab{a}})}]{Andreev1964}%
  \BibitemOpen
  \bibfield  {author} {\bibinfo {author} {\bibfnamefont {A.~F.}\ \bibnamefont
  {Andreev}},\ }\href@noop {} {\bibfield  {journal} {\bibinfo  {journal} {Zh.
  Eksp. Teor. Fiz.}\ }\textbf {\bibinfo {volume} {46}},\ \bibinfo {pages}
  {1823} (\bibinfo {year} {1964}{\natexlab{a}})}\BibitemShut {NoStop}%
\bibitem [{\citenamefont {Andreev}(1964{\natexlab{b}})}]{Andreev1964alt}%
  \BibitemOpen
  \bibfield  {author} {\bibinfo {author} {\bibfnamefont {A.~F.}\ \bibnamefont
  {Andreev}},\ }\href
  {http://www.jetp.ac.ru/cgi-bin/e/index/e/19/5/p1228?a=list} {\bibfield
  {journal} {\bibinfo  {journal} {J. Exp. Theor. Phys.}\ }\textbf {\bibinfo
  {volume} {19}},\ \bibinfo {pages} {1228} (\bibinfo {year}
  {1964}{\natexlab{b}})}\BibitemShut {NoStop}%
\bibitem [{\citenamefont {Andreev}(1966{\natexlab{a}})}]{Andreev1966}%
  \BibitemOpen
  \bibfield  {author} {\bibinfo {author} {\bibfnamefont {A.~F.}\ \bibnamefont
  {Andreev}},\ }\href@noop {} {\bibfield  {journal} {\bibinfo  {journal} {Zh.
  Eksp. Teor. Fiz.}\ }\textbf {\bibinfo {volume} {49}},\ \bibinfo {pages} {655}
  (\bibinfo {year} {1966}{\natexlab{a}})}\BibitemShut {NoStop}%
\bibitem [{\citenamefont {Andreev}(1966{\natexlab{b}})}]{Andreev1966alt}%
  \BibitemOpen
  \bibfield  {author} {\bibinfo {author} {\bibfnamefont {A.~F.}\ \bibnamefont
  {Andreev}},\ }\href
  {http://www.jetp.ac.ru/cgi-bin/e/index/e/22/2/p455?a=list} {\bibfield
  {journal} {\bibinfo  {journal} {J. Exp. Theor. Phys.}\ }\textbf {\bibinfo
  {volume} {22}},\ \bibinfo {pages} {455} (\bibinfo {year}
  {1966}{\natexlab{b}})}\BibitemShut {NoStop}%
\bibitem [{\citenamefont {Su}\ \emph {et~al.}(2017)\citenamefont {Su},
  \citenamefont {Tacla}, \citenamefont {Hocevar}, \citenamefont {Car},
  \citenamefont {Plissard}, \citenamefont {Bakkers}, \citenamefont {Daley},
  \citenamefont {Pekker},\ and\ \citenamefont {Frolov}}]{Su2017}%
  \BibitemOpen
  \bibfield  {author} {\bibinfo {author} {\bibfnamefont {Z.}~\bibnamefont
  {Su}}, \bibinfo {author} {\bibfnamefont {A.~B.}\ \bibnamefont {Tacla}},
  \bibinfo {author} {\bibfnamefont {M.}~\bibnamefont {Hocevar}}, \bibinfo
  {author} {\bibfnamefont {D.}~\bibnamefont {Car}}, \bibinfo {author}
  {\bibfnamefont {S.~R.}\ \bibnamefont {Plissard}}, \bibinfo {author}
  {\bibfnamefont {E.~P. A.~M.}\ \bibnamefont {Bakkers}}, \bibinfo {author}
  {\bibfnamefont {A.~J.}\ \bibnamefont {Daley}}, \bibinfo {author}
  {\bibfnamefont {D.}~\bibnamefont {Pekker}}, \ and\ \bibinfo {author}
  {\bibfnamefont {S.~M.}\ \bibnamefont {Frolov}},\ }\href
  {https://doi.org/10.1038/s41467-017-00665-7} {\bibfield  {journal} {\bibinfo
  {journal} {Nat. Commun.}\ }\textbf {\bibinfo {volume} {8}},\ \bibinfo {pages}
  {585} (\bibinfo {year} {2017})}\BibitemShut {NoStop}%
\bibitem [{\citenamefont {Fu}\ and\ \citenamefont {Kane}(2008)}]{Fu2008}%
  \BibitemOpen
  \bibfield  {author} {\bibinfo {author} {\bibfnamefont {L.}~\bibnamefont
  {Fu}}\ and\ \bibinfo {author} {\bibfnamefont {C.~L.}\ \bibnamefont {Kane}},\
  }\href {\doibase 10.1103/PhysRevLett.100.096407} {\bibfield  {journal}
  {\bibinfo  {journal} {Phys. Rev. Lett.}\ }\textbf {\bibinfo {volume} {100}},\
  \bibinfo {pages} {096407} (\bibinfo {year} {2008})}\BibitemShut {NoStop}%
\bibitem [{\citenamefont {Oreg}\ \emph {et~al.}(2010)\citenamefont {Oreg},
  \citenamefont {Refael},\ and\ \citenamefont {{von Oppen}}}]{Oreg2010}%
  \BibitemOpen
  \bibfield  {author} {\bibinfo {author} {\bibfnamefont {Y.}~\bibnamefont
  {Oreg}}, \bibinfo {author} {\bibfnamefont {G.}~\bibnamefont {Refael}}, \ and\
  \bibinfo {author} {\bibfnamefont {F.}~\bibnamefont {{von Oppen}}},\ }\href
  {http://link.aps.org/doi/10.1103/PhysRevLett.105.177002} {\bibfield
  {journal} {\bibinfo  {journal} {Phys. Rev. Lett.}\ }\textbf {\bibinfo
  {volume} {105}},\ \bibinfo {pages} {177002} (\bibinfo {year}
  {2010})}\BibitemShut {NoStop}%
\bibitem [{\citenamefont {Duckheim}\ and\ \citenamefont
  {Brouwer}(2011)}]{Duckheim2011}%
  \BibitemOpen
  \bibfield  {author} {\bibinfo {author} {\bibfnamefont {M.}~\bibnamefont
  {Duckheim}}\ and\ \bibinfo {author} {\bibfnamefont {P.~W.}\ \bibnamefont
  {Brouwer}},\ }\href {http://link.aps.org/doi/10.1103/PhysRevB.83.054513}
  {\bibfield  {journal} {\bibinfo  {journal} {Phys. Rev. B}\ }\textbf {\bibinfo
  {volume} {83}},\ \bibinfo {pages} {054513} (\bibinfo {year}
  {2011})}\BibitemShut {NoStop}%
\bibitem [{\citenamefont {Mourik}\ \emph {et~al.}(2012)\citenamefont {Mourik},
  \citenamefont {Zuo}, \citenamefont {Frolov}, \citenamefont {Plissard},
  \citenamefont {Bakkers},\ and\ \citenamefont {Kouwenhoven}}]{Mourik2012}%
  \BibitemOpen
  \bibfield  {author} {\bibinfo {author} {\bibfnamefont {V.}~\bibnamefont
  {Mourik}}, \bibinfo {author} {\bibfnamefont {K.}~\bibnamefont {Zuo}},
  \bibinfo {author} {\bibfnamefont {S.~M.}\ \bibnamefont {Frolov}}, \bibinfo
  {author} {\bibfnamefont {S.~R.}\ \bibnamefont {Plissard}}, \bibinfo {author}
  {\bibfnamefont {E.~P. A.~M.}\ \bibnamefont {Bakkers}}, \ and\ \bibinfo
  {author} {\bibfnamefont {L.~P.}\ \bibnamefont {Kouwenhoven}},\ }\href
  {\doibase 10.1126/science.1222360} {\bibfield  {journal} {\bibinfo  {journal}
  {Science}\ }\textbf {\bibinfo {volume} {336}},\ \bibinfo {pages} {1003}
  (\bibinfo {year} {2012})}\BibitemShut {NoStop}%
\bibitem [{\citenamefont {Rokhinson}\ \emph {et~al.}(2012)\citenamefont
  {Rokhinson}, \citenamefont {Liu},\ and\ \citenamefont
  {Furdyna}}]{Rokhinson2012}%
  \BibitemOpen
  \bibfield  {author} {\bibinfo {author} {\bibfnamefont {L.~P.}\ \bibnamefont
  {Rokhinson}}, \bibinfo {author} {\bibfnamefont {X.}~\bibnamefont {Liu}}, \
  and\ \bibinfo {author} {\bibfnamefont {J.~K.}\ \bibnamefont {Furdyna}},\
  }\href {http://dx.doi.org/10.1038/nphys2429} {\bibfield  {journal} {\bibinfo
  {journal} {Nat. Phys.}\ }\textbf {\bibinfo {volume} {8}},\ \bibinfo {pages}
  {795} (\bibinfo {year} {2012})}\BibitemShut {NoStop}%
\bibitem [{\citenamefont {Nadj-Perge}\ \emph {et~al.}(2014)\citenamefont
  {Nadj-Perge}, \citenamefont {Drozdov}, \citenamefont {Li}, \citenamefont
  {Chen}, \citenamefont {Jeon}, \citenamefont {Seo}, \citenamefont {MacDonald},
  \citenamefont {Bernevig},\ and\ \citenamefont {Yazdani}}]{Nadj-Perge2014}%
  \BibitemOpen
  \bibfield  {author} {\bibinfo {author} {\bibfnamefont {S.}~\bibnamefont
  {Nadj-Perge}}, \bibinfo {author} {\bibfnamefont {I.~K.}\ \bibnamefont
  {Drozdov}}, \bibinfo {author} {\bibfnamefont {J.}~\bibnamefont {Li}},
  \bibinfo {author} {\bibfnamefont {H.}~\bibnamefont {Chen}}, \bibinfo {author}
  {\bibfnamefont {S.}~\bibnamefont {Jeon}}, \bibinfo {author} {\bibfnamefont
  {J.}~\bibnamefont {Seo}}, \bibinfo {author} {\bibfnamefont {A.~H.}\
  \bibnamefont {MacDonald}}, \bibinfo {author} {\bibfnamefont {B.~A.}\
  \bibnamefont {Bernevig}}, \ and\ \bibinfo {author} {\bibfnamefont
  {A.}~\bibnamefont {Yazdani}},\ }\href
  {http://www.sciencemag.org/content/346/6209/602.abstract} {\bibfield
  {journal} {\bibinfo  {journal} {Science}\ }\textbf {\bibinfo {volume}
  {346}},\ \bibinfo {pages} {602} (\bibinfo {year} {2014})}\BibitemShut
  {NoStop}%
\bibitem [{\citenamefont {Yu}(1965)}]{Yu1965}%
  \BibitemOpen
  \bibfield  {author} {\bibinfo {author} {\bibfnamefont {L.}~\bibnamefont
  {Yu}},\ }\href {http://wulixb.iphy.ac.cn/CN/Y1965/V21/I1/75} {\bibfield
  {journal} {\bibinfo  {journal} {Acta Phys. Sin.}\ }\textbf {\bibinfo {volume}
  {21}},\ \bibinfo {pages} {75} (\bibinfo {year} {1965})}\BibitemShut {NoStop}%
\bibitem [{\citenamefont {Shiba}(1968)}]{Shiba1968}%
  \BibitemOpen
  \bibfield  {author} {\bibinfo {author} {\bibfnamefont {H.}~\bibnamefont
  {Shiba}},\ }\href {\doibase https://doi.org/10.1143/PTP.40.435} {\bibfield
  {journal} {\bibinfo  {journal} {Progr. Theoret. Phys.}\ }\textbf {\bibinfo
  {volume} {40}},\ \bibinfo {pages} {435} (\bibinfo {year} {1968})}\BibitemShut
  {NoStop}%
\bibitem [{\citenamefont {Shiba}\ and\ \citenamefont {Soda}(1969)}]{Shiba1969}%
  \BibitemOpen
  \bibfield  {author} {\bibinfo {author} {\bibfnamefont {H.}~\bibnamefont
  {Shiba}}\ and\ \bibinfo {author} {\bibfnamefont {T.}~\bibnamefont {Soda}},\
  }\href {\doibase 10.1143/PTP.41.25} {\bibfield  {journal} {\bibinfo
  {journal} {Progr. Theoret. Phys.}\ }\textbf {\bibinfo {volume} {41}},\
  \bibinfo {pages} {25} (\bibinfo {year} {1969})}\BibitemShut {NoStop}%
\bibitem [{\citenamefont {Rusinov}(1968)}]{Rusinov1968}%
  \BibitemOpen
  \bibfield  {author} {\bibinfo {author} {\bibfnamefont {A.~I.}\ \bibnamefont
  {Rusinov}},\ }\href@noop {} {\bibfield  {journal} {\bibinfo  {journal} {Zh.
  Eksp. Teor. Fiz.}\ }\textbf {\bibinfo {volume} {9}},\ \bibinfo {pages} {146}
  (\bibinfo {year} {1968})}\BibitemShut {NoStop}%
\bibitem [{\citenamefont {Rusinov}(1969)}]{Rusinov1968alt}%
  \BibitemOpen
  \bibfield  {author} {\bibinfo {author} {\bibfnamefont {A.~I.}\ \bibnamefont
  {Rusinov}},\ }\href
  {http://www.jetpletters.ac.ru/ps/1658/article_25295.shtml} {\bibfield
  {journal} {\bibinfo  {journal} {JETP Lett.}\ }\textbf {\bibinfo {volume}
  {9}},\ \bibinfo {pages} {85} (\bibinfo {year} {1969})}\BibitemShut {NoStop}%
\bibitem [{\citenamefont {Satori}\ \emph {et~al.}(1992)\citenamefont {Satori},
  \citenamefont {Shiba}, \citenamefont {Sakai},\ and\ \citenamefont
  {Shimizu}}]{Satori1992}%
  \BibitemOpen
  \bibfield  {author} {\bibinfo {author} {\bibfnamefont {K.}~\bibnamefont
  {Satori}}, \bibinfo {author} {\bibfnamefont {H.}~\bibnamefont {Shiba}},
  \bibinfo {author} {\bibfnamefont {O.}~\bibnamefont {Sakai}}, \ and\ \bibinfo
  {author} {\bibfnamefont {Y.}~\bibnamefont {Shimizu}},\ }\href {\doibase
  10.1143/JPSJ.61.3239} {\bibfield  {journal} {\bibinfo  {journal} {J. Phys.
  Soc. Jpn.}\ }\textbf {\bibinfo {volume} {61}},\ \bibinfo {pages} {3239}
  (\bibinfo {year} {1992})}\BibitemShut {NoStop}%
\bibitem [{\citenamefont {Simonin}\ and\ \citenamefont
  {Allub}(1995)}]{Simonin1995}%
  \BibitemOpen
  \bibfield  {author} {\bibinfo {author} {\bibfnamefont {J.}~\bibnamefont
  {Simonin}}\ and\ \bibinfo {author} {\bibfnamefont {R.}~\bibnamefont
  {Allub}},\ }\href {\doibase 10.1103/PhysRevLett.74.466} {\bibfield  {journal}
  {\bibinfo  {journal} {Phys. Rev. Lett.}\ }\textbf {\bibinfo {volume} {74}},\
  \bibinfo {pages} {466} (\bibinfo {year} {1995})}\BibitemShut {NoStop}%
\bibitem [{\citenamefont {Salkola}\ \emph {et~al.}(1997)\citenamefont
  {Salkola}, \citenamefont {Balatsky},\ and\ \citenamefont
  {Schrieffer}}]{Salkola1997}%
  \BibitemOpen
  \bibfield  {author} {\bibinfo {author} {\bibfnamefont {M.~I.}\ \bibnamefont
  {Salkola}}, \bibinfo {author} {\bibfnamefont {A.~V.}\ \bibnamefont
  {Balatsky}}, \ and\ \bibinfo {author} {\bibfnamefont {J.~R.}\ \bibnamefont
  {Schrieffer}},\ }\href {\doibase 10.1103/PhysRevB.55.12648} {\bibfield
  {journal} {\bibinfo  {journal} {Phys. Rev. B}\ }\textbf {\bibinfo {volume}
  {55}},\ \bibinfo {pages} {12648} (\bibinfo {year} {1997})}\BibitemShut
  {NoStop}%
\bibitem [{\citenamefont {Balatsky}\ \emph {et~al.}(2006)\citenamefont
  {Balatsky}, \citenamefont {Vekhter},\ and\ \citenamefont
  {Zhu}}]{Balatsky2006}%
  \BibitemOpen
  \bibfield  {author} {\bibinfo {author} {\bibfnamefont {A.~V.}\ \bibnamefont
  {Balatsky}}, \bibinfo {author} {\bibfnamefont {I.}~\bibnamefont {Vekhter}}, \
  and\ \bibinfo {author} {\bibfnamefont {J.-X.}\ \bibnamefont {Zhu}},\ }\href
  {\doibase 10.1103/RevModPhys.78.373} {\bibfield  {journal} {\bibinfo
  {journal} {Rev. Mod. Phys.}\ }\textbf {\bibinfo {volume} {78}},\ \bibinfo
  {pages} {373} (\bibinfo {year} {2006})}\BibitemShut {NoStop}%
\bibitem [{\citenamefont {Tsai}\ \emph {et~al.}(2009)\citenamefont {Tsai},
  \citenamefont {Zhang}, \citenamefont {Fang},\ and\ \citenamefont
  {Hu}}]{Tsai2009}%
  \BibitemOpen
  \bibfield  {author} {\bibinfo {author} {\bibfnamefont {W.-F.}\ \bibnamefont
  {Tsai}}, \bibinfo {author} {\bibfnamefont {Y.-Y.}\ \bibnamefont {Zhang}},
  \bibinfo {author} {\bibfnamefont {C.}~\bibnamefont {Fang}}, \ and\ \bibinfo
  {author} {\bibfnamefont {J.}~\bibnamefont {Hu}},\ }\href {\doibase
  10.1103/PhysRevB.80.064513} {\bibfield  {journal} {\bibinfo  {journal} {Phys.
  Rev. B}\ }\textbf {\bibinfo {volume} {80}},\ \bibinfo {pages} {064513}
  (\bibinfo {year} {2009})}\BibitemShut {NoStop}%
\bibitem [{\citenamefont {Ruby}\ \emph {et~al.}(2016)\citenamefont {Ruby},
  \citenamefont {Peng}, \citenamefont {von Oppen}, \citenamefont {Heinrich},\
  and\ \citenamefont {Franke}}]{Ruby2016}%
  \BibitemOpen
  \bibfield  {author} {\bibinfo {author} {\bibfnamefont {M.}~\bibnamefont
  {Ruby}}, \bibinfo {author} {\bibfnamefont {Y.}~\bibnamefont {Peng}}, \bibinfo
  {author} {\bibfnamefont {F.}~\bibnamefont {von Oppen}}, \bibinfo {author}
  {\bibfnamefont {B.~W.}\ \bibnamefont {Heinrich}}, \ and\ \bibinfo {author}
  {\bibfnamefont {K.~J.}\ \bibnamefont {Franke}},\ }\href {\doibase
  10.1103/PhysRevLett.117.186801} {\bibfield  {journal} {\bibinfo  {journal}
  {Phys. Rev. Lett.}\ }\textbf {\bibinfo {volume} {117}},\ \bibinfo {pages}
  {186801} (\bibinfo {year} {2016})}\BibitemShut {NoStop}%
\bibitem [{\citenamefont {Kaladzhyan}\ \emph {et~al.}(2017)\citenamefont
  {Kaladzhyan}, \citenamefont {Hoffman},\ and\ \citenamefont
  {Trif}}]{Kaladzhyan2017}%
  \BibitemOpen
  \bibfield  {author} {\bibinfo {author} {\bibfnamefont {V.}~\bibnamefont
  {Kaladzhyan}}, \bibinfo {author} {\bibfnamefont {S.}~\bibnamefont {Hoffman}},
  \ and\ \bibinfo {author} {\bibfnamefont {M.}~\bibnamefont {Trif}},\ }\href
  {\doibase 10.1103/PhysRevB.95.195403} {\bibfield  {journal} {\bibinfo
  {journal} {Phys. Rev. B}\ }\textbf {\bibinfo {volume} {95}},\ \bibinfo
  {pages} {195403} (\bibinfo {year} {2017})}\BibitemShut {NoStop}%
\bibitem [{\citenamefont {K\"orber}\ \emph {et~al.}(2018)\citenamefont
  {K\"orber}, \citenamefont {Trauzettel},\ and\ \citenamefont
  {Kashuba}}]{Koerber2018}%
  \BibitemOpen
  \bibfield  {author} {\bibinfo {author} {\bibfnamefont {S.}~\bibnamefont
  {K\"orber}}, \bibinfo {author} {\bibfnamefont {B.}~\bibnamefont
  {Trauzettel}}, \ and\ \bibinfo {author} {\bibfnamefont {O.}~\bibnamefont
  {Kashuba}},\ }\href {\doibase 10.1103/PhysRevB.97.184503} {\bibfield
  {journal} {\bibinfo  {journal} {Phys. Rev. B}\ }\textbf {\bibinfo {volume}
  {97}},\ \bibinfo {pages} {184503} (\bibinfo {year} {2018})}\BibitemShut
  {NoStop}%
\bibitem [{\citenamefont {\ifmmode~\check{Z}\else \v{Z}\fi{}itko}\ \emph
  {et~al.}(2015)\citenamefont {\ifmmode~\check{Z}\else \v{Z}\fi{}itko},
  \citenamefont {Lim}, \citenamefont {L\'opez},\ and\ \citenamefont
  {Aguado}}]{Zitko2015}%
  \BibitemOpen
  \bibfield  {author} {\bibinfo {author} {\bibfnamefont {R.}~\bibnamefont
  {\ifmmode~\check{Z}\else \v{Z}\fi{}itko}}, \bibinfo {author} {\bibfnamefont
  {J.~S.}\ \bibnamefont {Lim}}, \bibinfo {author} {\bibfnamefont
  {R.}~\bibnamefont {L\'opez}}, \ and\ \bibinfo {author} {\bibfnamefont
  {R.}~\bibnamefont {Aguado}},\ }\href {\doibase 10.1103/PhysRevB.91.045441}
  {\bibfield  {journal} {\bibinfo  {journal} {Phys. Rev. B}\ }\textbf {\bibinfo
  {volume} {91}},\ \bibinfo {pages} {045441} (\bibinfo {year}
  {2015})}\BibitemShut {NoStop}%
\bibitem [{\citenamefont {Jellinggaard}\ \emph {et~al.}(2016)\citenamefont
  {Jellinggaard}, \citenamefont {Grove-Rasmussen}, \citenamefont {Madsen},\
  and\ \citenamefont {Nyg\aa{}rd}}]{Jellinggaard2016}%
  \BibitemOpen
  \bibfield  {author} {\bibinfo {author} {\bibfnamefont {A.}~\bibnamefont
  {Jellinggaard}}, \bibinfo {author} {\bibfnamefont {K.}~\bibnamefont
  {Grove-Rasmussen}}, \bibinfo {author} {\bibfnamefont {M.~H.}\ \bibnamefont
  {Madsen}}, \ and\ \bibinfo {author} {\bibfnamefont {J.}~\bibnamefont
  {Nyg\aa{}rd}},\ }\href {\doibase 10.1103/PhysRevB.94.064520} {\bibfield
  {journal} {\bibinfo  {journal} {Phys. Rev. B}\ }\textbf {\bibinfo {volume}
  {94}},\ \bibinfo {pages} {064520} (\bibinfo {year} {2016})}\BibitemShut
  {NoStop}%
\bibitem [{\citenamefont {Kawabata}\ \emph {et~al.}(2012)\citenamefont
  {Kawabata}, \citenamefont {Tanaka}, \citenamefont {Golubov}, \citenamefont
  {Vasenko},\ and\ \citenamefont {Asano}}]{Kawabata2012}%
  \BibitemOpen
  \bibfield  {author} {\bibinfo {author} {\bibfnamefont {S.}~\bibnamefont
  {Kawabata}}, \bibinfo {author} {\bibfnamefont {Y.}~\bibnamefont {Tanaka}},
  \bibinfo {author} {\bibfnamefont {A.~A.}\ \bibnamefont {Golubov}}, \bibinfo
  {author} {\bibfnamefont {A.~S.}\ \bibnamefont {Vasenko}}, \ and\ \bibinfo
  {author} {\bibfnamefont {Y.}~\bibnamefont {Asano}},\ }\href {\doibase
  https://doi.org/10.1016/j.jmmm.2012.02.067} {\bibfield  {journal} {\bibinfo
  {journal} {J. Magn. Magn. Mater.}\ }\textbf {\bibinfo {volume} {324}},\
  \bibinfo {pages} {3467 } (\bibinfo {year} {2012})}\BibitemShut {NoStop}%
\bibitem [{\citenamefont {Sau}\ and\ \citenamefont {Demler}(2013)}]{Sau2013}%
  \BibitemOpen
  \bibfield  {author} {\bibinfo {author} {\bibfnamefont {J.~D.}\ \bibnamefont
  {Sau}}\ and\ \bibinfo {author} {\bibfnamefont {E.}~\bibnamefont {Demler}},\
  }\href {\doibase 10.1103/PhysRevB.88.205402} {\bibfield  {journal} {\bibinfo
  {journal} {Phys. Rev. B}\ }\textbf {\bibinfo {volume} {88}},\ \bibinfo
  {pages} {205402} (\bibinfo {year} {2013})}\BibitemShut {NoStop}%
\bibitem [{\citenamefont {Pientka}\ \emph {et~al.}(2015)\citenamefont
  {Pientka}, \citenamefont {Peng}, \citenamefont {Glazman},\ and\ \citenamefont
  {von Oppen}}]{Pientka2015}%
  \BibitemOpen
  \bibfield  {author} {\bibinfo {author} {\bibfnamefont {F.}~\bibnamefont
  {Pientka}}, \bibinfo {author} {\bibfnamefont {Y.}~\bibnamefont {Peng}},
  \bibinfo {author} {\bibfnamefont {L.}~\bibnamefont {Glazman}}, \ and\
  \bibinfo {author} {\bibfnamefont {F.}~\bibnamefont {von Oppen}},\ }\href
  {http://stacks.iop.org/1402-4896/2015/i=T164/a=014008} {\bibfield  {journal}
  {\bibinfo  {journal} {Phys. Scripta}\ }\textbf {\bibinfo {volume} {2015}},\
  \bibinfo {pages} {014008} (\bibinfo {year} {2015})}\BibitemShut {NoStop}%
\bibitem [{\citenamefont {Hatter}\ \emph {et~al.}(2015)\citenamefont {Hatter},
  \citenamefont {Heinrich}, \citenamefont {Ruby}, \citenamefont {Pascual},\
  and\ \citenamefont {Franke}}]{Hatter2015}%
  \BibitemOpen
  \bibfield  {author} {\bibinfo {author} {\bibfnamefont {N.}~\bibnamefont
  {Hatter}}, \bibinfo {author} {\bibfnamefont {B.~W.}\ \bibnamefont
  {Heinrich}}, \bibinfo {author} {\bibfnamefont {M.}~\bibnamefont {Ruby}},
  \bibinfo {author} {\bibfnamefont {J.~I.}\ \bibnamefont {Pascual}}, \ and\
  \bibinfo {author} {\bibfnamefont {K.~J.}\ \bibnamefont {Franke}},\ }\href
  {\doibase 10.1038/ncomms9988} {\bibfield  {journal} {\bibinfo  {journal}
  {Nat. Commun.}\ }\textbf {\bibinfo {volume} {6}},\ \bibinfo {pages} {8988}
  (\bibinfo {year} {2015})}\BibitemShut {NoStop}%
\bibitem [{\citenamefont {Sakurai}(1970)}]{Sakurai1970}%
  \BibitemOpen
  \bibfield  {author} {\bibinfo {author} {\bibfnamefont {A.}~\bibnamefont
  {Sakurai}},\ }\href {\doibase 10.1143/PTP.44.1472} {\bibfield  {journal}
  {\bibinfo  {journal} {Progr. Theor. Exp. Phys.}\ }\textbf {\bibinfo {volume}
  {44}},\ \bibinfo {pages} {1472} (\bibinfo {year} {1970})}\BibitemShut
  {NoStop}%
\bibitem [{\citenamefont {Andersen}\ \emph {et~al.}(2006)\citenamefont
  {Andersen}, \citenamefont {Bobkova}, \citenamefont {Hirschfeld},\ and\
  \citenamefont {Barash}}]{Andersen2006}%
  \BibitemOpen
  \bibfield  {author} {\bibinfo {author} {\bibfnamefont {B.~M.}\ \bibnamefont
  {Andersen}}, \bibinfo {author} {\bibfnamefont {I.~V.}\ \bibnamefont
  {Bobkova}}, \bibinfo {author} {\bibfnamefont {P.~J.}\ \bibnamefont
  {Hirschfeld}}, \ and\ \bibinfo {author} {\bibfnamefont {Y.~S.}\ \bibnamefont
  {Barash}},\ }\href {\doibase 10.1103/PhysRevLett.96.117005} {\bibfield
  {journal} {\bibinfo  {journal} {Phys. Rev. Lett.}\ }\textbf {\bibinfo
  {volume} {96}},\ \bibinfo {pages} {117005} (\bibinfo {year}
  {2006})}\BibitemShut {NoStop}%
\bibitem [{\citenamefont {Kawabata}\ \emph {et~al.}(2010)\citenamefont
  {Kawabata}, \citenamefont {Asano}, \citenamefont {Tanaka}, \citenamefont
  {Golubov},\ and\ \citenamefont {Kashiwaya}}]{Kawabata2010}%
  \BibitemOpen
  \bibfield  {author} {\bibinfo {author} {\bibfnamefont {S.}~\bibnamefont
  {Kawabata}}, \bibinfo {author} {\bibfnamefont {Y.}~\bibnamefont {Asano}},
  \bibinfo {author} {\bibfnamefont {Y.}~\bibnamefont {Tanaka}}, \bibinfo
  {author} {\bibfnamefont {A.~A.}\ \bibnamefont {Golubov}}, \ and\ \bibinfo
  {author} {\bibfnamefont {S.}~\bibnamefont {Kashiwaya}},\ }\href {\doibase
  10.1103/PhysRevLett.104.117002} {\bibfield  {journal} {\bibinfo  {journal}
  {Phys. Rev. Lett.}\ }\textbf {\bibinfo {volume} {104}},\ \bibinfo {pages}
  {117002} (\bibinfo {year} {2010})}\BibitemShut {NoStop}%
\bibitem [{\citenamefont {Bychkov}\ and\ \citenamefont
  {Rashba}(1984)}]{Bychkov1984}%
  \BibitemOpen
  \bibfield  {author} {\bibinfo {author} {\bibfnamefont {Y.~A.}\ \bibnamefont
  {Bychkov}}\ and\ \bibinfo {author} {\bibfnamefont {E.~I.}\ \bibnamefont
  {Rashba}},\ }\href
  {http://stacks.iop.org/0022-3719/17/i=33/a=015?key=crossref.2f8159f54a3070499f32bac53e23f947}
  {\bibfield  {journal} {\bibinfo  {journal} {J. Phys. C}\ }\textbf {\bibinfo
  {volume} {17}},\ \bibinfo {pages} {6039} (\bibinfo {year}
  {1984})}\BibitemShut {NoStop}%
\bibitem [{\citenamefont {De~Gennes}(1989)}]{DeGennes1989}%
  \BibitemOpen
  \bibfield  {author} {\bibinfo {author} {\bibfnamefont {P.~G.}\ \bibnamefont
  {De~Gennes}},\ }\href@noop {} {\emph {\bibinfo {title} {{Superconductivity of
  Metals and Alloys}}}}\ (\bibinfo  {publisher} {Addison Wesley, Redwood
  City},\ \bibinfo {year} {1989})\BibitemShut {NoStop}%
\bibitem [{\citenamefont {{de Jong}}\ and\ \citenamefont
  {Beenakker}(1995)}]{DeJong1995}%
  \BibitemOpen
  \bibfield  {author} {\bibinfo {author} {\bibfnamefont {M.~J.~M.}\
  \bibnamefont {{de Jong}}}\ and\ \bibinfo {author} {\bibfnamefont {C.~W.~J.}\
  \bibnamefont {Beenakker}},\ }\href
  {http://link.aps.org/doi/10.1103/PhysRevLett.74.1657} {\bibfield  {journal}
  {\bibinfo  {journal} {Phys. Rev. Lett.}\ }\textbf {\bibinfo {volume} {74}},\
  \bibinfo {pages} {1657} (\bibinfo {year} {1995})}\BibitemShut {NoStop}%
\bibitem [{\citenamefont {{\v{Z}}uti{\'{c}}}\ and\ \citenamefont
  {Valls}(1999)}]{Zutic1999}%
  \BibitemOpen
  \bibfield  {author} {\bibinfo {author} {\bibfnamefont {I.}~\bibnamefont
  {{\v{Z}}uti{\'{c}}}}\ and\ \bibinfo {author} {\bibfnamefont {O.~T.}\
  \bibnamefont {Valls}},\ }\href
  {http://link.aps.org/doi/10.1103/PhysRevB.60.6320} {\bibfield  {journal}
  {\bibinfo  {journal} {Phys. Rev. B}\ }\textbf {\bibinfo {volume} {60}},\
  \bibinfo {pages} {6320} (\bibinfo {year} {1999})}\BibitemShut {NoStop}%
\bibitem [{\citenamefont {{\v{Z}}uti{\'{c}}}\ and\ \citenamefont
  {Valls}(2000)}]{Zutic2000}%
  \BibitemOpen
  \bibfield  {author} {\bibinfo {author} {\bibfnamefont {I.}~\bibnamefont
  {{\v{Z}}uti{\'{c}}}}\ and\ \bibinfo {author} {\bibfnamefont {O.~T.}\
  \bibnamefont {Valls}},\ }\href
  {http://link.aps.org/doi/10.1103/PhysRevB.61.1555} {\bibfield  {journal}
  {\bibinfo  {journal} {Phys. Rev. B}\ }\textbf {\bibinfo {volume} {61}},\
  \bibinfo {pages} {1555} (\bibinfo {year} {2000})}\BibitemShut {NoStop}%
\bibitem [{\citenamefont {Zhi-Hong}\ \emph {et~al.}(2012)\citenamefont
  {Zhi-Hong}, \citenamefont {Yong-Hong},\ and\ \citenamefont {Jun}}]{Yang2012}%
  \BibitemOpen
  \bibfield  {author} {\bibinfo {author} {\bibfnamefont {Y.}~\bibnamefont
  {Zhi-Hong}}, \bibinfo {author} {\bibfnamefont {Y.}~\bibnamefont {Yong-Hong}},
  \ and\ \bibinfo {author} {\bibfnamefont {W.}~\bibnamefont {Jun}},\ }\href
  {http://stacks.iop.org/1674-1056/21/i=5/a=057402} {\bibfield  {journal}
  {\bibinfo  {journal} {Chin. Phys. B}\ }\textbf {\bibinfo {volume} {21}},\
  \bibinfo {pages} {057402} (\bibinfo {year} {2012})}\BibitemShut {NoStop}%
\bibitem [{Note1()}]{Note1}%
  \BibitemOpen
  \bibinfo {note} {See attached Supplemental Material, including Refs.~\cite
  {DeGennes1989,McMillan1968,Beenakker1991,Golubov2004,Oreg2010,Duckheim2011,Mourik2012,Rokhinson2012,Nadj-Perge2014,Fu2008,Yu1965,Shiba1968,Rusinov1968,*Rusinov1968alt,Costa2017,Carbotte1990,Andreev1991,Demler1997,Fulde1964,Larkin1964,*Larkin1964alt,Ryazanov2001,Brinkman2000,Kawabata2010},
  for more details.}\BibitemShut {Stop}%
\bibitem [{\citenamefont {Beenakker}(1991)}]{Beenakker1991}%
  \BibitemOpen
  \bibfield  {author} {\bibinfo {author} {\bibfnamefont {C.~W.~J.}\
  \bibnamefont {Beenakker}},\ }\href {\doibase 10.1103/PhysRevLett.67.3836}
  {\bibfield  {journal} {\bibinfo  {journal} {Phys. Rev. Lett.}\ }\textbf
  {\bibinfo {volume} {67}},\ \bibinfo {pages} {3836} (\bibinfo {year}
  {1991})}\BibitemShut {NoStop}%
\bibitem [{Note2()}]{Note2}%
  \BibitemOpen
  \bibinfo {note} {Since momentum is not a good quantum number, we rather use a
  general index $\protect \mathbf {n}$ that labels perturbed eigenmodes above
  and below the S gap.}\BibitemShut {Stop}%
\bibitem [{\citenamefont {Kulik}(1969)}]{Kulik1969}%
  \BibitemOpen
  \bibfield  {author} {\bibinfo {author} {\bibfnamefont {I.~O.}\ \bibnamefont
  {Kulik}},\ }\href@noop {} {\bibfield  {journal} {\bibinfo  {journal} {Zh.
  Eksp. Teor. Fiz.}\ }\textbf {\bibinfo {volume} {57}},\ \bibinfo {pages}
  {1745} (\bibinfo {year} {1969})}\BibitemShut {NoStop}%
\bibitem [{\citenamefont {Kulik}(1970)}]{Kulik1969alt}%
  \BibitemOpen
  \bibfield  {author} {\bibinfo {author} {\bibfnamefont {I.~O.}\ \bibnamefont
  {Kulik}},\ }\href {http://www.jetp.ac.ru/cgi-bin/e/index/e/30/5/p944?a=list}
  {\bibfield  {journal} {\bibinfo  {journal} {J. Exp. Theor. Phys.}\ }\textbf
  {\bibinfo {volume} {30}},\ \bibinfo {pages} {944} (\bibinfo {year}
  {1970})}\BibitemShut {NoStop}%
\bibitem [{\citenamefont {Nitta}\ \emph {et~al.}(1997)\citenamefont {Nitta},
  \citenamefont {Akazaki}, \citenamefont {Takayanagi},\ and\ \citenamefont
  {Enoki}}]{Nitta1997}%
  \BibitemOpen
  \bibfield  {author} {\bibinfo {author} {\bibfnamefont {J.}~\bibnamefont
  {Nitta}}, \bibinfo {author} {\bibfnamefont {T.}~\bibnamefont {Akazaki}},
  \bibinfo {author} {\bibfnamefont {H.}~\bibnamefont {Takayanagi}}, \ and\
  \bibinfo {author} {\bibfnamefont {T.}~\bibnamefont {Enoki}},\ }\href
  {\doibase 10.1103/PhysRevLett.78.1335} {\bibfield  {journal} {\bibinfo
  {journal} {Phys. Rev. Lett.}\ }\textbf {\bibinfo {volume} {78}},\ \bibinfo
  {pages} {1335} (\bibinfo {year} {1997})}\BibitemShut {NoStop}%
\bibitem [{\citenamefont {Koga}\ \emph {et~al.}(2002)\citenamefont {Koga},
  \citenamefont {Nitta}, \citenamefont {Akazaki},\ and\ \citenamefont
  {Takayanagi}}]{Koga2002}%
  \BibitemOpen
  \bibfield  {author} {\bibinfo {author} {\bibfnamefont {T.}~\bibnamefont
  {Koga}}, \bibinfo {author} {\bibfnamefont {J.}~\bibnamefont {Nitta}},
  \bibinfo {author} {\bibfnamefont {T.}~\bibnamefont {Akazaki}}, \ and\
  \bibinfo {author} {\bibfnamefont {H.}~\bibnamefont {Takayanagi}},\ }\href
  {\doibase 10.1103/PhysRevLett.89.046801} {\bibfield  {journal} {\bibinfo
  {journal} {Phys. Rev. Lett.}\ }\textbf {\bibinfo {volume} {89}},\ \bibinfo
  {pages} {046801} (\bibinfo {year} {2002})}\BibitemShut {NoStop}%
\bibitem [{\citenamefont {McMillan}(1968)}]{McMillan1968}%
  \BibitemOpen
  \bibfield  {author} {\bibinfo {author} {\bibfnamefont {W.~L.}\ \bibnamefont
  {McMillan}},\ }\href {\doibase 10.1103/PhysRev.175.559} {\bibfield  {journal}
  {\bibinfo  {journal} {Phys. Rev.}\ }\textbf {\bibinfo {volume} {175}},\
  \bibinfo {pages} {559} (\bibinfo {year} {1968})}\BibitemShut {NoStop}%
\bibitem [{\citenamefont {Carbotte}(1990)}]{Carbotte1990}%
  \BibitemOpen
  \bibfield  {author} {\bibinfo {author} {\bibfnamefont {J.~P.}\ \bibnamefont
  {Carbotte}},\ }\href
  {http://journals.aps.org/rmp/abstract/10.1103/RevModPhys.62.1027} {\bibfield
  {journal} {\bibinfo  {journal} {Rev. Mod. Phys.}\ }\textbf {\bibinfo {volume}
  {62}},\ \bibinfo {pages} {1027} (\bibinfo {year} {1990})}\BibitemShut
  {NoStop}%
\bibitem [{\citenamefont {Fulde}\ and\ \citenamefont
  {Ferrell}(1964)}]{Fulde1964}%
  \BibitemOpen
  \bibfield  {author} {\bibinfo {author} {\bibfnamefont {P.}~\bibnamefont
  {Fulde}}\ and\ \bibinfo {author} {\bibfnamefont {R.~A.}\ \bibnamefont
  {Ferrell}},\ }\href {\doibase 10.1103/PhysRev.135.A550} {\bibfield  {journal}
  {\bibinfo  {journal} {Phys. Rev.}\ }\textbf {\bibinfo {volume} {135}},\
  \bibinfo {pages} {A550} (\bibinfo {year} {1964})}\BibitemShut {NoStop}%
\bibitem [{\citenamefont {Larkin}\ and\ \citenamefont
  {Ovchinnikov}(1964)}]{Larkin1964}%
  \BibitemOpen
  \bibfield  {author} {\bibinfo {author} {\bibfnamefont {I.}~\bibnamefont
  {Larkin}}\ and\ \bibinfo {author} {\bibfnamefont {Y.~N.}\ \bibnamefont
  {Ovchinnikov}},\ }\href@noop {} {\bibfield  {journal} {\bibinfo  {journal}
  {Zh. Eksp. Teor. Fiz.}\ }\textbf {\bibinfo {volume} {47}},\ \bibinfo {pages}
  {1136} (\bibinfo {year} {1964})}\BibitemShut {NoStop}%
\bibitem [{\citenamefont {Larkin}\ and\ \citenamefont
  {Ovchinnikov}(1965)}]{Larkin1964alt}%
  \BibitemOpen
  \bibfield  {author} {\bibinfo {author} {\bibfnamefont {I.}~\bibnamefont
  {Larkin}}\ and\ \bibinfo {author} {\bibfnamefont {Y.~N.}\ \bibnamefont
  {Ovchinnikov}},\ }\href@noop {} {\bibfield  {journal} {\bibinfo  {journal}
  {Sov. Phys. JETP}\ }\textbf {\bibinfo {volume} {20}},\ \bibinfo {pages} {762}
  (\bibinfo {year} {1965})}\BibitemShut {NoStop}%
\bibitem [{\citenamefont {Brinkman}\ and\ \citenamefont
  {Golubov}(2000)}]{Brinkman2000}%
  \BibitemOpen
  \bibfield  {author} {\bibinfo {author} {\bibfnamefont {A.}~\bibnamefont
  {Brinkman}}\ and\ \bibinfo {author} {\bibfnamefont {A.~A.}\ \bibnamefont
  {Golubov}},\ }\href
  {http://journals.aps.org/prb/abstract/10.1103/PhysRevB.61.11297} {\bibfield
  {journal} {\bibinfo  {journal} {Phys. Rev. B}\ }\textbf {\bibinfo {volume}
  {61}},\ \bibinfo {pages} {11297} (\bibinfo {year} {2000})}\BibitemShut
  {NoStop}%
\end{thebibliography}%

\onecolumngrid
\newpage


\section*{Supplementary material}

# Connection between zero-energy Yu-Shiba-Rusinov states and 0-π transitions in magnetic Josephson junctions

Andreas Costa, Jaroslav Fabian, and Denis Kochan

*Institute for Theoretical Physics, University of Regensburg, 93040 Regensburg, Germany*

In this Supplemental Material, we present the computational details not included into the manuscript and additional model calculations for a richer set of parameter combinations, which are useful to get a deeper understanding of the physical background.

## I. THEORETICAL MODEL AND BOUND STATE SPECTRUM – COMPUTATIONAL DETAILS

We consider a vertical ballistic S/FI/S Josephson junction consisting of two semi-infinite S regions, spanning $z < 0$ and $z > 0$ half-spaces, respectively. They are separated by a thin deltalike ferromagnetic insulator layer. The junction is schematically depicted in Fig. S1. In order to analyze its spectral properties, we model the system in terms of stationary BdG equations [S1], which in particle-hole Nambu space $\Psi = \left[ \psi_\uparrow, \psi_\downarrow, \psi_\downarrow^\dagger, -\psi_\uparrow^\dagger \right]^\top$ read

$$\begin{bmatrix} \hat{H}_{\mathrm{e}} & \hat{\Delta}_{\mathrm{S}}(z) \\ \hat{\Delta}_{\mathrm{S}}^\dagger(z) & \hat{H}_{\mathrm{h}} \end{bmatrix} \Psi(\mathbf{r}) = E \Psi(\mathbf{r}) \,. \tag{S1}$$

The electron and hole Hamiltonians including perturbation terms and the superconducting order parameter are provided in the manuscript.

Since the wave vector $\mathbf{k}_\parallel = [k_x, k_y, 0]^\top$ parallel to the interface is conserved, the bound state solution ansatz for Eq. (S1) reads $\Psi(\mathbf{r}) = \Psi(z) \, \mathrm{e}^{\mathrm{i} \mathbf{k}_\parallel \cdot \mathbf{r}_\parallel}$, where $\mathbf{r}_\parallel = [x, y, 0]^\top$. The unknown transverse wave function $\Psi(z)$ satisfies a reduced one-dimensional BdG scattering problem, whose general solution can be written for the left ($z < 0$) and right ($z > 0$) S regions as

$$\Psi(z < 0) = \mathcal{A} \, \mathrm{e}^{-\mathrm{i} q_{\mathrm{ez}} z} \begin{bmatrix} u \\ 0 \\ v \\ 0 \end{bmatrix} + \mathcal{B} \, \mathrm{e}^{-\mathrm{i} q_{\mathrm{ez}} z} \begin{bmatrix} 0 \\ u \\ 0 \\ v \end{bmatrix} + \mathcal{C} \, \mathrm{e}^{+\mathrm{i} q_{\mathrm{hz}} z} \begin{bmatrix} v \\ 0 \\ u \\ 0 \end{bmatrix} + \mathcal{D} \, \mathrm{e}^{+\mathrm{i} q_{\mathrm{hz}} z} \begin{bmatrix} 0 \\ v \\ 0 \\ u \end{bmatrix} \tag{S2}$$

and

$$\Psi(z > 0) = \mathcal{E} \, \mathrm{e}^{+\mathrm{i} q_{\mathrm{ez}} z} \begin{bmatrix} u \mathrm{e}^{\mathrm{i} \phi_{\mathrm{S}}} \\ 0 \\ v \\ 0 \end{bmatrix} + \mathcal{F} \, \mathrm{e}^{+\mathrm{i} q_{\mathrm{ez}} z} \begin{bmatrix} 0 \\ u \mathrm{e}^{\mathrm{i} \phi_{\mathrm{S}}} \\ 0 \\ v \end{bmatrix} + \mathcal{G} \, \mathrm{e}^{-\mathrm{i} q_{\mathrm{hz}} z} \begin{bmatrix} v \mathrm{e}^{\mathrm{i} \phi_{\mathrm{S}}} \\ 0 \\ u \\ 0 \end{bmatrix} + \mathcal{J} \, \mathrm{e}^{-\mathrm{i} q_{\mathrm{hz}} z} \begin{bmatrix} 0 \\ v \mathrm{e}^{\mathrm{i} \phi_{\mathrm{S}}} \\ 0 \\ u \end{bmatrix} \,. \tag{S3}$$

Along with that ansatz yielding a solution $\Psi(z)$ with positive energy $E > 0$, one has also a second, functionally independent solution $\tilde{\Psi}(z)$ belonging to the negative energy manifold that can be obtained from $\Psi(z)$ when interchanging $u$ to $-v^*$ and $v$ to $u^*$ in the Nambu spinors. Above, $u = u(E)$ and $v = v(E)$ are the conventional BCS coherence factors that satisfy

$$u(E) = \sqrt{\frac{1}{2} \left( 1 + \sqrt{1 - \frac{|\Delta_{\mathrm{S}}|^2}{E^2}} \right)} = \sqrt{1 - v(E)^2} \,. \tag{S4}$$

Since we are looking for the bound state (evanescent) solutions only, the wave function ansatz in Eqs. (S2) and (S3) does not account for an incoming wave. Moreover, the $\hat{z}$ projections of the wave vectors for electronlike and holelike quasiparticles,

$$q_{\mathrm{ez}}(E) = \sqrt{q_{\mathrm{F}}^2 + \frac{2m}{\hbar^2} \sqrt{E^2 - |\Delta_{\mathrm{S}}|^2} - |\mathbf{k}_\parallel|^2} \quad \text{and} \quad q_{\mathrm{hz}}(E) = \sqrt{q_{\mathrm{F}}^2 - \frac{2m}{\hbar^2} \sqrt{E^2 - |\Delta_{\mathrm{S}}|^2} - |\mathbf{k}_\parallel|^2} \,, \tag{S5}$$

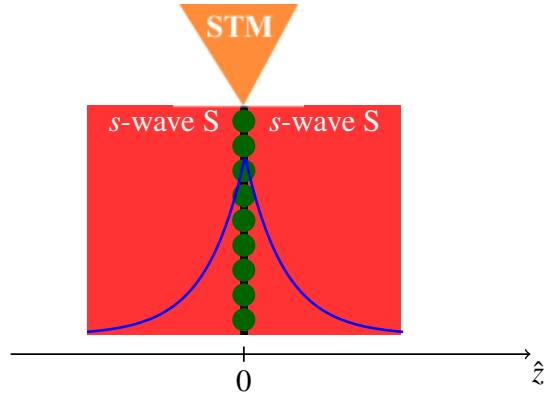

FIG. S1. Schematical sketch of the considered S/FI/S Josephson junction; STM spectroscopy can be used to study the Andreev and YSR bound states.

already contain an evanescent imaginary part while $-|\Delta_S| < E < |\Delta_S|$; $q_F = \sqrt{2m\mu}/\hbar$ stands for the Fermi momentum. Because of the similarities with the true scattering problem, the wave amplitudes $\mathcal{A}$ and $\mathcal{B}$ in Eq. (S2) resemble electronlike specular reflection with and without spin flip, and $\mathcal{C}$ and $\mathcal{D}$ their Andreev reflected holelike counterparts. In the same way, the coefficients $\mathcal{E}$, $\mathcal{F}$, $\mathcal{G}$, and $\mathcal{J}$ in Eq. (S3) stand for spin-resolved transmission amplitudes in the right superconductor. To obtain them, we need to fulfill the interfacial ($z = 0$) matching conditions

$$\Psi(z)\big|_{z=0_-} = \Psi(z)\big|_{z=0_+},$$
(S6)

$$\left\{\left[\frac{\hbar^2}{2m}\frac{d}{dz} + \lambda_{SC}\right]\boldsymbol{\eta} + \lambda_{MA}\boldsymbol{\omega}\right\}\Psi(z)\big|_{z=0_-} = \frac{\hbar^2}{2m}\frac{d}{dz}\boldsymbol{\eta}\Psi(z)\big|_{z=0_+},$$
(S7)

where $\boldsymbol{\eta}$ and $\boldsymbol{\omega}$ are diagonal four-by-four matrices; particularly $\boldsymbol{\eta} = \mathrm{diag}[\hat{\sigma}_0, -\hat{\sigma}_0]$ and $\boldsymbol{\omega} = \mathrm{diag}[\hat{\sigma}_z, \hat{\sigma}_z]$. It turns out that Eqs. (S6)–(S7) represent a homogeneous system of eight linear equations for the unknown wave function amplitudes $\mathcal{A}, \ldots, \mathcal{J}$. To obtain a nontrivial solution, the determinant of the system's coefficient matrix must be zero, what gives a secular equation for the bound state energies of Eq. (S1).

Numerically, we are in a position to compute the bound state spectrum for a generic S/FI/S Josephson junction. Nevertheless, to give full analytical solutions, we first restrict ourselves to quasi-one-dimensional junctions, for which the thickness in the $\hat{x}$ and $\hat{y}$ direction is much shorter than in the $\hat{z}$ direction, or equivalently that $\mathbf{k}_\parallel \to \mathbf{0}$, and Rashba SOC in the leads is absent. Under these assumptions, we obtain four different solutions $E_n^\pm$ ($n = 1, 2$) for the bound state energies inside the S gap,

$$E_n^\pm = \pm|\Delta_S|\left\{\frac{2\left(\bar{\lambda}_{SC}^2 - \bar{\lambda}_{MA}^2 + 4\right)\cos\phi_S + \bar{\lambda}_{SC}^4 + 2\bar{\lambda}_{MA}^2 + \bar{\lambda}_{MA}^4 - 2\bar{\lambda}_{SC}^2\left(\bar{\lambda}_{MA}^2 - 3\right) + 8}{\left(\bar{\lambda}_{SC}^2 - \bar{\lambda}_{MA}^2 + 4\right)^2 + 16\bar{\lambda}_{MA}^2}\right.$$
$$\left. + 4(-1)^n\frac{\sqrt{2\bar{\lambda}_{MA}^2\left[1 + \bar{\lambda}_{SC}^2 + \bar{\lambda}_{MA}^2 + \left(\bar{\lambda}_{MA}^2 - \bar{\lambda}_{SC}^2\right)\cos\phi_S - \cos(2\phi_S)\right]}}{\left(\bar{\lambda}_{SC}^2 - \bar{\lambda}_{MA}^2 + 4\right)^2 + 16\bar{\lambda}_{MA}^2}\right\}^{1/2}.$$
(S8)

By using basic trigonometric identities, one can immediately bring this result into the more compact form given by Eq. (2) in the manuscript. Obviously, the bound state spectrum $E_n^\pm$ is symmetric (superscript $\pm$) with respect to the center of the S gap and depends only on the macroscopic phase difference $\phi_S$ and the two dimensionless parameters $\bar{\lambda}_{SC} = 2m\lambda_{SC}/(\hbar^2 q_F)$ and $\bar{\lambda}_{MA} = 2m\lambda_{MA}/(\hbar^2 q_F)$, representing effective strengths of scalar and magnetic tunneling, respectively.

Let us note that the secular bound state equation originating from the matching conditions, Eqs. (S6)–(S7), is not a simple polynomial of eighth degree since the coherence factors $u(E)$ and $v(E)$ contain square roots and lead to a transcendental equation. Therefore, we only obtain four independent bound state energies instead of eight solutions one would typically expect for a polynomial of eighth order. Moreover, the interfacial impurity perturbations entering the BdG equation are diagonal in the spin space and therefore, one can introduce a reduced spin-resolved Nambu

basis (in the absence of Rashba SOC in the leads), labeled by $\sigma = \uparrow = +1$ and $\sigma = \downarrow = -1$. In this case, the original bound state ansatz can be recast as

$$\Psi_\sigma(z < 0) = \Psi_\sigma^e(z < 0) + \Psi_\sigma^h(z < 0) = a_\sigma \, \mathrm{e}^{-\mathrm{i}q_{ez}z} \begin{bmatrix} u \\ v \end{bmatrix} + b_\sigma \, \mathrm{e}^{+\mathrm{i}q_{hz}z} \begin{bmatrix} v \\ u \end{bmatrix} \tag{S9}$$

and

$$\Psi_\sigma(z > 0) = \Psi_\sigma^e(z > 0) + \Psi_\sigma^h(z > 0) = c_\sigma \, \mathrm{e}^{+\mathrm{i}q_{ez}z} \begin{bmatrix} u\mathrm{e}^{\mathrm{i}\phi_S} \\ v \end{bmatrix} + d_\sigma \, \mathrm{e}^{-\mathrm{i}q_{hz}z} \begin{bmatrix} v\mathrm{e}^{\mathrm{i}\phi_S} \\ u \end{bmatrix} \, , \tag{S10}$$

where we explicitly also separated the electronlike and holelike components for reasons which will become clear later. The matching conditions then read

$$\left. \Psi_\sigma(z)\right|_{z=0_-} = \left. \Psi_\sigma(z)\right|_{z=0_+} \, , \tag{S11}$$

$$\left\{ \left[ \frac{\hbar^2}{2m} \frac{\mathrm{d}}{\mathrm{d}z} + \lambda_{\mathrm{SC}} \right] \mathrm{diag}[1, -1] + \lambda_{\mathrm{MA}} \mathrm{diag}[\sigma, \sigma] \right\} \left. \Psi_\sigma(z)\right|_{z=0_-} = \frac{\hbar^2}{2m} \frac{\mathrm{d}}{\mathrm{d}z} \mathrm{diag}[1, -1] \left. \Psi_\sigma(z)\right|_{z=0_+} \, . \tag{S12}$$

Using the fact that $\mathrm{Im}[q_{ez}(E_{\mathrm{B}})] = -\mathrm{Im}[q_{hz}(E_{\mathrm{B}})]$ for bound state energies $|E_{\mathrm{B}}| < |\Delta_{\mathrm{S}}|$, see Eq. (S5), the normalization of the wave functions [see Eqs. (S9)–(S10)] requires

$$1 \overset{(!)}{=} \left[ |u(E_{\mathrm{B}})|^2 + |v(E_{\mathrm{B}})|^2 \right] \left\{ \left[ |a_\sigma(E_{\mathrm{B}})|^2 + |b_\sigma(E_{\mathrm{B}})|^2 \right] + \left[ |c_\sigma(E_{\mathrm{B}})|^2 + |d_\sigma(E_{\mathrm{B}})|^2 \right] \right\} \int\limits_0^\infty \mathrm{d}z \, \mathrm{e}^{-2\mathrm{Im}[q_{ez}(E_{\mathrm{B}})]z}$$

$$= \frac{|u(E_{\mathrm{B}})|^2 + |v(E_{\mathrm{B}})|^2}{2\mathrm{Im}[q_{ez}(E_{\mathrm{B}})]} \left[ |a_\sigma(E_{\mathrm{B}})|^2 + |b_\sigma(E_{\mathrm{B}})|^2 + |c_\sigma(E_{\mathrm{B}})|^2 + |d_\sigma(E_{\mathrm{B}})|^2 \right] \, . \tag{S13}$$

Continuity of the wave functions at the interface, as imposed by Eq. (S11), relates the coefficients $c_\sigma(E_{\mathrm{B}})$ and $d_\sigma(E_{\mathrm{B}})$ in the right superconductor to $a_\sigma(E_{\mathrm{B}})$ and $b_\sigma(E_{\mathrm{B}})$ in the left superconductor, particularly,

$$c_\sigma(E_{\mathrm{B}}) = \alpha_1(E_{\mathrm{B}}) \, a_\sigma(E_{\mathrm{B}}) + \beta_1(E_{\mathrm{B}}) \, b_\sigma(E_{\mathrm{B}}) \, , \tag{S14}$$

$$d_\sigma(E_{\mathrm{B}}) = \alpha_2(E_{\mathrm{B}}) \, a_\sigma(E_{\mathrm{B}}) + \beta_2(E_{\mathrm{B}}) \, b_\sigma(E_{\mathrm{B}}) \, , \tag{S15}$$

where

$$\alpha_1(E_{\mathrm{B}}) = \frac{\mathrm{e}^{-\mathrm{i}\phi_S} - v(E_{\mathrm{B}})^2/u(E_{\mathrm{B}})^2}{1 - v(E_{\mathrm{B}})^2/u(E_{\mathrm{B}})^2} \, , \qquad \beta_1(E_{\mathrm{B}}) = \frac{v(E_{\mathrm{B}})/u(E_{\mathrm{B}}) \cdot \left[ \mathrm{e}^{-\mathrm{i}\phi_S} - 1 \right]}{1 - v(E_{\mathrm{B}})^2/u(E_{\mathrm{B}})^2} \, , \tag{S16}$$

$$\alpha_2(E_{\mathrm{B}}) = \frac{v(E_{\mathrm{B}})/u(E_{\mathrm{B}}) \cdot \left[ 1 - \mathrm{e}^{-\mathrm{i}\phi_S} \right]}{1 - v(E_{\mathrm{B}})^2/u(E_{\mathrm{B}})^2} \, , \qquad \beta_2(E_{\mathrm{B}}) = \frac{1 - v(E_{\mathrm{B}})^2/u(E_{\mathrm{B}})^2 \cdot \mathrm{e}^{-\mathrm{i}\phi_S}}{1 - v(E_{\mathrm{B}})^2/u(E_{\mathrm{B}})^2} \, . \tag{S17}$$

Applying the remaining boundary condition, Eq. (S12), we get $b_\sigma(E_{\mathrm{B}})$ in terms of $a_\sigma(E_{\mathrm{B}})$, i.e.,

$$b_\sigma(E_{\mathrm{B}}) = \gamma_\sigma(E_{\mathrm{B}}) \, a_\sigma(E_{\mathrm{B}}), \tag{S18}$$

where $\gamma_\sigma(E_{\mathrm{B}})$ reads

$$\gamma_\sigma(E_{\mathrm{B}}) = -\frac{\left[ -\frac{q_{ez}(E_{\mathrm{B}})}{q_F} - \mathrm{i}\overline{\lambda}_{\mathrm{SC}} - \mathrm{i}\sigma\overline{\lambda}_{\mathrm{MA}} - \frac{q_{ez}(E_{\mathrm{B}})}{q_F} \cdot \alpha_1(E_{\mathrm{B}})\mathrm{e}^{\mathrm{i}\phi_S} \right] u(E_{\mathrm{B}}) + \frac{q_{hz}(E_{\mathrm{B}})}{q_F} \cdot \alpha_2(E_{\mathrm{B}})\mathrm{e}^{\mathrm{i}\phi_S} \, v(E_{\mathrm{B}})}{\left[ \frac{q_{hz}(E_{\mathrm{B}})}{q_F} - \mathrm{i}\overline{\lambda}_{\mathrm{SC}} - \mathrm{i}\sigma\overline{\lambda}_{\mathrm{MA}} + \frac{q_{hz}(E_{\mathrm{B}})}{q_F} \cdot \gamma_2(E_{\mathrm{B}})\mathrm{e}^{\mathrm{i}\phi_S} \right] v(E_{\mathrm{B}}) - \frac{q_{ez}(E_{\mathrm{B}})}{q_F} \cdot \gamma_1(E_{\mathrm{B}})\mathrm{e}^{\mathrm{i}\phi_S} \, u(E_{\mathrm{B}})} \, . \tag{S19}$$

Eliminating the coefficients $b_\sigma(E_{\mathrm{B}})$, $c_\sigma(E_{\mathrm{B}})$, and $d_\sigma(E_{\mathrm{B}})$ in the normalization condition, Eq. (S13), by inserting Eqs. (S14), (S15), and (S18), we finally obtain

$$|a_\sigma(E_{\mathrm{B}})|^2 = \frac{2\mathrm{Im}[q_{ez}(E_{\mathrm{B}})] \Big/ \left[ |u(E_{\mathrm{B}})|^2 + |v(E_{\mathrm{B}})|^2 \right]}{\left[ 1 + |\gamma_\sigma(E_{\mathrm{B}})|^2 + |\alpha_1(E_{\mathrm{B}}) + \beta_1(E_{\mathrm{B}})\gamma_\sigma(E_{\mathrm{B}})|^2 + |\alpha_2(E_{\mathrm{B}}) + \beta_2(E_{\mathrm{B}})\gamma_\sigma(E_{\mathrm{B}})|^2 \right]} \tag{S20}$$

for the square of the absolute value of $a_\sigma(E_{\mathrm{B}})$. Going back with the same set of equations, we consecutively get all remaining coefficients $b_\sigma(E_{\mathrm{B}})$, $c_\sigma(E_{\mathrm{B}})$, and $d_\sigma(E_{\mathrm{B}})$.

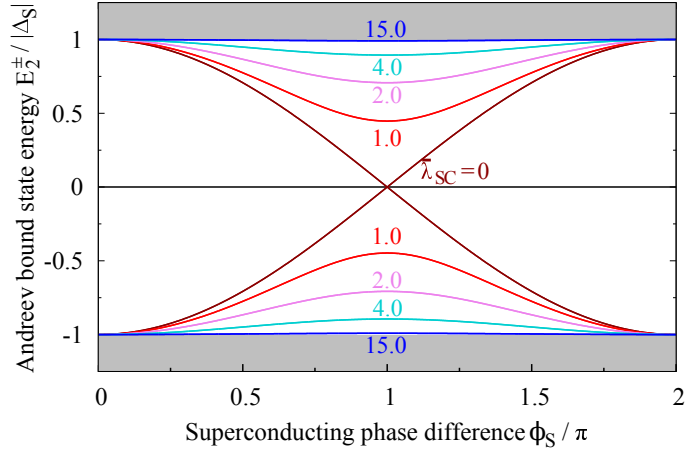

FIG. S2. Andreev bound state energies as a function of $\phi_S$ for junctions without interfacial magnetic tunneling ($\overline{\lambda}_{MA} = 0$) and Rashba SOC. Different scalar tunnelings $\overline{\lambda}_{SC}$ are color coded; all states are doubly spin-degenerate.

In the manuscript, we introduced electron-hole density differences to explain the connection between zero-energy YSR states and the Josephson current reversing $0$-$\pi$ transitions. To obtain those differences, $|\Psi_\sigma^e(z)|^2 - |\Psi_\sigma^h(z)|^2$, at an energy $E_B$ corresponding either to an Andreev or a YSR state, we use

$$\left|\Psi_\sigma^e(z \leq 0)\right|^2 = \left[|u(E_B)|^2 + |v(E_B)|^2\right] |a_\sigma(E_B)|^2 \, e^{2\text{Im}[q_{ez}(E_B)]z} \tag{S21}$$

and

$$\left|\Psi_\sigma^h(z \leq 0)\right|^2 = \left[|u(E_B)|^2 + |v(E_B)|^2\right] |b_\sigma(E_B)|^2 \, e^{2\text{Im}[q_{ez}(E_B)]z} \tag{S22}$$

valid in the left superconductor, and

$$\left|\Psi_\sigma^e(z \geq 0)\right|^2 = \left[|u(E_B)|^2 + |v(E_B)|^2\right] |c_\sigma(E_B)|^2 \, e^{-2\text{Im}[q_{ez}(E_B)]z} \tag{S23}$$

and

$$\left|\Psi_\sigma^h(z \geq 0)\right|^2 = \left[|u(E_B)|^2 + |v(E_B)|^2\right] |d_\sigma(E_B)|^2 \, e^{-2\text{Im}[q_{ez}(E_B)]z} \tag{S24}$$

valid in the right superconductor, respectively.

As mentioned, for each solution $\Psi$ with positive energy $E_B$, there is simultaneously also a second solution of the BdG equation with negative energy $-E_B$. The corresponding wave function $\tilde{\Psi}$ can be obtained from Eqs. (S9) and (S10) by interchanging $u$ to $-v^*$ and $v$ to $u^*$; for the corresponding tilde probabilities, $|\tilde{a}_\sigma|^2$, $|\tilde{b}_\sigma|^2$, $|\tilde{c}_\sigma|^2$, and $|\tilde{d}_\sigma|^2$, one obtains

$$\begin{aligned} |\tilde{a}_\sigma|^2 &= |a_{-\sigma}|^2, & |\tilde{b}_\sigma|^2 &= |b_{-\sigma}|^2, \\ |\tilde{c}_\sigma|^2 &= |c_{-\sigma}|^2, & |\tilde{d}_\sigma|^2 &= |d_{-\sigma}|^2. \end{aligned} \tag{S25}$$

Thus, the negative energy eigenstates $\tilde{\Psi}$ have not only opposite spin, but due to $u \mapsto -v^*$ and $v \mapsto u^*$ also the opposite relative sign between the electron and hole components when compared to the positive energy eigenstates $\Psi$.

## II. DERIVATION OF THE QUASIPARTICLE DOS

To compute the quasiparticle DOS in the system, we adapt McMillan's method [S2], which allows us to construct the spin-resolved retarded Green's function $G_\sigma^R(z, z'; E)$ directly from the incoming and scattered states. One can use

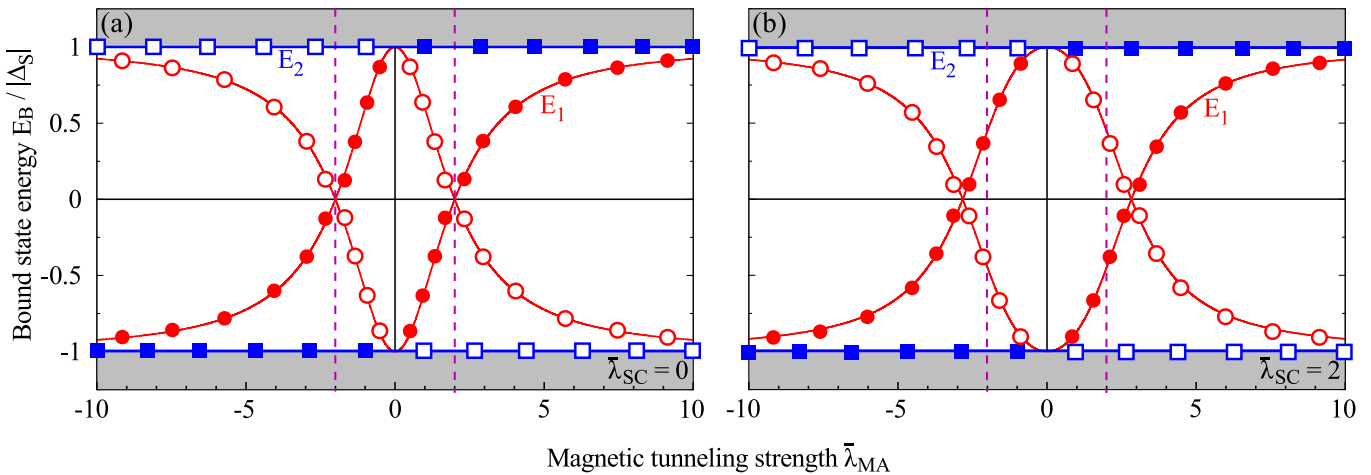

FIG. S3. Spin-resolved bound state energies $E_1$ and $E_2$ of the YSR (red) and the Andreev (blue) states as functions of the magnetic tunneling strength $\bar{\lambda}_{\text{MA}}$ at zero phase difference. Values of the interfacial scalar tunneling are $\bar{\lambda}_{\text{SC}} = 0$, panel (a), and $\bar{\lambda}_{\text{SC}} = 2$, panel (b); Rashba SOC is absent. The dashed vertical lines signify the positions of zero-energy YSR states for the junction in panel (a). For additional scalar tunneling at the interface, panel (b), the zero-energy states shift to larger values of $\bar{\lambda}_{\text{MA}}$, in accordance with Eq. (5) in the manuscript. Filled (empty) boxes indicate spin up (down) Andreev states and filled (empty) circles spin up (down) YSR states.

Eqs. (S9)–(S10) and perform their analytical continuation to get the true incoming and outgoing (scattered) states. After some tedious calculations, we end up with

$$
\begin{aligned}
G_\sigma^{\text{R}}(z, z'; E) = {} & -\frac{\mathrm{i}m}{\hbar^2 q_{\text{ez}} \left[ |u|^2 - |v|^2 \right]} \left\{ \left[ \mathrm{e}^{\mathrm{i}q_{\text{ez}}|z-z'|} + a_\sigma \mathrm{e}^{-\mathrm{i}q_{\text{ez}}(z+z')} \right] \begin{bmatrix} |u|^2 & uv^* \\ u^*v & |v|^2 \end{bmatrix} + b_\sigma \mathrm{e}^{\mathrm{i}(q_{\text{hz}}z - q_{\text{ez}}z')} \begin{bmatrix} u^*v & |v|^2 \\ |u|^2 & uv^* \end{bmatrix} \right\} \\
& -\frac{\mathrm{i}m}{\hbar^2 q_{\text{hz}} \left[ |u|^2 - |v|^2 \right]} \left\{ \left[ \mathrm{e}^{-\mathrm{i}q_{\text{hz}}|z-z'|} + b'_\sigma \mathrm{e}^{\mathrm{i}q_{\text{hz}}(z+z')} \right] \begin{bmatrix} |v|^2 & u^*v \\ uv^* & |u|^2 \end{bmatrix} + a'_\sigma \mathrm{e}^{\mathrm{i}(q_{\text{hz}}z' - q_{\text{ez}}z)} \begin{bmatrix} uv^* & |u|^2 \\ |v|^2 & u^*v \end{bmatrix} \right\}, \quad \text{(S26)}
\end{aligned}
$$

where $a_\sigma$ and $b_\sigma$ are the previously introduced reflection coefficients for an incoming electron-like quasiparticle, and $a'_\sigma$ as well as $b'_\sigma$ their counterparts in case of an incident hole-like quasiparticle.

Finally, the quasiparticle DOS (at energies $|E| \geq |\Delta_{\text{S}}|$) in the superconducting state is given by

$$
\text{DOS}_{\text{S}}(E) = -\frac{1}{\pi} \text{Im} \left\{ \text{Tr} \left[ G_\sigma^{\text{R}}(z, z; E) \right] \right\}. \quad \text{(S27)}
$$

This expression was evaluated in the manuscript close to the interface ($z = 0$), where we also normalized the quasiparticle DOS to the related one in the normal-conducting case, $\text{DOS}_{\text{N}}(E)$. The latter is obtained from the above expressions when setting $u(E) = 1$ and $v(E) = 0$.

## III. BOUND STATE SPECTRUM ANALYSIS AND ZERO-ENERGY STATES

In this section, we briefly analyze how further parameter regimes—not studied in the manuscript—would affect the bound state energies. We again distinguish the two important limiting cases: the *Andreev* and the *YSR* limit.

### A. Andreev limit

In the absence of magnetic tunneling, the subgap bound states are solely formed by the conventional Andreev states; see Eq. (3) in the manuscript. Generally, the energies of those states can be tuned either by changing the strength of scalar tunneling $\bar{\lambda}_{\text{SC}}$—which effectively corresponds to changing the degree of interfacial transparency—or by altering the superconducting phase difference $\phi_{\text{S}}$ by means of external magnetic fields. The Andreev subgap spectrum is shown in Fig. S2 as a function of $\phi_{\text{S}}$ and for different strengths of $\bar{\lambda}_{\text{SC}}$; since there is no magnetic tunneling and SOC is not present, each state is doubly spin degenerate. For perfectly transparent junctions, $\bar{\lambda}_{\text{SC}} = 0$, Eq. (3) in the manuscript

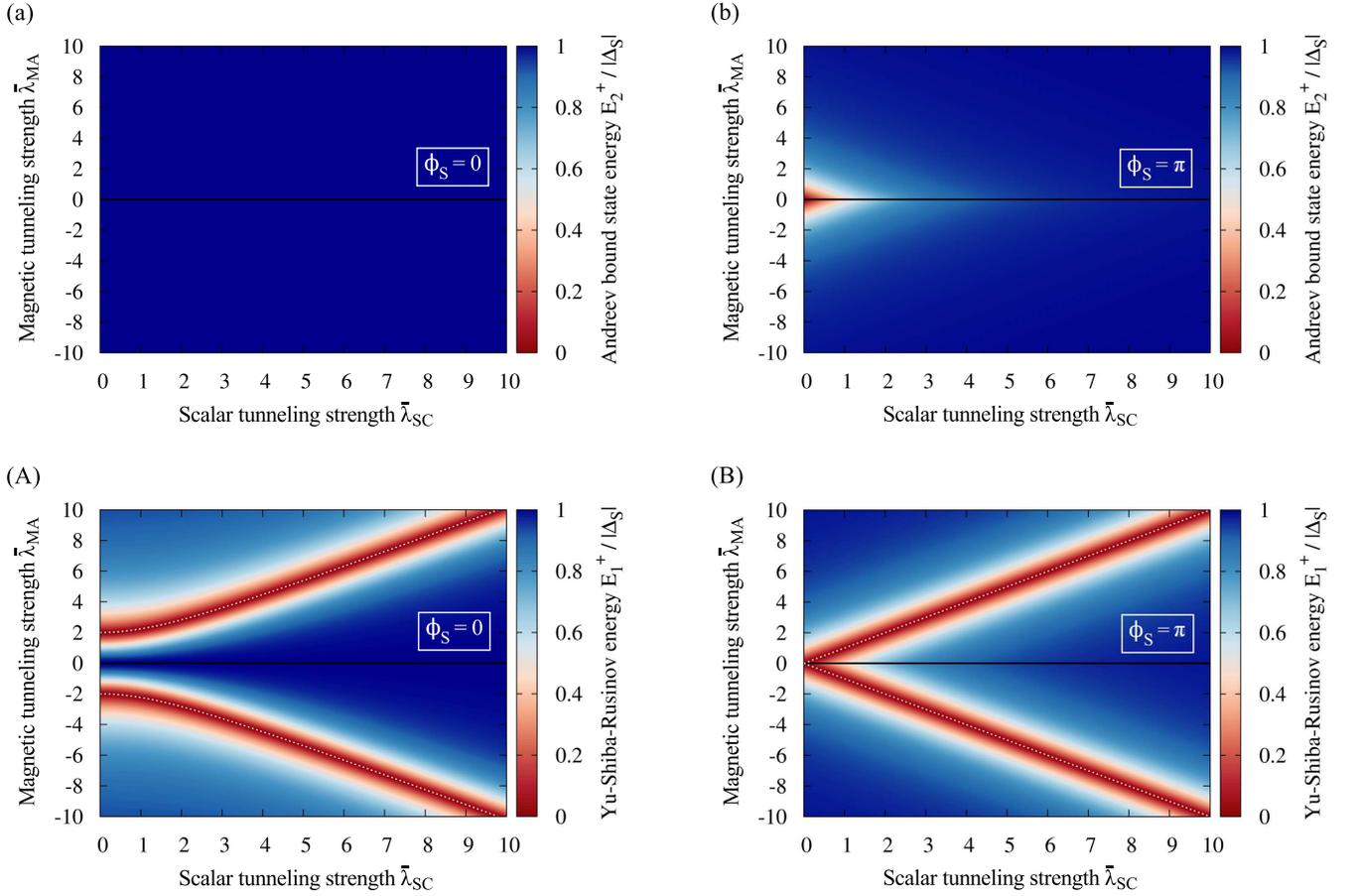

FIG. S4. (a) Bound state energies of the Andreev states as functions of $\bar{\lambda}_{SC}$ and $\bar{\lambda}_{MA}$ at fixed phase difference $\phi_S = 0$. The color code alters from darkred (gap center) to darkblue (gap edge) and we only show the positive bound state energies. (b) Same calculations for phase difference $\phi_S = \pi$. A clear formation of zero-energy Andreev states is visible for a small parameter region around the origin, referring to nearly perfectly transparent Josephson junctions. Panels (A) & (B) show similar calculations for the YSR states. Generally, the zero-energy YSR parameter space is much richer than the one for zero-energy Andreev states.

further simplifies to [S3, S4]

$$E_1^{\pm} = E_2^{\pm} = \pm |\Delta_S| \cos(\phi_S/2).$$ (S28)

As a consequence, the two Andreev branches cross at zero energy for all $\phi_S = \pi \,(\text{mod}\, 2\pi)$. The formation of such zero-energy states attracted recently lots of attention, mainly in connection with the emergent Majorana modes [S5–S9]. However, the main experimental obstacle acting against these zero-energy Andreev modes is the required perfect junction transparency, which is still challenging to achieve in practice. Already for small, but finite $\bar{\lambda}_{SC}$, the zero-energy states disappear and the ABSs shift towards energies close to the gap edges, approaching $\pm |\Delta_S|$ in the limit of $\bar{\lambda}_{SC} \to \infty$. Nevertheless, the situation can become different for Josephson junctions on top of topological insulators. In such systems, Fu and Kane predicted [S10] the existence of topologically protected zero-energy Majorana states, which show robustness against disorder and also extraordinary transport properties like, for example, a $4\pi$-periodic Josephson current.

## B. YSR limit

If there is no interfacial scalar tunneling and the S phase difference is tuned to zero, we effectively deal with a uniform bulk superconductor with an independent magnetic tunneling barrier in its center. In that case, we expect YSR bound states to form. Setting $\bar{\lambda}_{SC}$ and $\phi_S$ to zero, the bound state spectrum in Eq. (S8) reduces to the expression

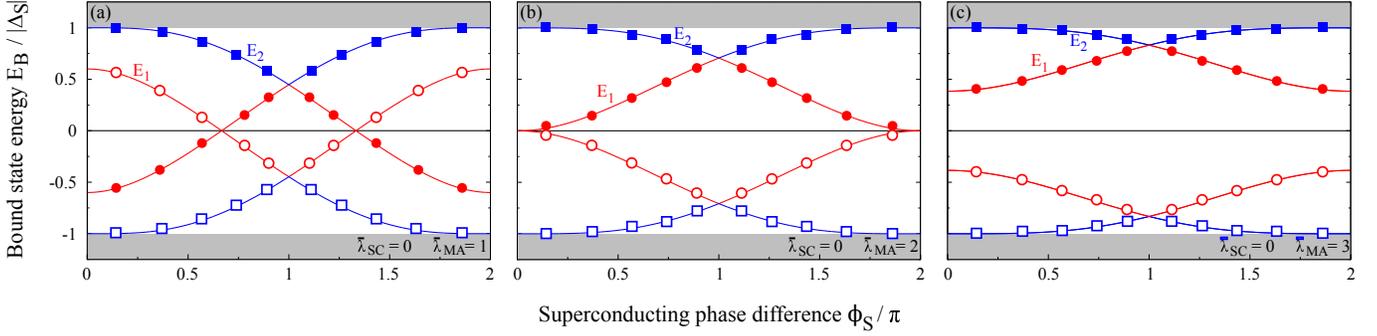

FIG. S5. Spin-resolved bound state energies $E_1$ and $E_2$ of the YSR (red) and the Andreev (blue) states as functions of $\phi_S$ for scalar tunneling strength $\overline{\lambda}_{SC} = 0$ and various magnetic tunnelings $\overline{\lambda}_{MA}$ given in the plots; Rashba SOC is absent. Filled (empty) boxes indicate spin up (down) Andreev states and filled (empty) circles spin up (down) YSR states.

given in the manuscript,

$$E_1^\pm = \pm |\Delta_S| \, \frac{\overline{\lambda}_{MA}^2 - 4}{\overline{\lambda}_{MA}^2 + 4} \quad \text{and} \quad E_2^\pm = \pm |\Delta_S| \, . \tag{S29}$$

Returning to dimensional $\lambda_{MA}$ and identifying $\rho(\mu) = 2m/(\pi\hbar^2 q_F)$ with the one-dimensional density of states per unit length at the Fermi level, one recovers in $E_1^\pm$ exactly the bound state energies obtained by Yu, Shiba, and Rusinov; see Refs. [S11–S13]. Simultaneously, there is still the $E_2^\pm$ branch, which is in this limit located at the gap edges, similarly as in the Andreev limit at $\phi_S = 0$. Because of that analogy, we introduced in the manuscript the convention to call the $E_1^\pm$ branch of the subgap spectrum the *YSR* and the $E_2^\pm$ branch the *Andreev* branch—even in the more general case when $\overline{\lambda}_{SC}$ and $\phi_S$ are nonzero.

The bound state spectrum corresponding to the YSR limit is displayed in Fig. S3(a). It shows the characteristic spin structure, whose qualitative discussion is provided in the manuscript; consult Fig. 3 there. The chosen strengths of $\overline{\lambda}_{MA}$ vary from $-10$ to $10$ and cover both antiferromagnetic and ferromagnetic coupling scenarios. For $\overline{\lambda}_{MA} = 0$, we recover the (conventional) doubly spin-degenerate ABSs located at the gap edges, while the energy branches start to split at finite $\overline{\lambda}_{MA}$. The YSR branches approach the center of the gap with increasing $\overline{\lambda}_{MA}$ and cross zero energy at $\overline{\lambda}_{MA}^{\text{crit.}} = 2$, where the YSR eigenstates individually 'flip' their spin. For stronger magnetic tunnelings, the YSR branches shift again towards the gap edges, resembling the situation in Fig. S2 for large $\overline{\lambda}_{SC}$. The Andreev branches are not affected in energy by a change of $\overline{\lambda}_{MA}$ at zero phase difference; also their spin does not change with an increase of $\overline{\lambda}_{MA}$ in sharp contrast to the YSR part of the spectrum.

## C. Interplay between Andreev and YSR states

After exploring the two limiting cases, we now concentrate on the combined situation. Motivated by Fig. S3(a), we again start with the case $\phi_S = 0$, i.e., an effective single bulk superconductor with scalar and magnetic tunneling in the center. In this case, Eq. (S8) reduces to

$$E_1^\pm = \pm |\Delta_S| \sqrt{\frac{\left(\overline{\lambda}_{SC}^2 - \overline{\lambda}_{MA}^2 + 4\right)^2}{\left(\overline{\lambda}_{SC}^2 - \overline{\lambda}_{MA}^2 + 4\right)^2 + 16\overline{\lambda}_{MA}^2}} \tag{S30}$$

for the YSR branch and

$$E_2^\pm = \pm |\Delta_S| \tag{S31}$$

for the Andreev branch. The calculated bound state spectrum in the case of generally nonzero $\overline{\lambda}_{SC}$'s and $\overline{\lambda}_{MA}$'s is displayed in Fig. S3(b) for $\overline{\lambda}_{SC} = 2$. Comparing Fig. S3(b) with the pure YSR limit in Fig. S3(a), we see that the additional presence of interfacial scalar tunneling does not qualitatively change the spectral characteristics and the spin structure of the underlying bound states. However, the presence of scalar tunneling naturally shifts the YSR

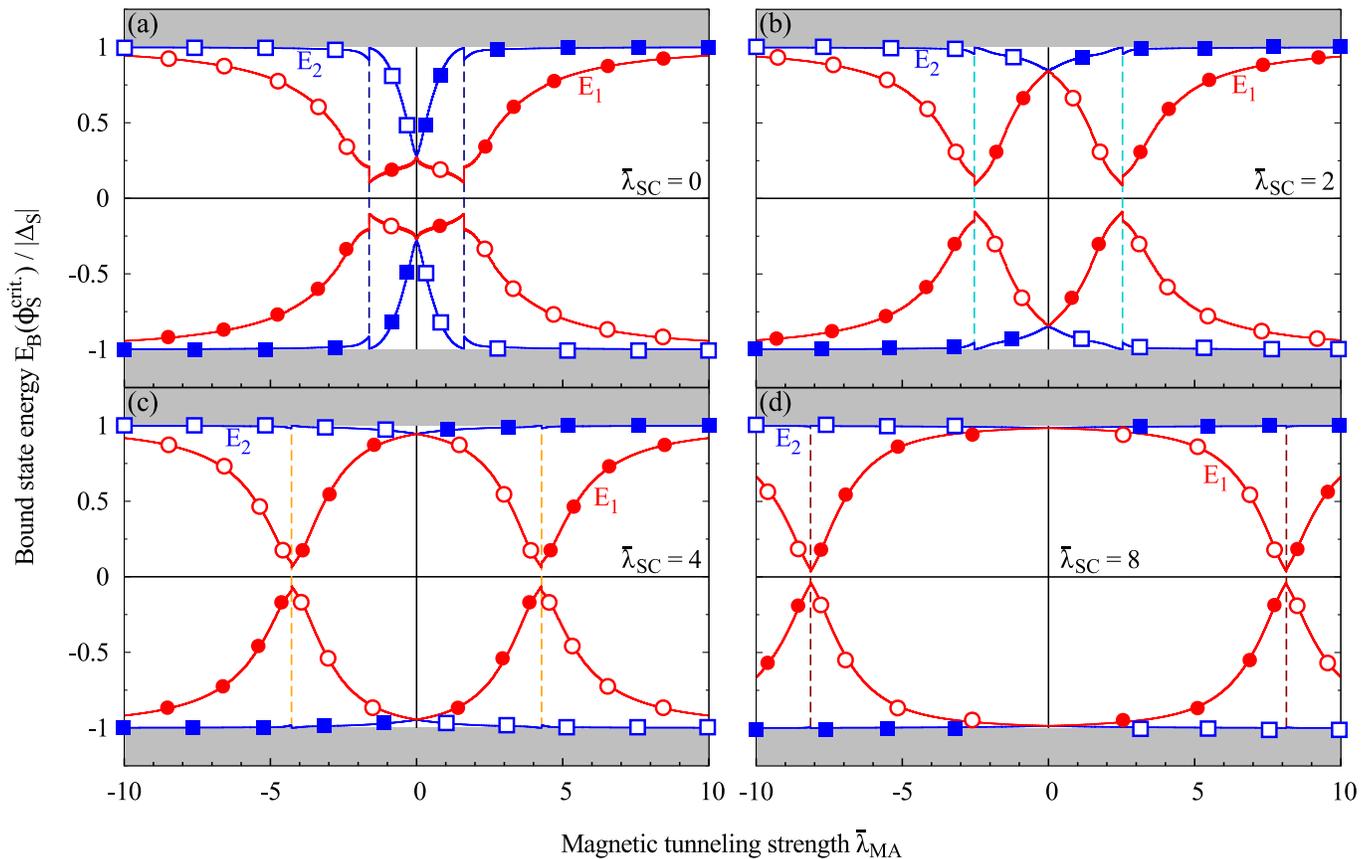

FIG. S6. Spin-resolved bound state energies $E_1$ and $E_2$ of the YSR (red) and the Andreev (blue) states as functions of the magnetic tunneling strength $\bar{\lambda}_{\mathrm{MA}}$ at the critical phase difference $\phi_{\mathrm{S}}^{\mathrm{crit.}}$. Values of the interfacial scalar tunneling are (a) $\bar{\lambda}_{\mathrm{SC}} = 0$, (b) $\bar{\lambda}_{\mathrm{SC}} = 2$, (c) $\bar{\lambda}_{\mathrm{SC}} = 4$, and (d) $\bar{\lambda}_{\mathrm{SC}} = 8$; Rashba SOC is absent. The dashed vertical lines emphasize the positions of magnetic tunneling-induced transitions from 0 to $\pi$ states. Filled (empty) boxes indicate spin up (down) Andreev states and filled (empty) circles spin up (down) YSR states.

branch more to the gap edges [S12] and, as another consequence, the points of zero-energy YSR crossings are shifted to larger $\bar{\lambda}_{\mathrm{MA}}$'s—in the presented case to $\bar{\lambda}_{\mathrm{MA}}^{\mathrm{crit.}} = 2\sqrt{2}$ in accordance with Eq. (5) in the manuscript. In comparison with the Andreev limit, we see a general and experimentally positive trend: the presence of magnetic barriers gives rise to the formation of zero-energy YSR states, which are much more robust against "tunneling disorder" than their Andreev counterparts.

To illustrate the robustness of the zero-energy subgap states, Fig. S4 shows the spectrum for a wide range of tunneling parameters. To be specific, we show the positive Andreev and YSR bound state energies $E_2^+$ and $E_1^+$ as functions of $\bar{\lambda}_{\mathrm{SC}}$ and $\bar{\lambda}_{\mathrm{MA}}$ for $\phi_{\mathrm{S}} = 0$ and $\phi_{\mathrm{S}} = \pi$, respectively. The corresponding negative counterparts of the spectrum (not shown) are simply $E_1^- = -E_1^+$ and $E_2^- = -E_2^+$. As noticed in Sec. III A, the ABSs are always located at the gap edges if there is no phase difference across the junction, independently of the actual strength of $\bar{\lambda}_{\mathrm{SC}}$. The continuously darkblue colored region in Fig. S4(a) shows the same behavior, now for any scalar and magnetic tunneling strength. If $\phi_{\mathrm{S}}$ gets tuned to $\pi$, see Fig. S4(b), and both scalar and magnetic perturbations are absent, we recover the mentioned zero-energy Andreev states (darkred region), which are themselves extremely sensitive to a weak perturbation in terms of finite $\bar{\lambda}_{\mathrm{SC}}$ or $\bar{\lambda}_{\mathrm{MA}}$. The situation turns out to become more interesting for the YSR states, see Figs. S4(A) and (B). From Eq. (5) in the manuscript, we deduce that one can always find a fixed $\bar{\lambda}_{\mathrm{MA}}$ for fixed values of scalar tunneling strength and S phase difference such that the YSR branches cross at the gap center, giving rise to zero-energy YSR states. The related parameter space for zero-energy YSR states is thus indeed much richer than for the conventional Andreev states as indicated by the darkred regions in Figs. S4(A) and (B). For better visualization, the analytical condition for the formation of zero-energy YSR states, see Eq. (5) in the manuscript, is included as white dashed lines.

When discussing the phase dependence of the spectrum in the manuscript, we did not show the situation for perfectly transparent junctions, $\bar{\lambda}_{\mathrm{SC}} = 0$, as this is experimentally rather hard achievable. The related results are

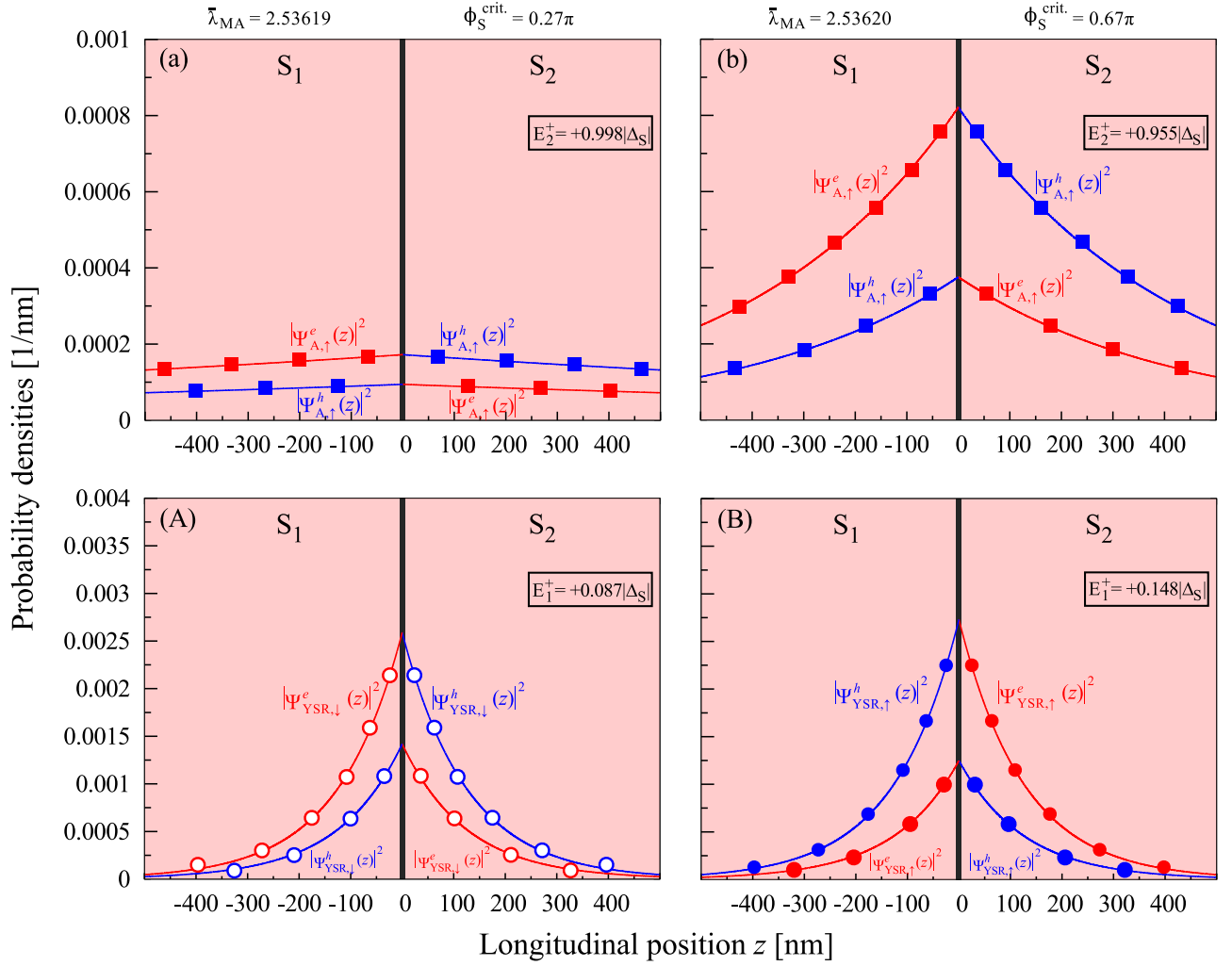

FIG. S7. Spatial profiles of spin-resolved and electron-hole separated probability densities for the Andreev (upper panel) and YSR (lower panel) states with positive energies. Scalar tunneling is fixed at $\bar{\lambda}_{SC} = 2$ and the values of $\bar{\lambda}_{MA}$ and $\phi_S$ are tuned to points just before and after the 0-$\pi$ transition; Rashba SOC is not present. The figures at the left, panels (a) and (A), show the corresponding densities around the interface up to 500 nm for the 0 state, whereas the figures at the right, panels (b) and (B), show analogous data for the $\pi$ state. The related junction parameters are listed at the upper margin; filled (empty) boxes indicate spin up (down) Andreev and filled (empty) circles spin up (down) YSR states. Clear signatures for the 0-$\pi$ transition can be observed in the wave function behavior of the YSR states: *the change of the spin ordering and the reversal of electron-hole content.*

displayed in Fig. S5. Both the qualitative characteristics of the subgap states and their spin structure are the same as described in the manuscript. However, it is worth to mention that the Andreev and YSR states always cross at $\phi_S = \pi$ as long as scalar tunneling is absent. The energies at which the two branches cross follow from the general spectrum in Eq. (S8) and are given by

$$E_{cross}^{\pm} = \pm |\Delta_S| \sqrt{\frac{\bar{\lambda}_{MA}^2}{\bar{\lambda}_{MA}^2 + 4}}. \tag{S32}$$

## IV. SPECTRUM AT THE CRITICAL PHASE DIFFERENCE

As briefly mentioned in the manuscript, it is more common from the experimental point of view to investigate the bound state spectrum by means of STM/STS not at zero superconducting phase difference, but instead at the critical phase difference, at which the maximal Josephson current flows through the system. Therefore, we present in Fig. S6

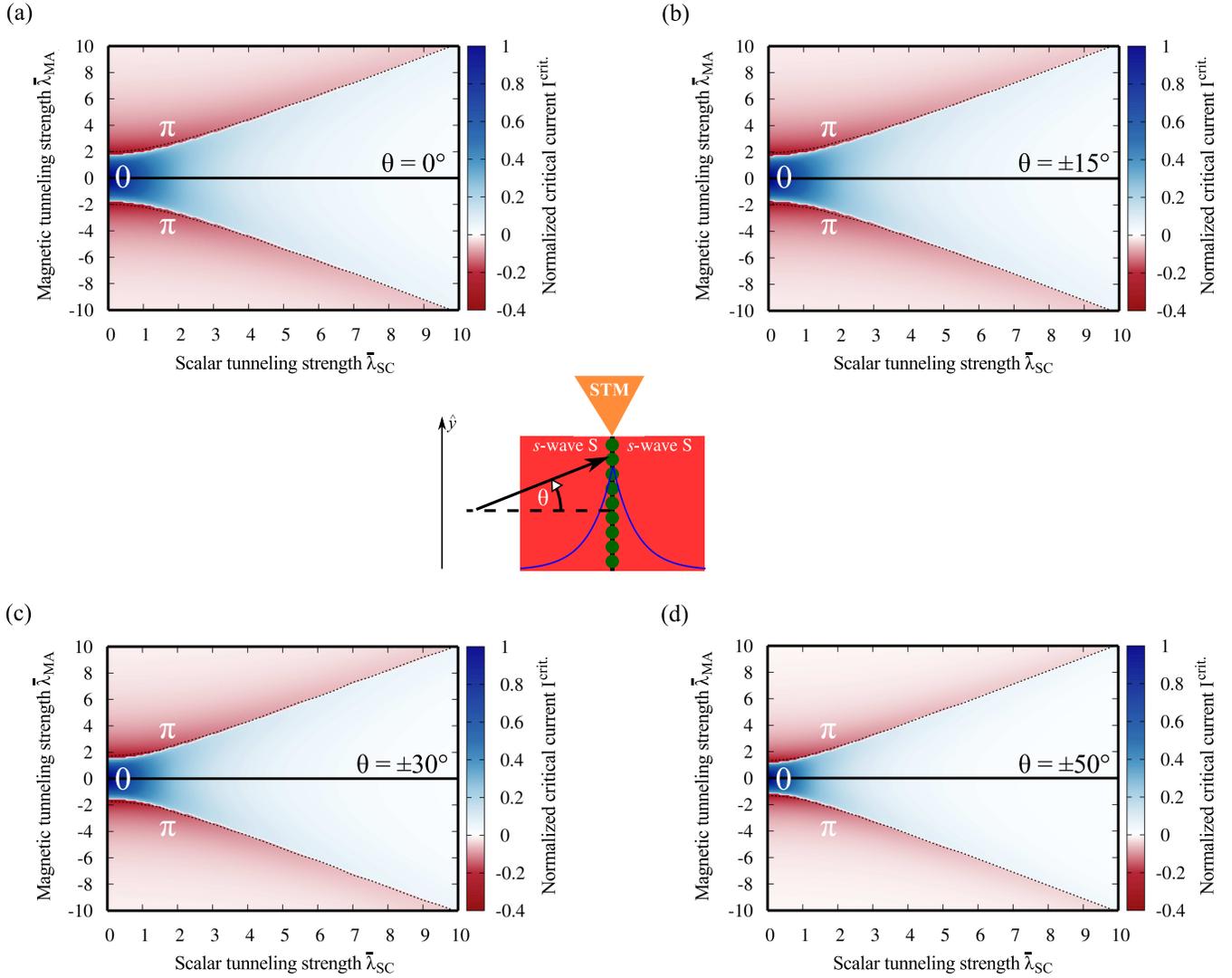

FIG. S8. Contour plot of the normalized critical current as a function of $\bar{\lambda}_{SC}$ and $\bar{\lambda}_{MA}$ in the absence of Rashba SOC at zero temperature. Blue and red regions represent 0 and $\pi$ junction regimes. The transition lines separating the two phases are displayed by white borderlines; the parameters at which zero-energy YSR states emerge are shown by dashed lines. We consider effective two-dimensional junctions in which the incident quasiparticles enclose the angles (a) $\theta = 0°$, (b) $\theta = \pm15°$, (c) $\theta = \pm30°$, and (d) $\theta = \pm50°$ with the interface normal; for the definition of $\theta$, see the illustration in the center.

the energy level calculations as a function of the magnetic tunneling strength also for that particular scenario and for various scalar tunneling strengths. The qualitative features are the same as explained at zero phase difference. The 'spin flips' at the 0-$\pi$ transitions points (additionally emphasized by dashed lines) again fingerprint the quantum phase transition between the junctions' 0 and $\pi$ states.

## V. WAVE FUNCTION ANALYSIS

In the manuscript, we qualitatively explain the physical connection between the formation of zero-energy YSR states—indicating a quantum phase transition in the systems' ground states—and the Josephson current reversing 0-$\pi$ transitions. Therefore, we analyzed the difference between electronlike and holelike densities in the bound state wave functions, see Fig. 3 in the manuscript, and directly connected them to the Cooper pair transporting channels in the junction. For completeness, we here show the spatial profiles of the wave functions' electron and hole components. The procedure how to obtain $|\Psi_{A,\sigma}^{e\,(h)}(z)|^2$ and $|\Psi_{YSR,\sigma}^{e\,(h)}(z)|^2$ for the Andreev and YSR states, respectively, is summarized

in Sec. I. We checked various parameter regimes, observing the same qualitative features; therefore, we just present the results for one particular scalar tunneling strength, i.e., for $\bar{\lambda}_{SC} = 2$; see Fig. S7. The values of $\bar{\lambda}_{MA}$ are chosen slightly below and above the 0-$\pi$ transition point, which for the regarded junction at zero temperature is numerically estimated to lie between 2.53619 and 2.53620. The ABSs (upper panel) are spatially less localized than their YSR counterparts (lower panel) since their energies are typically closer to the gap edges ($0 \lesssim E_1^+ < E_2^+ \lesssim |\Delta_S|$) and the plotted densities decay with the decay length $\kappa = 1/\{2\mathrm{Im}[q_{ez}(E_B)]\}$, which increases with increasing energy. The second important observation is that the spatial electron-hole content of $\Psi_{A,\sigma}(z)$ and $\Psi_{YSR,\sigma}(z)$ changes. While the electronlike character dominates at the left and the holelike character at the right of the interface for both the Andreev and YSR states in the 0 state, the YSR states 'flip' their spin and—since the effective longitudinal momentum particles experience at the magnetic interface is spin-dependent through the $\bar{\lambda}_{MA}$-terms in our model Hamiltonian—also the electron-hole content of the states' wave functions reverses in the $\pi$ state; see calculations provided in Sec. I. This reversal leads to the reversal of the Josephson current flow as shown in Fig. 3 in the manuscript.

## VI. TWO-DIMENSIONAL AND THREE-DIMENSIONAL JUNCTIONS

In the main text, we focused for simplicity on quasi-one-dimensional junctions grown along the $\hat{z}$ direction, which allowed us to find closed analytical formulas. Here, we emphasize that the discussed characteristics—the connection between zero-energy YSR states and 0-$\pi$ transitions—persist also in two-dimensional or three-dimensional junctions. Figure S8 shows the normalized critical current for a two-dimensional junction with a nonzero incident transverse momentum along the $\hat{y}$ direction. This transverse momentum is described by an angle $\theta$, which the incoming electronlike quasiparticle forms with the interface normal. We present numerical calculations for $\theta = 0°$, $\theta = \pm 15°$, $\theta = \pm 30°$, and $\theta = 50°$, respectively. With increasing $\theta$, the critical $\bar{\lambda}_{MA}$ supporting zero-energy YSR states slightly decreases. This can be anticipated from the analytical formulas for the one-dimensional case provided in the manuscript. The tunneling parameters there need to be replaced by effective ones, $\bar{\lambda}_{SC} \mapsto \bar{\lambda}_{SC}/\sqrt{1 - \sin^2\theta}$ and $\bar{\lambda}_{MA} \mapsto \bar{\lambda}_{MA}/\sqrt{1 - \sin^2\theta}$, to treat two-dimensional junctions (and analogously for three-dimensional junctions). As mentioned, the coincidence between forming zero-energy YSR states and the 0-$\pi$ transitions remains clearly visible also for nonnormal incidence.

## VII. COMPARISON WITH S/F/S JUNCTIONS

To anticipate 0-$\pi$ transition differences in conventional S/F/S Josephson junctions with finite F layer thickness and the ultrathin ballistic S/FI/S Josephson junctions investigated in the main text, we compare the critical currents in both systems for comparable model parameters.

To model S/F/S junctions, we adapted the approach elaborated in Ref. [S14]. The system is additionally shown in the inset of Fig. S9. It consists of a F layer grown in the $\hat{z}$ direction with a fixed longitudinal thickness $d_{XC}$ (= 15 nm in the forthcoming calculation to mimic the thickness of realistic junctions) and the variable exchange splitting $\Delta_{XC}$; the exchange part is modeled by the Hamiltonian $\hat{H}_{XC}(z) = (\Delta_{XC}/2)\,\Theta(z)\,\Theta(d_{XC} - z)$. Tunneling at S/F and F/S interfaces, located at $z = 0$ and $z = d_{XC}$, is assumed to be symmetric and governed by the Hamiltonian $\hat{H}_{SC}(z) = V_0 d_0 [\delta(z) + \delta(d_{XC} - z)]$; we assume fixed $d_0$ (= 2.5 nm in the forthcoming calculation to mimic realistic tunnel barriers) and scalar tunneling with varying barrier height $V_0$. To adjust the ranges of $\Delta_{XC}$ and $V_0$ to those that describe quantitatively the S/FI/S Josephson junctions, $\bar{\lambda}_{MA} \in [-10, 10]$ and $\bar{\lambda}_{SC} \in [0, 10]$ [see Fig. 4 in the main text], we need to fulfill

$$-10 \leq \frac{2m}{\hbar^2 q_F}\frac{\Delta_{XC}}{2}\,d_{XC} \leq 10 \qquad \text{and} \qquad 0 \leq \frac{2m}{\hbar^2 q_F} V_0\,d_0 \leq 10\,. \tag{S33}$$

In what follows, we use for $m$ the free electron mass and for $q_F$ the Fermi wave vector of iron [S15]. The parameter ranges of $\Delta_{XC}$ and $V_0$ in units of eV that are set up by the above inequalities are highlighted by the green box in Fig. S9; the contour plot represents the normalized critical current of the S/F/S junction as a function of $\Delta_{XC}$ and $V_0$ and at zero temperature.

As expected from the general analysis of S/F/S junctions—see, for example, Refs. [S16, S17]—a leakage of Cooper pairs along the F layer brings nontrivial superconducting order into the F region. Since the Cooper pairs experience a spin imbalance in the F due to the exchange splitting, the related order parameter therein spatially oscillates—a famous FFLO effect [S18, S19]. In the most simple situation (no scalar tunneling and zero temperature), the oscillation length $\xi_p$ solely depends on the strength of the exchange splitting, $\xi_p \sim \hbar v_F/\Delta_{XC}$ (where $v_F$ stands for the Fermi velocity) [S16, S17, S20]; if $d_{XC}/\xi_p \approx 2n$, where $n$ is a positive integer, the phase difference accumulates an additional $\pi$ shift. Consequently, the order parameter has opposite relative sign at the two F/S interfaces and

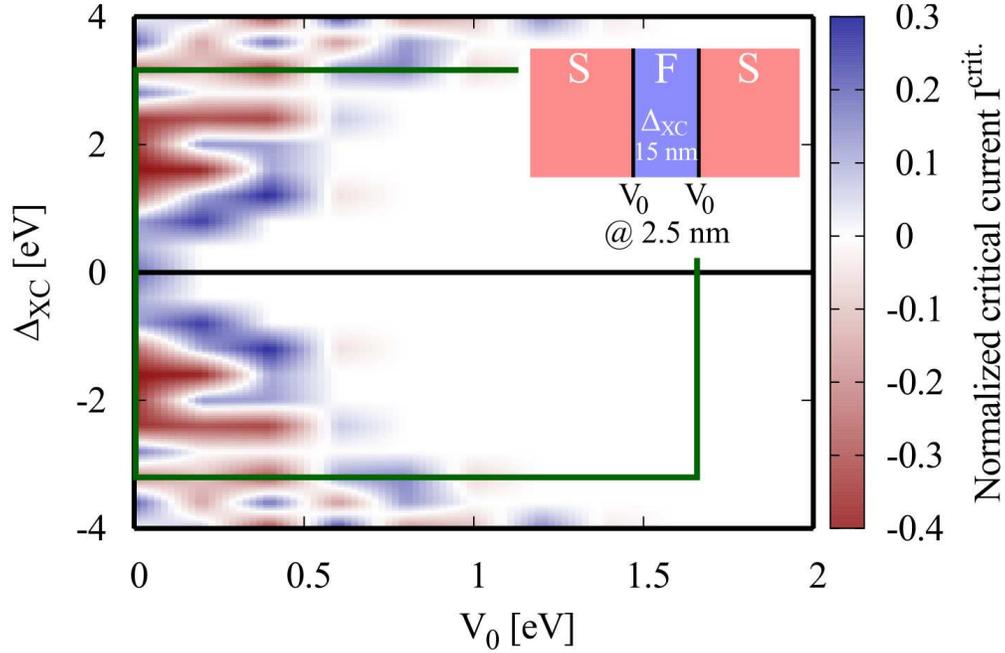

FIG. S9. Contour plot of the normalized critical current in the S/F (15 nm)/S junction as a function of the tunneling barrier height $V_0$ and the exchange splitting $\Delta_{\mathrm{XC}}$ in the F layer. The system is schematically shown in the inset; for details of the calculation, consult Ref. [S14]. To compare the results with the S/FI/S junction, see Fig. 4 in the manuscript, the equivalent ranges of $V_0$ and $\Delta_{\mathrm{XC}}$ [limited by the inequalities in Eq. (S33)] are highlighted by the darkgreen box. In contrast to the S/FI/S scenario, there are several distinct 0 and $\pi$ state "islands" in the S/F/S junctions' parameter space, indicated by blue and red colors, respectively.

the junction switches from the 0 into the $\pi$ state regime. We indeed observe in our numerical calculation already at $V_0 = 0$ the expected characteristic oscillations of the critical current when modulating the strength of the exchange splitting; see Fig. S9. Finite $V_0$ gives rise to additional geometrical oscillations [S21] in the current, additionally impacting the 0-$\pi$ transition regions. As $\Delta_{\mathrm{XC}}$ increases, $\xi_{\mathrm{p}}$ decreases and the junction switches into the $\pi$ regime as soon as $d_{\mathrm{XC}}$ and $\xi_{\mathrm{p}}$ satisfy the aforementioned criterion. In strong contrast to the S/FI/S case studied in the main text, where the 0 state regime at low $\overline{\lambda}_{\mathrm{MA}}$ and the $\pi$ state regime at large $\overline{\lambda}_{\mathrm{MA}}$ become separated by the separatrix consisting of zero-energy YSR states, the 0-$\pi$ transitions in S/F/S systems are controlled by the unique interplay of the exchange splitting and the thickness of the F layer. Consequently, 0 and $\pi$ state regimes form (the blue and red) "islands" in the $(V_0; \Delta_{\mathrm{XC}})$ parameter space. These "islands" are separated by complex 0-$\pi$ transition lines, which must be attributed to the nontrivial proximity effect in the F layer and are presumably not related to the bound state spectroscopic features. These considerations clearly indicate that the physical mechanism behind the 0-$\pi$ transitions in S/FI/S junctions is indeed distinct from that established in the conventional S/F/S scenario [S22].

[S1]  P. G. De Gennes, *Superconductivity of Metals and Alloys* (Addison Wesley, Redwood City, 1989).

[S2]  W. L. McMillan, Phys. Rev. **175**, 559 (1968).

[S3]  C. W. J. Beenakker, Phys. Rev. Lett. **67**, 3836 (1991).

[S4]  A. A. Golubov, M. Y. Kupriyanov, and E. Il'ichev, Rev. Mod. Phys. **76**, 411 (2004).

[S5]  Y. Oreg, G. Refael, and F. von Oppen, Phys. Rev. Lett. **105**, 177002 (2010).

[S6]  M. Duckheim and P. W. Brouwer, Phys. Rev. B **83**, 054513 (2011).

[S7]  V. Mourik, K. Zuo, S. M. Frolov, S. R. Plissard, E. P. A. M. Bakkers, and L. P. Kouwenhoven, Science **336**, 1003 (2012).

[S8]  L. P. Rokhinson, X. Liu, and J. K. Furdyna, Nat. Phys. **8**, 795 (2012).

[S9]  S. Nadj-Perge, I. K. Drozdov, J. Li, H. Chen, S. Jeon, J. Seo, A. H. MacDonald, B. A. Bernevig, and A. Yazdani, Science **346**, 602 (2014).

[S10]  L. Fu and C. L. Kane, Phys. Rev. Lett. **100**, 096407 (2008).

[S11]  L. Yu, Acta Phys. Sin. **21**, 75 (1965).

[S12]  H. Shiba, Progr. Theoret. Phys. **40**, 435 (1968).

[S13] A. I. Rusinov, Zh. Eksp. Teor. Fiz. **9**, 146 (1968); JETP Lett. **9**, 85 (1969).

[S14] A. Costa, P. Högl,  and J. Fabian, Phys. Rev. B **95**, 024514 (2017).

[S15] J. P. Carbotte, Rev. Mod. Phys. **62**, 1027 (1990).

[S16] A. V. Andreev, A. I. Buzdin,  and R. M. Osgood, Phys. Rev. B **43**, 10124 (1991).

[S17] E. A. Demler, G. B. Arnold,  and M. R. Beasley, Phys. Rev. B **55**, 15174 (1997).

[S18] P. Fulde and R. A. Ferrell, Phys. Rev. **135**, A550 (1964).

[S19] I. Larkin and Y. N. Ovchinnikov, Zh. Eksp. Teor. Fiz. **47**, 1136 (1964); Sov. Phys. JETP **20**, 762 (1965).

[S20] V. V. Ryazanov, V. A. Oboznov, A. Y. Rusanov, A. V. Veretennikov, A. A. Golubov,  and J. Aarts, Phys. Rev. Lett. **86**, 2427 (2001).

[S21] A. Brinkman and A. A. Golubov, Phys. Rev. B **61**, 11297 (2000).

[S22] S. Kawabata, Y. Asano, Y. Tanaka, A. A. Golubov,  and S. Kashiwaya, Phys. Rev. Lett. **104**, 117002 (2010).



\end{document}